\pdfoutput=1

\documentclass{article}

\newcommand{\ours}{NCCL-BOT}
\newcommand{\ourslong}{Bandit Online Tuning}
\newcommand{\oursintro}{\ours{} (\ourslong{})}

\providecommand*{\paperbuild}{arxiv}  %

\PassOptionsToPackage{hyphens}{url}

\usepackage{graphicx}
\usepackage{subfigure}
\usepackage{booktabs}
\usepackage{array}
\usepackage{tabularx}
\usepackage{supertabular}
\usepackage{amsmath,amssymb,amsthm}
\usepackage{xcolor}
\usepackage{enumitem}
\usepackage{algorithm}
\usepackage{algpseudocode}
\usepackage{hyperref}
\usepackage{tikz}
\usetikzlibrary{arrows.meta, positioning, fit, calc, backgrounds}
\makeatletter
\def\@paperbuild@arxiv{arxiv}
\def\@paperbuild@blind{blind}
\def\@paperbuild@camera{camera}
\newif\ifarxivbuild
\newif\ifblindbuild
\ifx\paperbuild\@paperbuild@blind
  \blindbuildtrue
  \usepackage[nohyperref]{mlsys2025}
\else
  \usepackage[accepted,nohyperref]{mlsys2025}
  \ifx\paperbuild\@paperbuild@arxiv
    \arxivbuildtrue
    \renewcommand{\Notice@String}{Preprint.}
    \renewcommand{\mlsysEqualContribution}{\textsuperscript{*}Equal contribution. }
    \renewcommand{\@pa}[1]{%
      \ifcsname the@affil#1\endcsname
      \else
        \ifcsname @icmlsymbol#1\endcsname
        \else
          \stepcounter{@affiliationcounter}%
          \newcounter{@affil#1}%
          \setcounter{@affil#1}{\value{@affiliationcounter}}%
        \fi
      \fi
      \ifcsname @icmlsymbol#1\endcsname
        \textsuperscript{\csname @icmlsymbol#1\endcsname\,}%
      \fi
    }
    \renewcommand{\printAffiliationsAndNotice}[1]{%
      \stepcounter{@affiliationcounter}%
      {\let\thefootnote\relax\footnotetext{\hspace*{-\footnotesep}\ifdefined\isaccepted #1\fi
      \ifdefined\mlsyscorrespondingauthor@text
      Correspondence to: \mlsyscorrespondingauthor@text.
      \else
      {\bf AUTHORERR: Missing \textbackslash{}mlsyscorrespondingauthor.}
      \fi
      \ \\
      \Notice@String
      }%
      }%
    }
  \else
    \ifx\paperbuild\@paperbuild@camera
    \else
      \PackageError{main}{Unknown \string\paperbuild=\paperbuild; use arxiv, blind, or camera}{}
    \fi
  \fi
\fi
\makeatother
\ifblindbuild
  \newcommand{\pagefooter}{\small\thepage}
\else
  \newcommand{\pagefooter}{\small\thepage\,/\,\pageref{LastPage}}
\fi
\AtBeginDocument{%
  \setlength{\footskip}{30pt}%
  \fancyfoot[C]{\pagefooter}%
  \fancypagestyle{plain}{%
    \fancyhf{}%
    \fancyfoot[C]{\pagefooter}%
  }%
}
\hypersetup{
  colorlinks=true,
  linkcolor=blue!60!black,
  citecolor=blue!60!black,
  urlcolor=blue!60!black,
}
\setcitestyle{numbers,sort&compress}

\newcommand{\dashfill}{\leavevmode\cleaders\hbox{\rule[0.35ex]{3pt}{0.4pt}\hskip2.5pt}\hfill\kern0pt}
\newcommand{\RLTlong}{Relative Lost Time}
\newcommand{\RLT}{\emph{RLT}}
\newcommand{\RLTintro}{\RLTlong{} (\RLT)}

\newcommand{\CRLT}{\emph{CRLT}}

\newcommand{\TRLTlong}{Total Relative Lost Time}
\newcommand{\TRLT}{\emph{TRLT}}
\newcommand{\TRLTintro}{\TRLTlong{} (\TRLT)}
\newcommand{\AUTOlong}{Automatic Cost-Model Selection}
\newcommand{\NCCLAUTO}{NCCL \emph{AUTO} (\AUTOlong{})}
\newcommand{\AUTO}{\emph{AUTO}}
\newcommand{\MABlong}{Multi-Armed Bandit}
\newcommand{\MAB}{\emph{MAB}}
\newcommand{\MABintro}{\MABlong{} (\MAB)}

\newcommand{\KPI}{\emph{KPI}}

\newcommand{\SEM}{\emph{SEM}}
\newcommand{\PBest}{$P(\text{best arm})$}
\newcommand{\capsplit}{scale-out-of-distribution split: tuned or fit on the 4/8-node contexts, scored on the 154
  held-out contexts (all 16/32-node contexts plus eight trace contexts; a context is a collective, node count and
  message size).}
\newcommand{\capTRLT}{\TRLT{} (total relative lost time, \%) is the latency a policy serves over $T{=}100$ steps of a
  context in excess of the measured-best arm's, relative to the latter.}
\newcommand{\capreal}{Realized \TRLT{} charges the latencies actually served, first-use transients included; selection-priced
  \TRLT{} charges each step the chosen arm's steady-state mean.}
\newcommand{\capptail}{p99 is the 99th percentile over contexts and $n_{>10}$ the number of contexts whose \TRLT{} exceeds
  10\%.}
\newcommand{\capTRLTany}{\TRLT{} (total relative lost time) is the latency a policy serves in excess of the
  measured-best arm's, relative to the latter.}
\newcommand{\capkey}{Key: \TRLT{} is the latency served over $T{=}100$ steps of a context in
  excess of the measured-best arm's, relative to it; realized pricing charges the latencies actually served, selection
  pricing the chosen arm's steady-state mean; test is the 154 held-out contexts (16/32 nodes plus eight trace contexts)
  of policies tuned on 4/8 nodes; p99 is over contexts; pp: percentage points.}
\newcommand{\capbold}{Bold: best value per column (lower is better).}

\mlsystitlerunning{\ours{}: Closing the Loop on NCCL's Cost Model}

\graphicspath{{figures/}}
\newcommand{\cclBenchCalls}{719}

\newcommand{\cclBenchContextsWord}{nine}
\newcommand{\cclBenchMeanReuse}{80}
\newcommand{\cclBenchTopTwoTrafficPct}{90}
\newcommand{\cclBenchMeanReuseDef}{mean reuse ($\#\text{calls}/\#\text{contexts}$)}

\newcommand{\CollectiveThirtyTwoNcclMeanTrltPct}{4.67}
\newcommand{\CollectiveThirtyTwoNcclPNineNineTrltPct}{78.0}
\newcommand{\CollectiveThirtyTwoBanditMeanTrltPct}{1.09}
\newcommand{\CollectiveThirtyTwoBanditPNineNineTrltPct}{9.39}
\newcommand{\CollectiveThirtyTwoBanditCaption}{UCB1 champion}

\begin{document}

\twocolumn[
\mlsystitle{\ours{}: Closing the Loop on NCCL's Cost Model with Online Bandits}

\mlsyssetsymbol{equal}{*}

\begin{mlsysauthorlist}
\mlsysauthor{Hanan Shteingart}{nvidia,equal}
\mlsysauthor{Amedeo Sapio}{nvidia,equal}
\mlsysauthor{Chen Avin}{nvidia}
\mlsysauthor{Devendar Bureddy}{nvidia}
\mlsysauthor{Manjunath Gorentla Venkata}{nvidia}
\mlsysauthor{Srivatsan Krishnan}{nvidia}
\mlsysauthor{Shie Mannor}{nvidia}
\mlsysauthor{Sreeram Potluri}{nvidia}
\mlsysauthor{Gil Shabat}{nvidia}
\mlsysauthor{Zsolt-Alon Wertheimer}{nvidia}
\end{mlsysauthorlist}

\ifarxivbuild
  {\centering\large NVIDIA\par}
\fi

\mlsysaffiliation{nvidia}{NVIDIA}

\mlsyscorrespondingauthor{Hanan Shteingart}{hshteingart@nvidia.com}

\mlsyskeywords{NCCL, GPU collectives, multi-armed bandits, production ML systems, distributed training}

\vskip 0.3in

\begin{abstract}
NCCL picks an algorithm--protocol pair for every collective from an analytic cost model (\AUTO{}) that never sees a
measured latency.
The model is a strong prior but a brittle final decision: priced on forced-arm runs, its choice is exact on most contexts and badly wrong in a few.
We built \ours{}, a tuner plugin that closes the loop: a per-context bandit, warm-started from the model, learns from
measured latencies.
We replay a forced-arm H100 sweep, tuned at 4--8 nodes and scored at 16--32, and price the model's choice on the same
forced arms; that sweep has no stock-NCCL run.
The bandit cuts the contexts where the model's choice loses more than 10\% from 19 to 1 (p99 103\% to 9.9\%, mean 6.1\% to
1.6\%; $p{=}0.047$ over nine scenarios after Holm correction, three of them unshifted trace scenarios; on the six
scale-shifted ones the three bandit comparisons reach the smallest attainable $p{=}0.031$, $0.125$ after Holm).
The prior and continued learning both matter: cold-started bandits, and committing after one probe of each arm, lose more
on the mean in every scenario.
Supervised selectors fit once on the same data and tuned with the same budget match the mean, and their heavier tail is
not significant (7 against 1 contexts above 10\%, $p{=}0.06$); at seen scales they match or beat online selection.
With hyperparameters frozen from H100, UCB1 cuts \AUTO{}'s p99 from 60\% to 23\% and from 56\% to 15\% on two GB200
fabrics, but its contexts above 10\% go from 16 to 8 on one and from 10 to 13 on the other, and scenario-level tests do
not separate the two.
Live, on two latency-emulated node sets of one H100 system with one run per policy, the learners repair most of a gross
\AUTO{} error on one (selection loss 31\% to 6--8\% for deterministic Greedy and UCB1; average all-reduce bandwidth
$1.18$--$1.24\times$ stock) but average $0.94$--$0.99\times$ stock on the other, and on both lose to stock below 1\,MiB (on
the first through plugin overhead); on H100 the cost model is exact there offline, so small sizes can be left to it.
We do not show a net end-to-end win.
\end{abstract}
]

\printAffiliationsAndNotice{\mlsysEqualContribution}

\section{Introduction}
\label{sec:intro}
Distributed training and inference rest on \emph{collective} operations (all-reduce,
all-gather, reduce-scatter), which the NVIDIA Collective Communications Library
(NCCL)~\cite{hu2025,nccl,woolley2015} implements on GPU clusters.
For every collective NCCL chooses, in real time, among many algorithm--protocol pairs
(Ring, Tree, and NVLink-SHARP variants crossed with the LL, LL128, and Simple
protocols)~\cite{hu2025}.
The default selector is an analytic cost model, \NCCLAUTO{}~\cite{nccl,williams2025nccltuning},
that predicts the elapsed time of each pair from GPU, topology, network, and protocol
properties and plays the argmin.
It is calibrated before deployment and then frozen: a \emph{Day-0} policy in the
Day-0/1/2 taxonomy of Section~\ref{sec:related}.

On our measurements \AUTO{} is exact on most contexts, but when the target fabric,
topology, scale, or contention differs from its calibration it mis-ranks the arms on a
minority of contexts, and the loss on those contexts is not small.
We call a \emph{context} $x=(\text{collective}, P, n)$ a collective type, rank count, and
message size.
In our replay on a held-out, larger-scale split, the model's choice, priced on forced-arm runs, loses 6.08\% of latency
on average but 103\% at the 99th percentile: the mean hides a thin catastrophic tail (Section~\ref{sec:failure}).
Neither \AUTO{} nor the upstream tuner-plugin interface can repair this, because NCCL
never returns a chosen pair's measured latency to the plugin~\cite{williams2025nccltuning}.

We therefore keep the cost model as a \emph{prior} and add online feedback
(Section~\ref{sec:method}).
\oursintro{} runs one independent \MABintro{} per context.
The cost model plays three roles: it \emph{filters} implausible arms, \emph{warm-starts}
each arm's statistics, and \emph{scales} the reward.
We implement the loop inside NCCL with a tuner plugin whose profiler returns measured
latency to the policy and whose state is synchronized across ranks.

\paragraph{Thesis and claim boundaries.}
\emph{On a forced-arm sweep, NCCL's cost-model argmin is exact on most contexts and badly wrong on a few.
A per-context bandit seeded by the model removes most of that tail in replay; against fit-once
supervised selectors tuned with the same budget it has a similar mean and a narrow tail edge under scale extrapolation,
and with hyperparameters frozen it cuts \AUTO{}'s
p99 by $2.6$--$3.7\times$ on two GB200 fabrics, though not its exceedance count on one, at a mean cost on that fabric.
We do not establish that stock NCCL has this tail (forcing an arm can change its latency), a mean advantage over supervised selection, a tail advantage over it
at seen scales, or that the
deployed system is a net win.}
Seven boundaries apply wherever the result is quoted.
(i)~\AUTO{}'s tail is 19 of 154 held-out contexts in 7 of 9 (collective, nodes) scenarios; contexts within a
scenario are not independent, so we test at the scenario level, where nine clusters limit the resolution
(Sections~\ref{sec:failure} and~\ref{sec:h100}).
(ii)~The mean difference to \AUTO{} ($3.99$\,pp over scenarios, $4.5$ pooled) passes our primary scenario-level test, but on the contexts where \AUTO{}
is already exact UCB1 pays a small exploration tax (median $0.98\%$), and on GB200 with MNNVL on the mean is worse than \AUTO{}'s
(Sections~\ref{sec:h100} and~\ref{sec:gb200}).
(iii)~Supervised selection is matched on the mean, and the tail edge over it is narrow.
The strongest supervised selector (gradient boosting with the cost-model prediction as a feature, tuned with the
learners' 151-trial budget; lowest held-out mean and p99 of the eight fits) exceeds $10\%$ on 7 contexts to UCB1's 1
($p{=}0.06$ at the scenario level); the edge depends on the threshold and holds against UCB1 only (Section~\ref{sec:h100}).
On same-scale and leave-one-collective-out splits, with a small labelled target-scale sweep, or fit on 4/8/16 nodes and
tested at 32, the static fits match or beat UCB1 on the mean (Section~\ref{sec:h100}, Appendices~\ref{app:split-robustness}
and~\ref{app:label-budget}).
(iv)~The prior claim is about the leading cold bandits at $T{=}100$ under pull-count replay, and about the mean
(Section~\ref{sec:prior}).
(v)~The offline replay does not charge the measured NVLS setup stall (5.7--7.8\,s on H100)
or the plugin's profiling and synchronization cost
(Sections~\ref{sec:live}--\ref{sec:limitations}).
(vi)~The live evidence is two latency-emulated closed-loop campaigns on H100 configurations (one collective, one run
per policy), not independent systems.
The tuned UCB1 champion ran live and lost several times its offline loss, and in both campaigns the learners lose to
stock below 1\,MiB, partly to state sync (Section~\ref{sec:live}).
(vii)~Every offline number is one hardware campaign under a deliberate scale shift, priced on forced-arm jobs (one per
arm), so between-job variation is not measured; on H100-B, forced prices were within $1.18\times$ of tuner selection except NVLS and stock differed
from both, so the \AUTO{} row is not shown to be stock's loss (Section~\ref{sec:limitations}).

\paragraph{Contributions.}
\begin{enumerate}[noitemsep,leftmargin=*]
  \item \textbf{A closed-loop tuner plugin} returning profiler latency to a per-context bandit that reuses the cost
    model as filter, warm start, and reward scale (Section~\ref{sec:method}).
  \item \textbf{A failure mode}: on forced arms, \AUTO{} is exact on about 80\% of contexts but has a thin tail of
    misplaced protocol and algorithm crossovers (Section~\ref{sec:failure}).
  \item \textbf{A realized-latency, scale-OOD evaluation} against fixed, supervised, and online baselines, showing
    online selection matching equally tuned supervised selectors on the mean with a narrow tail edge under scale
    extrapolation, beating commit-once probing (ETC), and a frozen-hyperparameter GB200 replication (Sections~\ref{sec:protocol}--\ref{sec:gb200}).
  \item \textbf{Deployment lessons from two live H100 campaigns}: selection quality, plugin overhead, setup cost, and
    per-size speed against stock (Sections~\ref{sec:live} and~\ref{sec:lessons}).
\end{enumerate}

\section{Background and related work}
\label{sec:related}
We organize prior work by \emph{when} the selection policy is fixed.
\emph{Day~0} is the shipped default (\AUTO{}); \emph{Day~1} is a one-time calibration
fit at deployment and then frozen; \emph{Day~2} is continuous adaptation from runtime
feedback.
We target Day~2 and reuse the Day-0 model as its prior (Appendix Table~\ref{tab:tuning-comparison}).
Our evidence covers short horizons ($T{=}100$, one synthetic change), so it supports in-job repair of a mis-placed crossover more than
long-run adaptation.

\paragraph{NCCL cost model and tuner plugins.}
\AUTO{} ranks pairs by predicted elapsed time from topology and protocol terms at
initialization~\cite{nccl,woolley2015,hu2025,williams2025nccltuning}.
Williams et al.~\cite{williams2025nccltuning} note that the shipped model is calibrated
for isolated collectives and recommend external tuner plugins when defaults
underperform; vendor plugins do so with fixed rules, e.g.\ the AWS OFI tuner picks a pair from measured regions over
message size and rank count~\cite{awsofinccl}.
Since NCCL~2.22 the tuner interface (v3) passes the callback \texttt{getCollInfo} the collective type, message size, and a
per-collective cost table of predicted times for every algorithm--protocol pair, which the plugin edits to select a
pair~\cite{nccl222blog}.
It passes no measured latency, so a plugin can keep state but NCCL never reports the realized cost of its choice.
By design our plugin selects through the cost table and changes only algorithm and protocol, leaving the channel count the v3
interface also returns at NCCL's default.
NCCL also exposes operator-level controls over the same arm set (\texttt{NCCL\_ALGO},
\texttt{NCCL\_PROTO}, including exclusion lists such as \texttt{allreduce:\^{}nvlstree}),
a profiler plugin interface, and \texttt{NCCL\_RUNTIME\_CONNECT}, which connects peers at
init rather than on first use~\cite{nccl}; the profiler and tuner are separate plugins.
We have not established which parts of our reward channel and state synchronization could be
reproduced without modifying NCCL (Appendix~\ref{app:deployment}).

\paragraph{MPI and GPU collective autotuning.}
Offline MPI autotuners freeze a lookup table (Open MPI's \texttt{tuned} decision rules~\cite{pjesivac2006,pjesivac2007perf}),
a task-level cost model (HAN~\cite{luo2020han}), or a supervised predictor after a benchmark
campaign~\cite{hunold2020,wilkins2021,wilkins2022}; our fit-once regressors
are Day-1 baselines of this kind, including a variant that takes the cost model's own
prediction as a feature.
Online systems tuners adapt from runtime feedback: STAR-MPI~\cite{faraj2006starmpi}
selects among MPI collective algorithms by measured performance inside a running
application, OMPICollTune~\cite{hunold2022ompicolltune,steiner2023} does so incrementally,
and AutoCCL~\cite{xu2025} tunes NCCL's algorithm, protocol, and resource parameters
(channels, chunk size, threads) during the early iterations of a training job.
None seeds a regret-minimizing policy from a shipped cost model.
We search only algorithm--protocol pairs, so composing with AutoCCL's resource parameters is untested.
TACCL~\cite{shah2023}, MSCCLang~\cite{cowan2023mscclang}, TE-CCL~\cite{liu2024teccl}, ForestColl~\cite{zhao2024forestcoll},
and MSCCL++~\cite{hwang2025mscclpp} synthesize or program schedules and TCCL~\cite{kim2024tccl} profiles ring paths, which enlarges the arm set rather than selecting among
NCCL's fixed pairs, and production stacks such as NCCLX~\cite{si2025ncclx} extend NCCL itself at 100k-GPU scale.
Conservative bandits~\cite{wu2016conservative} formalize ``never fall far below a default
policy,'' which is the role \AUTO{} plays here; our warm start and filter are a pragmatic
version of it, without that guarantee.
Seeding a learner from a prior has theory for supervised-data warm starts~\cite{zhang2019warmstart}
and for hints of unknown quality~\cite{bhaskara2020hints}; our pseudo-count is the corresponding
trust knob, tuned and not derived.
Our claim is about the tail of the loss, a risk criterion in the sense of risk-averse
bandits~\cite{sani2012risk,cassel2018risk}, yet every policy here is tuned on the mean loss, not on the tail.
The Day-0 prior descends from Hockney/LogP models~\cite{hockney1994,culler1993logp}.

\paragraph{Explore-then-commit for NCCL.}
Madan et al.~\cite{madan2026}, in a course project report that we cite as prior art and not
as a peer-reviewed comparison, describe a live NCCL plugin that explores pairs round-robin, commits to
the IQR-trimmed fastest pair, and falls back to \AUTO{} through a 5\% safety gate.
They report 25--42\% latency gains on A100.
Their design is explore-then-commit with a permanent commit, and the gate is not a warm start.
We include a re-implemented, search-tuned round-robin row (Table~\ref{tab:realized}); it is not a
matched head-to-head with their live plugin, and our results neither reproduce nor validate theirs.

\paragraph{Bandits.}
We use UCB1~\cite{auer2002}, KL-UCB~\cite{garivier2011klucb}, Softmax, Gaussian Thompson
sampling~\cite{thompson1933}, and adversarial variants~\cite{auer2002exp3,seldin2014exp3pp,zimmert2021,honda2023},
and discuss discounting for non-stationarity~\cite{kocsis2006discounted,garivier2011switching}
in Section~\ref{sec:lessons}.
Appendix~\ref{app:policies} specifies every family.

\section{Failure mode: on forced arms, AUTO is usually exact, with a thin tail}
\label{sec:failure}
We audit \AUTO{}'s argmin against measured sweeps on the H100 grid ($n{=}300$ contexts
with at least two measured arms).
\AUTO{} picks the measured-best arm on about 80\% of contexts (pooled error rate
$20.0\%$), so its typical loss is exactly zero.
The errors are not benign: over all contexts \AUTO{}'s selection-priced \TRLT{} is
$5.04\%$ on average and $75.4\%$ at p99, and a wrong choice can cost more than the best
arm's entire latency.

The cost model exposes a decision-time confidence signal, the predicted runner-up margin
$m=\hat{\mu}_{(2)}/\hat{\mu}_{(1)}-1$ between its two best predictions.
Appendix Figure~\ref{fig:cost-model-margin-error-rate} audits it as a gate.
Retaining only contexts with $m\ge0.50$ keeps $45.3\%$ of them and cuts the error rate to
$3.7\%$, the mean to $0.49\%$, and the p99 to $16.4\%$.
So the model \emph{knows} when it is unsure, and the low-margin contexts carry
disproportionate tail loss.
But a static selector must still decide on those contexts, and its margin cannot reveal
the penalty of a wrong choice; only a measurement can.

Learning a static selector offline closes the gap only in part: a gradient-boosted selector fit with fixed settings on the training contexts
with the cost model's prediction as a feature nearly eliminates training error ($0.04\%$ \TRLT{}) yet pays $1.76\%$ mean and
$36.9\%$ p99 on the larger-scale test (selection-priced; Appendix~\ref{app:extended-results}).

\paragraph{Where the tail lives.}
\AUTO{} loses more than $10\%$ on 19 of the 154 held-out contexts, in 7 of the 9 (collective, nodes) scenarios
(Appendix Tables~\ref{tab:tail-regions} and~\ref{tab:scenario-breakdown}).
On all 19 the forced-arm sweep measures a protocol or algorithm crossover at a different message size than the cost
model predicts (predicted margins $0.06$--$0.57$).
These are forced-arm prices, not stock's (Section~\ref{sec:limitations}).
Within-pool sampling noise does not explain them (Appendix~\ref{app:tail-anatomy}).

\section{Method}
\label{sec:method}

\paragraph{Problem.}
Each collective request arrives with context $x=(\text{collective},\,P,\,n)$.
The action space is a finite set of algorithm--protocol arms $\mathcal{A}(x)$;
pulling arm~$a$ yields latency $T_a$ with unknown mean $\mu_a(x)$.
A policy $\pi$ maps $x$ to an arm $A\in\mathcal{A}(x)$.
Fixed selectors (\AUTO{}, supervised regressors) ignore feedback; online bandits
consume measured latency and revise future choices.
Production jobs revisit few contexts: a Llama3-8B torchtitan training trace from
CCL-Bench~\cite{cclbench2026} collapses \cclBenchCalls{} collective calls onto only
\cclBenchContextsWord{} contexts ($\cclBenchMeanReuse{}\times$ mean reuse), two of which
carry $\cclBenchTopTwoTrafficPct{}\%$ of the calls
(Appendix~\ref{app:empirical-motivation}).
We therefore run an independent bandit per context with no parameter sharing,
instantiated lazily on the first visit; contextual alternatives are in
Appendix~\ref{app:contextual}.

\paragraph{Reward and cost-model priors.}
\label{sec:reward}
The cost model plays three roles: arm filter, warm-start prior, and reward scale.
Bandits maximize the scale-free ratio
\begin{equation}
  \label{eq:ratio-reward}
  r(t) = \frac{\hat{\mu}_{\min}}{t},
\end{equation}
where $\hat{\mu}_{\min}=\min\{\hat{\mu}(a):\hat{\mu}(a)\ \text{defined}\}$
is the best cost-model prediction on the offered set, so $r{=}1$ means
``as fast as \AUTO{} expected,'' not oracle-optimal.
The arm filter $\rho_{\max}{=}2$ drops arms with
$\hat{\mu}(a)>\rho_{\max}\,\hat{\mu}_{\min}$ (undefined predictions are kept):
\begin{equation}
  \label{eq:filter}
  \mathcal{A}' =
  \left\{
    a \in \mathcal{A} :
    \hat{\mu}(a) \le \rho_{\max}\,\hat{\mu}_{\min}
    \;\;\text{or}\;\;
    \hat{\mu}(a)\ \text{undefined}
  \right\}.
\end{equation}
Warm start seeds each arm's statistics from $\{r(\hat{\mu}(a))\}$ with a tuned
pseudo-count before the first pull; cold start omits it.

\begin{algorithm}[t]
\caption{Per-context bandit loop for fixed $x=(\text{collective}, P, n)$}
\label{alg:loop}
\scriptsize
\begin{algorithmic}[1]
\Require $\mathcal{A}$, $\{\hat{\mu}(a)\}$, $\rho_{\max}$, $T$, policy $\pi$, reward $r(\cdot)$
\State $\hat{\mu}_{\min} \gets \min\{\hat{\mu}(a) : \hat{\mu}(a)\ \text{defined}\}$
\State $\mathcal{A}' \gets \{a \in \mathcal{A} : \hat{\mu}(a)\ \text{undefined}\ \textbf{or}\ \hat{\mu}(a) \le \rho_{\max}\,\hat{\mu}_{\min}\}$
  \Comment{Eq.~\eqref{eq:filter}}
\State Initialize $\pi$ from $\{r(\hat{\mu}(a))\}_{a\in\mathcal{A}'}$ \Comment{warm start}
\For{$t = 1, \ldots, T$}
  \State $A_t \gets \pi.\textsc{Select}(\mathcal{A}')$;\quad observe $X_t$;\quad $\pi.\textsc{Update}(A_t, r(X_t))$
\EndFor
\end{algorithmic}
\end{algorithm}

\paragraph{Featured policy: deterministic UCB1.}
The policy plays $A_t=\arg\max_{a\in\mathcal{A}'}\big[\hat{r}_a+\alpha\sqrt{2\ln N_t/n_a}\big]$
(Appendix~\ref{app:policies}), with $\hat{r}_a$ a clipped-SGD mean reward and $n_a$ its
count, seeded by warm start.
It departs from textbook UCB1 (tuned $\alpha$, pseudo-count, and decaying step size), so the
regret theory of Appendix~\ref{app:policies} does not apply to it.
With a warm-start pseudo-count no arm is unexplored, so the choice is a deterministic
function of the replicated statistics.
Every rank therefore selects the same arm without shared randomness, provided the statistics stay replicated and the update predicate is collectively agreed
(Appendix~\ref{app:deployment}; fallback untested, and not verified live), and replay has zero seed variance.
The tuned champion has $\alpha{=}0.093$ and a pseudo-count of $2.89$ (Appendix~\ref{app:hyperopt}).
Train-mean selection ranks KL-UCB first ($0.845\%$), then UCB1 ($0.864\%$) and Softmax
($0.868\%$): a tight cluster, so we feature UCB1 on systems grounds (deterministic, simplest to
reason about in a plugin), not because it won; the choice was made before realized accounting existed and is not a
pre-registration.
Softmax and KL-UCB are reported beside it in every table, and re-tuning on the realized objective leaves the three
within noise (Appendix~\ref{app:realized-tuned}).

\paragraph{Plugin.}
Appendix Figure~\ref{fig:spcx-arch} shows the deployed loop.
NCCL invokes the tuner with the cost table and job dimensions; the bandit selects an arm
and its parameters flow back to the collective; the profiler returns the measured latency
as the reward; and state is synchronized across ranks.
Upstream NCCL provides the per-call cost-table override; the
profiler$\rightarrow$tuner reward channel, durable per-context state, and a cross-rank
synchronization API are our additions in SPCX, the unreleased NVIDIA Spectrum-X NCCL environment plugin
(Appendix~\ref{app:deployment}).
The offline evaluation updates the policy after every collective and treats synchronization as free; its measured live
cost is reported separately (Sections~\ref{sec:live} and~\ref{sec:lessons}).

\section{Evaluation protocol}
\label{sec:protocol}

\paragraph{Data and replay.}
The primary ground truth is an open-loop, forced-arm NCCL sweep (one job per arm, the arm forced through
\texttt{NCCL\_ALGO} and \texttt{NCCL\_PROTO}, every other NCCL parameter at its default) on Dell HGX H100 nodes with
Spectrum-X Ethernet (8 GPUs per node; NCCL~2.30.4): three collectives (all-gather,
all-reduce, reduce-scatter) at 4, 8, 16, and 32 nodes over 26 message sizes from 512\,B to
16\,GiB, with 100 measured iterations for every offered arm (Appendix Tables~\ref{tab:datasets} and~\ref{tab:fabric}); its cost tables come from the same system.
Each iteration's latency is the per-collective time that GPU~0 of the job reports in the sweep's \texttt{nccl-tests}
output (not the maximum over ranks); which nodes each per-arm job ran on, and when, was not recorded.
Whether forcing an arm prices it as tuner selection does is untested on this sweep (on H100-B they agreed except
NVLS; Appendix~\ref{app:tail-anatomy}); the live campaigns (Section~\ref{sec:live}) select through the
cost table.
The cost model marks CollNet, pure NVLS, and PAT not applicable on this fabric, so the arms
are Ring$\times$\{LL, LL128, Simple\} on every collective plus Tree$\times$\{LL, LL128,
Simple\} and NVLSTree/Simple on all-reduce; \AUTO{} and the bandits draw from the same set.
We replay each arm's measurement pool sequentially in iteration order, one pull per step
for $T{=}100$ steps (the pool depth).
Sequential replay preserves NCCL's lazy-initialization transient, which makes the first
$\sim$10 samples of an arm heavy-tailed (Appendix~\ref{app:empirical-motivation}), so the
first pulls of an arm replay it.

\paragraph{Metrics.}
\label{sec:metrics}
We score in the time domain. With $\bar{\mu}_a$ the steady-state mean latency of arm~$a$
(iterations $11$--$99$) and $a^\star=\arg\min_a\bar{\mu}_a$, the per-step
\RLTintro{} and path \TRLTintro{} of a selection trace $(A_1,\dots,A_T)$ are
\begin{equation}
  \label{eq:rlt-i}
  \mathrm{RLT}_i =
  \frac{\bar{\mu}_{A_i} - \bar{\mu}_{a^\star}}{\bar{\mu}_{a^\star}},
  \qquad
  \mathrm{TRLT} =
  \frac{\sum_{i=1}^{T}\bar{\mu}_{A_i} - \sum_{i=1}^{T}\bar{\mu}_{a^\star}}
       {\sum_{i=1}^{T}\bar{\mu}_{a^\star}}.
\end{equation}
This \emph{selection-priced} \TRLT{} charges every step at the chosen arm's steady mean,
so it under-charges the cost of switching to a cold arm.
The \emph{realized} \TRLT{} instead sums the latencies the policy was actually served,
$\sum_t X_t$, against a realized sequential oracle, the measured-best arm replaying its
own pool for the same $T$ steps (so both sides pay the transient once):
$\mathrm{TRLT}^{\mathrm{real}}=\sum_t X_t/\sum_t X^{\star}_t-1$.
The two biases point in opposite directions: selection pricing under-charges arm switches,
and realized replay over-charges them.
In realized replay a policy's $k$-th pull of an arm is served that arm's $k$-th swept sample
however late it occurs, so a rarely played arm keeps paying lazy-init samples that production
would have warmed.
\emph{Realized \TRLT{} is the headline}; selection-priced values are secondary columns.
The learners maximize a reward proportional to $1/X$ while \TRLT{} ranks arms by mean latency; over all 306 grid
contexts the two pick a different best arm on 2, at most $0.3\%$ apart.
Neither accounting charges the one-time NVLS setup stall that the live campaigns measured
(5.7--7.8\,s on H100, Section~\ref{sec:live}); we re-price the
H100 replay separately with a sensitivity range
(Section~\ref{sec:lessons}).

\paragraph{Split and replication.}
The split is deliberately scale-out-of-distribution.
Hyperparameters are tuned only on the 146 contexts at 4 and 8 nodes; the 154 test contexts
are all 16/32-node contexts plus eight Llama3 trace contexts mapped onto 4-node pools.
Optuna's TPE sampler tunes each of 14 online families for 151 trials on mean train
selection-priced \TRLT{}, and the frozen winner is replayed once on test
(Appendix~\ref{app:hyperopt}).
The tuning objective is selection-priced train \TRLT{}; we do not re-tune on the realized
metric we report, and we show both.
The unit of replication is the context, not a training job.
Stochastic policies are averaged over 20 seeds per context; tables report mean $\pm$ \SEM{} over contexts,
p99 is the percentile across the 154 contexts, and intervals are bootstrap resamples of
contexts (each context its own cluster).
The 154 contexts are not independent, however: they come from nine (collective, nodes)
scenarios with adjacent message sizes, so we also resample whole scenarios and blocks of
four adjacent sizes (Appendix Table~\ref{tab:scenario-ci}).
Our inferential test is an exact sign-flip test over the nine scenarios (Section~\ref{sec:h100}).
Context-level paired statistics (Wilcoxon signed-rank, exact McNemar on exceedance counts, CVaR$_{5\%}$,
quantile profiles; Appendix Table~\ref{tab:tail-stats}) treat contexts as exchangeable, so every $p$-value from them
in this paper is descriptive.
We report them, with interval estimates and full distributions, in the spirit
of~\cite{agarwal2021precipice}.

\paragraph{Baselines.}
\AUTO{}; uniform Random; fit-once supervised regressors (linear, random forest, gradient
boosting) trained on the train contexts, with and without the cost model's own prediction
$\log\hat{\mu}(a)$ as a feature (``+cost''); the tree models appear with fixed sklearn settings (300 depth-4 trees at
rate $0.08$, 400 forest trees with leaf size 2) and tuned like the learners, with 151 TPE trials (trial 0 the fixed
settings) minimizing selection-priced loss on validation folds inside the train partition that copy the held-out shift
(here, leave one train node count out; Appendix~\ref{app:supervised-tuning}); explore-then-commit (ETC, $m{=}1$) and the round-robin variant of
Madan et al.; and the 14 tuned online families.

\section{Results}
\label{sec:results}

\subsection{H100: AUTO, supervised, and online selection}
\label{sec:h100}
\begin{figure}[t]
  \centering
  \includegraphics[width=\linewidth]{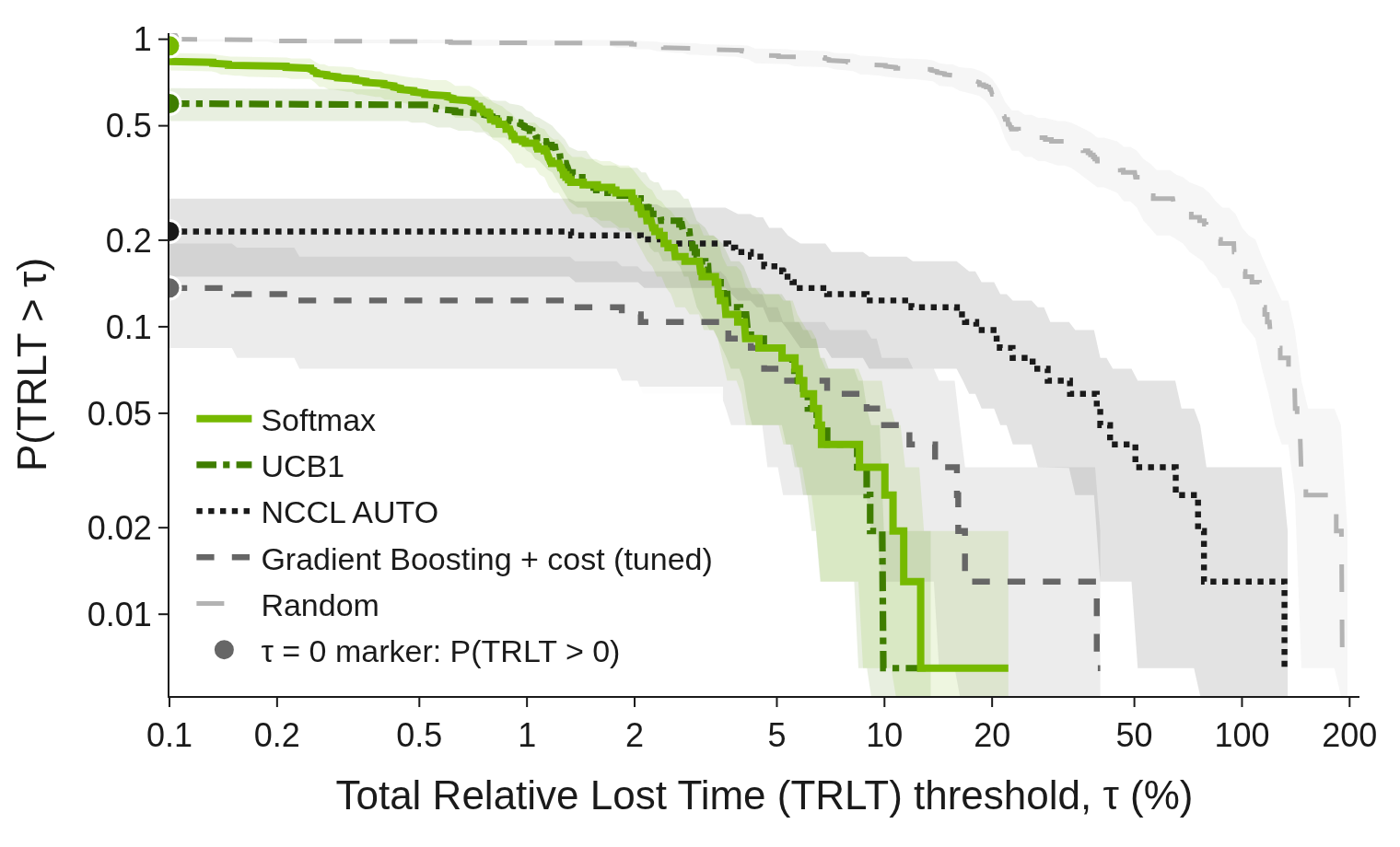}
  \caption{Realized \TRLT{} survival function $P(\TRLT{}>\tau)$ over the 154 held-out
    contexts (logarithmic axes; seeds averaged per context; shaded: 95\% bootstrap
    CI over contexts). Circles on the left axis give $P(\TRLT{}>0)$, counting values above $10^{-9}\%$ so that
    floating-point round-off is not counted as loss; a realized \TRLT{} can be negative (a non-oracle arm luckier than the
    oracle's pool; 13\% of \AUTO{}'s contexts) and then counts as no loss. The online policies pay a small exploration cost
    on many contexts but cap the tail, while \AUTO{}'s tail passes 100\%.}
  \label{fig:trlt-ccdf}
\end{figure}

Table~\ref{tab:realized} and Figure~\ref{fig:trlt-ccdf} give the headline.
\AUTO{} has no realized loss on 79\% of the 154 held-out contexts ($85\%$ for GBM+cost; a realized
\TRLT{} can be slightly negative when a non-oracle arm happens to serve less time than the oracle's pool), but its realized
\TRLT{} is $6.08\%$ on average and $103.3\%$ at p99, and it loses more than $10\%$, $50\%$,
and $100\%$ on 19, 6, and 2 contexts, in 7 of the 9 (collective, nodes) scenarios.
Warm-started UCB1 lowers the mean to $1.61\%$ and the p99 to $9.9\%$ and never loses more
than $50\%$; it exceeds $10\%$ on one context, in 1 of 9 scenarios (Appendix Table~\ref{tab:exceedance-ci}).
Softmax reaches $1.75\%$ and $11.9\%$.

\begin{table}[t]
  \centering
  \scriptsize
  \caption{H100 sweep, \capsplit{} \capTRLT{} \capreal{} \AUTO{} is the cost model's choice priced on forced-arm runs.
    The realized columns are the headline. Mean $\pm$ \SEM{}
    over contexts; \capptail{} Supervised rows are fit once on the train contexts, with fixed settings or tuned with the
    learners' 151-trial budget (Appendix Table~\ref{tab:supervised-tuning}); ``+cost'' adds the cost model's
    predicted latency as a feature. \capbold{} Random is a reference and is not bolded.}
  \label{tab:realized}
  \resizebox{\columnwidth}{!}{\begin{tabular}{lrrrr}
\toprule
& \multicolumn{1}{c}{Selection \TRLT{}} & \multicolumn{3}{c}{Realized \TRLT{}} \\
\cmidrule(lr){2-2}\cmidrule(lr){3-5}
Policy & mean (\%) & mean (\%) & p99 (\%) & $n_{>10\%}$ \\
\midrule
NCCL \AUTO{} & 5.99$\pm$1.58 & 6.08$\pm$1.58 & 103.3 & 19 \\
Supervised GBM+cost, fixed & 1.76$\pm$0.51 & 1.78$\pm$0.50 & 32.2 & 10 \\
Supervised GBM+cost, tuned & 1.30$\pm$0.42 & \textbf{1.37}$\pm$0.43 & 27.3 & 7 \\
ETC ($m{=}1$, $k{=}1$) & 2.70$\pm$0.49 & 4.74$\pm$0.75 & 63.0 & 11 \\
Softmax & \textbf{0.93}$\pm$0.11 & 1.75$\pm$0.22 & 11.9 & 5 \\
UCB1 & 1.00$\pm$0.13 & 1.61$\pm$0.19 & \textbf{9.9} & \textbf{1} \\
KL-UCB & 1.18$\pm$0.24 & 2.11$\pm$0.36 & 22.7 & 4 \\
Random & 45.22$\pm$3.72 & 46.85$\pm$3.79 & 190.3 & 125 \\
\bottomrule
\end{tabular}
}
\end{table}

Contexts are not independent units, so our primary test procedure is the exact sign-flip test over the nine
(collective, nodes) scenarios, for the mean difference and for the difference in contexts above $10\%$.
It rejects for both. The mean difference \AUTO{}$-$UCB1 is $4.46$\,pp over contexts (scenario-clustered bootstrap CI
$[2.8, 6.3]$) and $3.99$\,pp over the nine equally weighted scenario means, which the sign-flip tests ($p{=}0.016$; a
wild-cluster bootstrap-$t$~\cite{cameron2008} on the context-weighted mean agrees), and the exceedance difference has the
same $p{=}0.016$; the $10\%$ threshold was chosen after seeing the results, and at $20\%$ the counts are 15 against 0
($p{=}0.016$), at $5\%$ 25 against 13 ($p{=}0.062$; Appendix Table~\ref{tab:exceedance-thresholds}).
Over the four comparisons our claims rest on (these two, warm against cold-tuned UCB1, and tuned GBM+cost against
UCB1 on exceedance) Holm's correction gives $p{=}0.047$, $0.047$, $0.016$ and $0.062$
(Appendix Table~\ref{tab:confirmatory}).
Three of the nine scenarios are the 4-node trace contexts, which are not scale-shifted; restricted to the six 16/32-node
scenarios, the first three each reach the smallest $p$ six clusters allow ($0.031$), $0.125$ after Holm, so no
scale-shifted comparison survives correction, although all six scenarios favor UCB1 on the mean.
Choosing the oracle on odd iterations and pricing on even ones leaves \AUTO{} at $6.0\%$ and 18 contexts above $10\%$
and UCB1 at $1.0\%$ and none. The same test does not separate \AUTO{} from ETC ($p{\approx}0.4$; Appendix Table~\ref{tab:tail-region-tests}).
On individual contexts \AUTO{} is usually the better policy: it has no loss on most of them, so UCB1's median exploration
tax of $0.98\%$ makes \AUTO{} better on 40\% of contexts and UCB1 on 16\% (62 against 24 contexts; descriptive, since contexts within a scenario are not independent).
McNemar on the counts and CVaR$_{5\%}$ ($77.2\%$ against $9.1\%$) assume exchangeable contexts and are descriptive
(Appendix Table~\ref{tab:tail-stats}).
The same scenario-level test finds no UCB1-over-\AUTO{} difference on either GB200 fabric ($p{=}0.91$, $0.81$), so the
cross-platform support for the tail claim is thin.
UCB1 is close to \AUTO{} plus a safeguard: it plays \AUTO{}'s arm on $85\%$ of steps, leaves
it on $56\%$ of contexts (a mean of $1.6$ distinct arms), and plays the measured-best arm
over its last 20 steps on $82\%$ of contexts against \AUTO{}'s $79\%$.
Continued learning still matters: the train-best of 24 probe-and-commit settings (one transient sample discarded, one
probe per arm, no margin) coincides with ETC below when NVLS is offered, $4.74\%$ mean and 11 contexts above $10\%$, worse than UCB1 in all nine
scenarios ($p{=}0.004$; $3.21\%$ against $1.23\%$ with NVLS excluded; $3.26\%$ with ETC's adaptive-observed-max reward wrapper), and variants that keep \AUTO{}'s arm unless another
is $5\%$ faster do no better (Appendix Table~\ref{tab:probe}).

Supervised selection is the harder comparison.
With fixed settings the strongest fit, GBM+cost, ties the mean ($+0.17$\,pp over contexts, $+0.05$ over scenarios, $p{=}0.88$), is exact on 85\% of contexts,
and exceeds $10\%$ on 10 contexts to UCB1's one ($p{=}0.031$; Appendix Table~\ref{tab:scenario-primary}).
Tuned like the learners, GBM+cost improves to $1.37\%$ mean, $27.3\%$ p99, and 7 contexts above $10\%$: $0.24$\,pp below
UCB1 on the context mean but $0.07$\,pp above it on the scenario mean the primary test uses ($p{=}0.91$), with an exceedance difference the primary test only borderline separates
($p{=}0.06$; Appendix Table~\ref{tab:supervised-tuning}).
Resampling scenarios, its p99 and CVaR$_{5\%}$ exceed UCB1's by $17.5$ ($[2.6, 35.9]$) and $11.3$\,pp ($[3.3, 21.6]$), but
the edge is specific to UCB1: against Softmax and KL-UCB these intervals include zero
(Appendix Tables~\ref{tab:tuned-gbm-tail} and~\ref{tab:exceedance-thresholds}).
With NVLS excluded (an exploratory comparison outside the Holm family), tuned GBM+cost has $1.46\%$ mean, $38.6\%$ p99 and 9 contexts above $10\%$ against UCB1's
$1.23\%$, $11.2\%$ and 2 (exceedance $p{=}0.031$, mean $p{=}0.42$; Appendix Table~\ref{tab:tuned-gbm-nvls}).
Tuning mostly drops the node-count feature, which trees cannot extrapolate; the other fits are worse
(Appendix Table~\ref{tab:realized-full}).
On H100, then, online selection matches supervised selection on the mean, with a narrow tail edge at the held-out scales.
Once they see near-target scale, static fits match or beat the online policy, which needs no target-scale labels: one
sample of every arm at every fourth 16/32-node size gives GBM$+$cost $1.01\%$ mean against UCB1's $1.68\%$ (p99 $25\%$
against $10\%$; Appendix~\ref{app:label-budget}), and fit on 4/8/16 nodes and tested at 32 it reaches $0.47\%$ and $7.1\%$
against $1.89\%$ and $11.0\%$ (Appendix Table~\ref{tab:largest-scale}).

The commit-once policies do less well on the realized tail.
ETC ($m{=}1$ with the first sample discarded; the tuned Madan round-robin champion coincides with it) is cheap
selection-priced ($2.70\%$ mean, $8.7\%$ p99) but costs $4.74\%$ mean and $63.0\%$ p99 on
realized latency with 11 contexts above $10\%$ (paired p99 difference to UCB1 $53.1$\,pp, context-bootstrap CI
$[4.8, 61.7]$, scenario-bootstrap $[0.6, 61.4]$).
It is the train-best of an 18-point $(m,k)$ grid, and Madan-style round-robin does no better ($5.9$--$13.3\%$ mean;
Appendix Table~\ref{tab:transient}).
With NVLS excluded, ETC's tail falls to $13.8\%$ (Section~\ref{sec:lessons}).
The three leading online policies are mutually within noise (other families: Appendix Table~\ref{tab:realized-full}).

\paragraph{Robustness to the split and to tuning.}
Re-tuning on nine further splits and seeds (Appendix Table~\ref{tab:split-robustness}) keeps UCB1 well below \AUTO{}
($0.76$--$2.68\%$ against $3.96$--$6.08\%$ mean).
On same-scale random splits, however, the best tuned static fit has the lower mean ($0.64$--$0.76\%$ against UCB1's
$1.24$--$1.69\%$) and a comparable tail ($9.6$--$13.3\%$ p99 against $9.5$--$15.2\%$), and it also has the lower mean when
one collective is held out (Appendix Table~\ref{tab:supervised-tuning}).
The tail edge over supervised selection is therefore a scale-shift result.
Re-tuning on the realized objective leaves the leaders within noise (Appendix~\ref{app:realized-tuned}).

\subsection{The prior: matched, independently tuned, and transient-controlled cold ablations}
\label{sec:prior}
Table~\ref{tab:warm-cold} toggles only the cost-model prior for each tuned family on
realized \TRLT{}, at the warm-tuned knobs.
That handicaps the cold arm: UCB1's $\alpha{=}0.093$ and pseudo-count $2.89$ only make sense
with a prior, so ``cold $8.82\%$'' partly measures an untuned policy.
We therefore also tuned every warm-capable family \emph{cold}: a separate Optuna study with
warm start pinned off, on the same split, objective, sampler seed, and 151-trial budget
(Appendix Table~\ref{tab:cold-tuned}).
Tuning helps the cold arm (UCB1 $8.82\%\to4.84\%$, Softmax $6.71\%\to4.77\%$), but the
warm champions stay better.
Warm UCB1 has $1.61\%$ against $4.84\%$ (paired $\Delta{=}-3.22$\,pp, CI $[-4.76,-1.97]$) and is lower in all nine
scenarios, so the primary sign-flip test gives its minimum $p{=}0.004$, as it does for Softmax and KL-UCB
(Appendix Table~\ref{tab:scenario-primary}).
The five leading cold families keep $4.27$--$4.97\%$ mean and $33.9$--$64.5\%$ p99 against $1.61$--$2.17\%$ and
$9.9$--$23.3\%$ warm.
The cold-tuned UCB1 is near-greedy after one pull per arm and shares ETC's $63.0\%$ p99, and a $50$--$100\times$ wider search
over three seeds finds the same champion (Appendix Table~\ref{tab:cold-wide}).
So the warm start, on top of the arm filter that cold policies also use, is necessary at $T{=}100$ for these families, but the untuned-cold gap
overstated it: the paired warm advantage is about $3$\,pp, not $7$\,pp.

Realized replay serves a cold policy's early pulls each arm's lazy-init samples, so we controlled for that artifact
three ways (Appendix Tables~\ref{tab:transient} and~\ref{tab:transient-paired}).
(i)~Selection-priced accounting, which charges no transient, still favors the warm champions by
$1.6$--$1.8$\,pp (UCB1 $1.00\%$ against the cold-tuned $2.78\%$; paired CIs exclude zero).
(ii)~With NVLS excluded, the realized advantage is $1.6$--$2.1$\,pp (UCB1 $1.23\%$ against $3.31\%$,
lower in all nine scenarios, $p{=}0.004$), but the \emph{tail} gap disappears.
Cold UCB1's and Softmax's p99 falls from $63\%$ to $12\%$, and both warm and cold have two contexts above $10\%$, so the cold tail of
$34$--$65\%$ above is largely NVLS first-touch contamination, not worse decisions.
(iii)~Discarding $k{\ge}2$ first samples is worse than $k{\le}1$ for ETC and every cold family, because a discarded pull is still
a served step.
So the prior is necessary for the mean at $T{=}100$, not a replay artifact, while its realized-tail effect is mostly an
NVLS effect.
This holds for the leading families: FPL, EXP3, Tsallis-Inf, UCB-V and UCB-tuned are nearly insensitive to the prior
(at most $0.72$\,pp; Appendix Table~\ref{tab:warm-start-ablation}).

\begin{table}[t]
  \centering
  \scriptsize
  \setlength{\tabcolsep}{3pt}
  \caption{Warm versus cold start for the three champion families on the H100 sweep, \capsplit{} Only the cost-model
    prior is toggled, at the warm champion's hyperparameters. \capTRLT{} Realized \TRLT{} charges the latencies actually
    served, first-use transients included. Mean $\pm$ \SEM{} and p99 (99th percentile) over contexts; $\Delta$ is the
    paired per-context warm$-$cold mean in percentage points, with a Wilcoxon signed-rank $p$ over contexts (descriptive,
    since contexts are not independent; the scenario-level test is in Appendix Table~\ref{tab:scenario-primary}).
    All families, selection-priced: Appendix Table~\ref{tab:warm-start-ablation}. \capbold{}}
  \label{tab:warm-cold}
  \resizebox{\columnwidth}{!}{\begin{tabular}{lrrrrrr}
\toprule
& \multicolumn{2}{c}{Cold (no prior)} & \multicolumn{2}{c}{Warm (prior)} & \multicolumn{2}{c}{Warm$-$cold} \\
\cmidrule(lr){2-3}\cmidrule(lr){4-5}\cmidrule(lr){6-7}
Family & mean (\%) & p99 (\%) & mean (\%) & p99 (\%) & $\Delta$ mean (pp) & Wilcoxon $p$ \\
\midrule
UCB1 & 8.82$\pm$1.42 & 98.7 & \textbf{1.61}$\pm$0.19 & \textbf{9.9} & -7.20 & $<$0.001 \\
Softmax & 6.71$\pm$0.90 & 49.5 & 1.75$\pm$0.22 & 11.9 & -4.96 & $<$0.001 \\
KL-UCB & \textbf{4.94}$\pm$0.73 & \textbf{40.8} & 2.11$\pm$0.36 & 22.7 & -2.83 & $<$0.001 \\
\bottomrule
\end{tabular}
}
\end{table}

\subsection{GB200: cross-architecture replication with frozen hyperparameters}
\label{sec:gb200}
Table~\ref{tab:gb200} reruns the policy ladder on GB200 sweeps with every hyperparameter
frozen from the H100 tuning (train: 4/8 nodes; test: 16/32 nodes), once per fabric
configuration, now with realized accounting and paired tests.
After the same filter the arms are Ring and Tree pairs, plus PAT/Simple (all-gather, reduce-scatter; MNNVL off) and
NVLSTree/Simple (all-reduce; MNNVL on).
What reproduces is weaker than on H100.
With MNNVL off, UCB1 has $2.59\%$ realized mean and $23.1\%$ p99 against \AUTO{}'s $3.80\%$ and
$60.1\%$ (8 against 16 contexts above $10\%$, McNemar $p{=}0.057$; paired mean difference $1.2$\,pp,
CI $[-0.3, 3.1]$; these context-level $p$-values are descriptive).
With MNNVL on, \AUTO{} is \emph{not} worse: its realized mean ($2.65\%$) is below UCB1's ($3.54\%$)
and its count of contexts above $10\%$ is lower (10 against 13).
Only the p99 favors UCB1
($15.2\%$ against $56.3\%$), and Softmax ($2.61\%$, 5 contexts) is the only learner that is not
worse than \AUTO{} on the mean.
Commit-once ETC is worse than UCB1 with MNNVL off (20 against 8 contexts above $10\%$, McNemar $p{=}0.008$, descriptive)
and indistinguishable with MNNVL on.
Against \AUTO{}, online selection is a tail safeguard with MNNVL off; with MNNVL on it is worse on the mean and on
contexts above $10\%$, with a smaller p99.
At scenario level UCB1 vs.\ \AUTO{} is null on both fabrics (mean $p{=}0.91$, $0.81$; $n_{>10\%}$ $p{=}0.13$, $0.72$).

Two caveats bound this result.
First, the GB200 \AUTO{} row is also priced on forced arms; the dataset's notes record contexts where an unpinned \AUTO{}
run and the forced arm differ by more than 50\%, but those runs were excluded from the processed data and are not analysed here.
Second, each fabric has a single sweep.

\begin{table}[t]
  \centering
  \scriptsize
  \setlength{\tabcolsep}{2.5pt}
  \caption{GB200 sweeps, scale-out-of-distribution split (train 4/8 nodes, test 16/32 nodes: 164 held-out contexts with
    multi-node NVLink (MNNVL) off, 158 with it on). Online hyperparameters are frozen from the H100 tuning. \capTRLT{} \capreal{} Columns: selection-priced (sel.) and realized (real.)
    mean, realized p99 (99th percentile over contexts), and $n_{>10}$, the number of contexts with realized \TRLT{} above
    10\%. \AUTO{} is a cost-model-decision replay over forced-arm measurements. \capbold{}}
  \label{tab:gb200}
  \resizebox{\columnwidth}{!}{\begin{tabular}{@{}lrrrrrrrr@{}}
\toprule
& \multicolumn{4}{c}{MNNVL-off} & \multicolumn{4}{c}{MNNVL-on} \\
\cmidrule(lr){2-5}\cmidrule(lr){6-9}
Policy & sel.\ mean & real.\ mean & real.\ p99 & $n_{>10}$ & sel.\ mean & real.\ mean & real.\ p99 & $n_{>10}$ \\
\midrule
NCCL \AUTO{} & 3.43 & 3.80 & 60.1 & 16 & 2.02 & 2.65 & 56.3 & 10 \\
ETC ($m{=}1$) & 3.42 & 5.47 & 40.5 & 20 & 2.55 & 3.60 & 27.1 & 9 \\
Softmax & 1.22 & 2.86 & 22.8 & 7 & \textbf{1.63} & \textbf{2.61} & 25.8 & \textbf{5} \\
UCB1 & 1.07 & 2.59 & 23.1 & 8 & 2.46 & 3.54 & \textbf{15.2} & 13 \\
KL-UCB & \textbf{1.03} & \textbf{2.58} & \textbf{20.3} & \textbf{5} & 2.21 & 3.35 & 23.2 & 7 \\
\bottomrule
\end{tabular}
}
\end{table}

\subsection{Workload weighting: Llama3 trace replay}
\label{sec:trace-replay}
Replaying two cycles of the Llama3 CCL-Bench trace (nine contexts), warm UCB1 finishes at $0.22\%$ cumulative relative
lost time against $1.08\%$ for \AUTO{} (selection-priced, one 16-GPU trace; Appendix~\ref{app:workload}).
The per-context tail above is unweighted, whereas a real job weights contexts by call count.
Its 4-GPU groups are mapped onto 4-node pools by rank count, so the replay weights contexts realistically but not
their intra-node arm landscape, and its test contexts are not scale-shifted.

\subsection{Two closed-loop campaigns on H100}
\label{sec:live}
The live evaluation is limited to H100: two node sets of the Spectrum-X H100 system that also ran the primary sweep, so
they are not independent systems (SPCX plugin, AllReduce only, one run per policy).
\emph{H100-A} is 8$\times$HGX H100 (64 GPUs) on Spectrum-X Ethernet with a development build of NCCL~2.30.7;
\emph{H100-B} is the same hardware over RoCE with NCCL~2.32.0a4, on the nodes of a forced-arm sweep.
Both ran NCCL's long-haul latency emulation (\texttt{NCCL\_IB\_LH\_LATENCY=40}), which the primary sweep's metadata do not
record (Appendix~\ref{app:live-setup}).
UCB1, Softmax and Thompson ran the warm-started offline champions (UCB1 with its clipped-SGD mean); Greedy
($\varepsilon{=}0$ from the cost-model prior) and ETC with one sample per arm are implemented in the plugin itself, and
$\varepsilon$-greedy ran an outdated $\varepsilon{=}0.033$ (decaying) instead of its tuned $\varepsilon{=}0$.
No run used the constant-gain mean or excluded NVLS; Appendix Table~\ref{tab:live-decomp} separates selection, overhead and setup cost.

\begin{figure*}[t]
  \centering
  \includegraphics[width=0.8\textwidth]{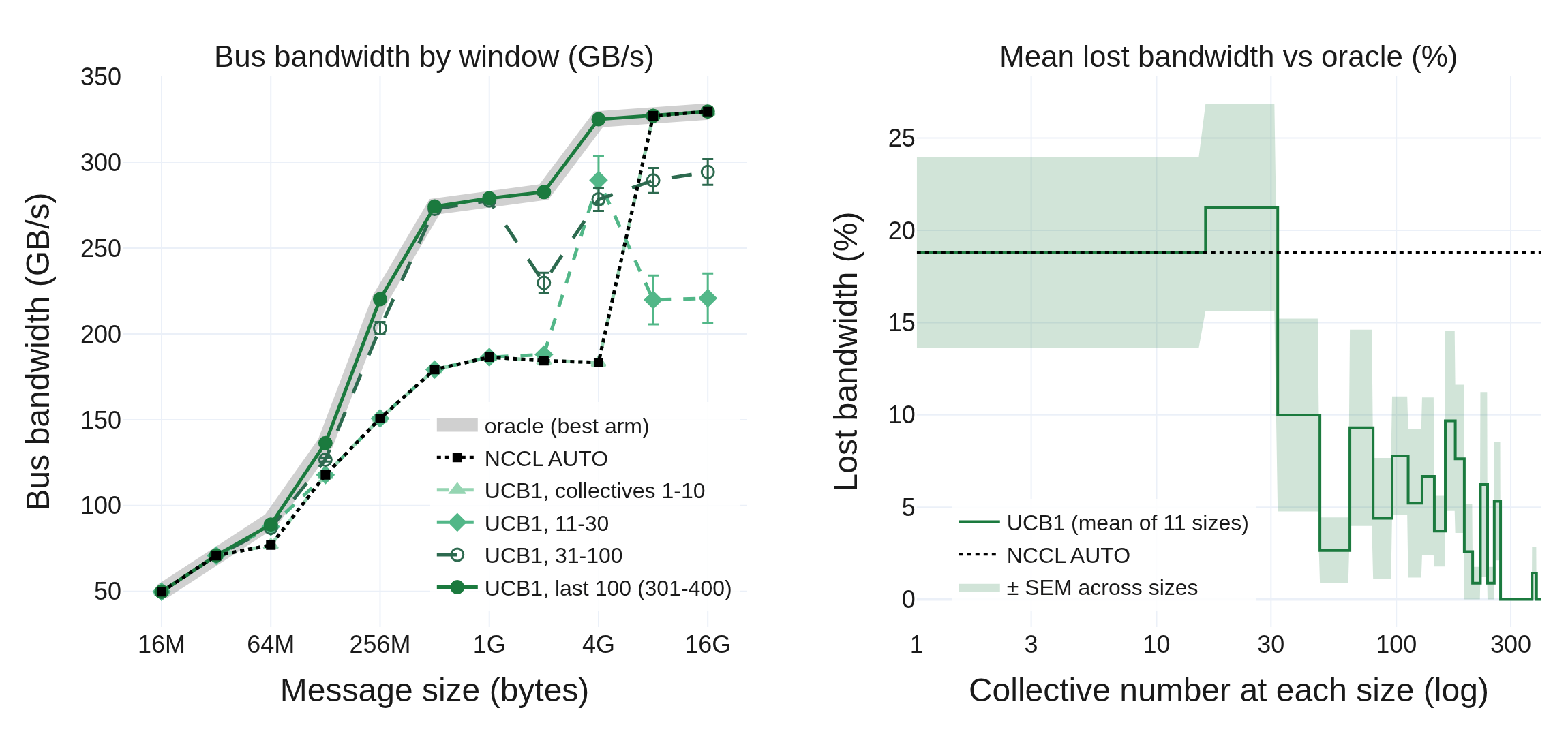}
  \caption{The offline UCB1 champion learning live (H100-A, 8 nodes, AllReduce, back-to-back, the 11 sizes from 16\,MiB,
    where \AUTO{}'s error lies; one run). Left: bus bandwidth per window of the 400 collectives at each size (error bars:
    \SEM{} over the window's collectives), beside the oracle and \AUTO{}. Right: lost bandwidth against the oracle per
    collective, averaged over the 11 sizes (band: \SEM{} across sizes). Collectives 1--10 replay \AUTO{}'s arm (warm start);
    the loss averages $20.6\%$ over collectives 11--30, $6.9\%$ over 31--100 and $0.2\%$ over the last 100, against $18.8\%$ for
    \AUTO{} throughout. Each collective is charged its arm's profiler mean latency pooled over all runs, the learners'
    included; ``\AUTO{}'' is the plugin run with \AUTO{}'s selection. Over all 26 sizes this
    run ranks third of six learners (Appendix Table~\ref{tab:live-ranking}).}
  \label{fig:live-busbw}
\end{figure*}

\paragraph{Selection quality.}
Priced by a profiler against a reconstructed oracle (setup excluded; Appendix Table~\ref{tab:live-ranking}),
\AUTO{} pays $31.4\%$ mean selection \TRLT{} on H100-A (worst size $77.3\%$), committing to
\texttt{NVLS\_TREE\_SIMPLE} across 256\,MiB--4\,GiB where ring pairs win.
On H100-B, with the same flag but one long-haul queue pair per connection instead of ten, the same choice at
256\,MiB--2\,GiB is within $8.2\%$ of the best forced arm, so the flag alone does not produce the error.
Every learner but ETC ($22.0\%$) repairs most of it; UCB1 leaves \AUTO{}'s loss after about 30 collectives per size and
plays the oracle arm over the last 100 (Figure~\ref{fig:live-busbw}).
Over all 26 sizes UCB1 reaches $8.28\%$, the third-lowest mean of six learners (mean rank $4.0$), behind $\varepsilon$-greedy
($2.85\%$; stochastic, with an untuned $\varepsilon$) and Greedy ($6.43\%$), with Thompson and Softmax at $8.49$--$9.00\%$.
On H100-B (no \AUTO{} profiler trace) UCB1 is fourth of six ($2.18\%$; best Greedy, $0.59\%$).
So the offline champion kept its relative cut over \AUTO{} ($74\%$ live, $80\%$ offline) but not its loss level ($1.53\%$
against \AUTO{}'s $7.83\%$ on the primary sweep's 8-node contexts of the same sizes) or its rank.

\paragraph{Against stock.}
On H100-A, wall-clock bus bandwidth averaged over the 26 sizes rises $1.18$--$1.24\times$ over stock for the adaptive
learners ($1.28\times$ for ETC), carried by the sizes of \AUTO{}'s error.
Below 1\,MiB every adaptive learner is slower than stock at all 11 sizes (median $+11$ to $+14\%$), and not by
selection: there they play \AUTO{}'s arm on $93$--$100\%$ of timed collectives with no higher
kernel time than the plugin run with \AUTO{}'s selection, and a state sync every 16 collectives, their inferred period,
alone adds $13\%$ host time ($4\%$ every 64; Appendix~\ref{app:deployment}).
On H100-B, against its only stock run (a separate job), every learner is faster at 1--8\,MiB (median $1.21$--$1.31\times$),
on par from 16\,MiB and slower below 1\,MiB ($0.59$--$0.65\times$; Appendix Table~\ref{tab:stock-band}); the 1--8\,MiB gain is
not selection either, since the forced run of the argmin the learners play is itself $1.26$--$1.51\times$ faster than stock.

\paragraph{Overhead and setup.}
With \AUTO{}'s selection untouched, profiling alone costs $2.1\%$ of average bus bandwidth on H100-A and a sync every four
collectives $7.5\%$ (one run each; Appendix Tables~\ref{tab:live-overhead} and~\ref{tab:live-overhead-size}).
Both campaigns stall once on the first use of \texttt{NVLS\_TREE\_SIMPLE} (median $6.6$\,s on H100-A, $5.8$\,s on H100-B),
which the offline harness under-charges about 580-fold or more (Appendix~\ref{app:live-setup}).

\paragraph{Reading.}
Good selection is needed but is not enough on its own, and the live evidence does not show a net win.
On H100-A, where \AUTO{} makes a gross error at large sizes on an emulated configuration, the learners beat stock on average
bandwidth while losing at every small size to plugin cost; on H100-B they win at 1--8\,MiB, where stock is itself slow.
These are deployment lessons, not a replication; we have no end-to-end step-time measurement and no test
under CUDA graphs, multiple communicators, or overlap.

\section{Design lessons}
\label{sec:lessons}
\paragraph{Do not learn where the model is right.}
Our offline results assume synchronization is free, but a live sync costs about $0.5$--$0.9$\,ms regardless of message size
(Section~\ref{sec:live}), which matters where collectives are short, also where learning has least to find.
The cheapest update is none: on the H100 sweep \AUTO{} is exact on every held-out context below 1\,MiB, and a size gate
that leaves those contexts to \AUTO{} (threshold chosen on train) lowers UCB1's test mean from $1.61\%$ to $1.15\%$ at a
p99 of $9.5\%$ (better in 5 of 9 scenarios, worse in none, $p{=}0.06$).
Below the gate the plugin would skip timing and sync, a saving not measured live.
The gate is platform-specific: on GB200 \AUTO{} errs at small sizes too (Appendix~\ref{app:size-gate}).

\paragraph{Filter, forget, and the filter's cost.}
The $\rho_{\max}{=}2$ arm filter never removed a measured-best arm and leaves test \TRLT{}
unchanged ($1.00\%$ vs.\ $1.01\%$) while nearly halving exploration
(Appendix Table~\ref{tab:filter-ablation}).
Under a synthetic best-to-worst pool swap, a frozen cost model pays $45\%$ dynamic \TRLT{}, ETC
$50.9\%$, plain Softmax $31.6\%$, discounted Softmax $5.96\%$, and UCB1 is already robust ($6.60\%$).
This is bound to $T{=}100$ and one change (Appendix Table~\ref{tab:drift-ablation}).

\paragraph{Price setup and transition costs.}
The NVLS stall is the clearest deployment lesson, and we price it three ways
(Table~\ref{tab:nvls-policy}).
Charged once per context replay, the ranking inverts (realized mean $546\%$ for \AUTO{}, $1{,}210\%$ for UCB1, $61\%$ for
GBM+cost), an upper bound because a communicator pays the stall once.
Charged once per communicator size, every NVLS-touching policy pays the same two tolls and the ranking becomes that of the
traffic-weighted pooled \TRLT{}: at $V{=}10^3$ visits per context it is $1.8$--$1.9\%$ for UCB1, Softmax, and KL-UCB, $2.1\%$
for ETC, $2.4\%$ for GBM+cost, and $4.1\%$ for \AUTO{}.
Excluding the setup arm from the plugin's arm set is the recommended configuration (\texttt{NCCL\_ALGO=allreduce:\^{}nvlstree}
would remove it from selection; whether that also avoids the setup is untested), chosen after seeing the test results.
\AUTO{} stays at $6.3\%$ mean and $111\%$ p99, so its tail is not an NVLS effect; UCB1 reaches $1.23\%$ and $11.2\%$,
Softmax $1.22\%$ and $8.6\%$, and ETC $3.26\%$, its tail falling from $63.0\%$ to $13.8\%$ (ETC$-$UCB1 $2.0$\,pp, all nine
scenarios, $p{=}0.004$; 3 against 2 contexts above $10\%$).
With a constant-gain mean $\hat r_a\leftarrow(1-\eta)\hat r_a+\eta r$ ($\eta{=}0.1$ chosen on train), the estimator we
recommend for a long-lived tuner, UCB1 has $1.21\%$ mean and $5.7\%$ p99, not separated from the champion here (mean
$p{=}0.89$); its case is drift (Appendix~\ref{app:recommended}). Train preferred a 1\,ms flush ($1.11\%$, $8.0\%$); we keep the per-collective schedule.
Against commit-once, learning lowers the mean but not the tail; the realized-tail gap in Table~\ref{tab:realized} is largely NVLS first observation.
Pre-warming such arms instead (perhaps with \texttt{NCCL\_RUNTIME\_CONNECT=0}) is untested.

\section{Limitations and artifact availability}
\label{sec:limitations}
\paragraph{Evidence scope.}
The primary replay is one hardware campaign (H100, Spectrum-X Ethernet, three collectives) chosen as a domain-shift
stress test of the shipped model; the finding may be specific to the NCCL~2.30.4 model on Spectrum-X.
Each arm's pool is 100 iterations from \emph{one} isolated forced-arm job, so job-to-job variation is not measured, and the \AUTO{} row is the cost model's choice priced on
forced arms.
Stock and repeated jobs, repeated or non-emulated live runs, and a live run of the recommended configuration or a size
gate need new runs we have not done.
On H100-B (all-reduce, 8 nodes, separate jobs) the learners started every size on the exported argmin, and arms they kept for a whole size cost $1.02$--$1.18\times$ their forced runs except NVLSTree/Simple ($0.70$--$0.98\times$), consistent with forcing slowing only NVLS among the arms checked; stock (no plugin) differed from both (unlogged choice or channels, or job variation). So there the \AUTO{} row is the cost model's choice applied through the tuner, not necessarily stock's, and an operator should compare a pass-through plugin with stock before crediting gains to learning (Appendix~\ref{app:tail-anatomy}).
This qualifies the offline magnitudes, not the mechanism: a deployed learner scores the arms it actually runs, so if forcing
changes an arm's latency the loop learns the deployed latency, while the cost model's choice stays fixed.
The horizon is a data limit ($T{=}100$), the ranking is horizon-bound (Appendix Table~\ref{tab:horizon-rtl}), and the drift
evidence is one synthetic swap, so behavior at $T{\ge}10^4$ is untested.
The search space is algorithm--protocol pairs only (a median of two arms after filtering).

\paragraph{Replay fidelity.}
The sweeps are isolated micro-benchmarks: no cross-arm connection reuse, no
compute--communication overlap or concurrent communicators, and measurement noise independent
across arms, whereas real noise is correlated in time.
The reward is collective latency, not step time, so we cannot say how selection gains translate
into job throughput.
Pull-count replay is a simulator built from forced-arm pools, not an unbiased replay
estimator in the sense of Li et al.~\cite{li2011replay}.

\paragraph{Live campaigns and deployability.}
The two H100 campaigns each cover one collective at one scale, one run per policy, and latency emulation, with oracles
pooled across the scored runs.
The plugin builds used live were not recorded with the runs, so we cannot tie them to a source revision; its interface version, sync rule and transport, how the stochastic learners shared draws, and how UCB1's clipped-SGD state was reconciled
across ranks are unknown; with data checking off, no learner's rank consistency is verified, and the stochastic Softmax and Thompson rows may not have been rank-consistent.
The live reward is the cross-rank maximum time, the offline one GPU~0's; we have not quantified how they differ.
Action consistency across ranks rests on a deterministic policy over replicated statistics with an
\AUTO{} fallback (Appendix~\ref{app:deployment}); we have not fault-injected the fallback or verified
bitwise-identical reductions.
Under CUDA-graph capture the choice is fixed at capture, so learning applies only to eager collectives or across recaptures.

\paragraph{Accounting and comparators.}
Realized \TRLT{} over-charges exploration and selection-priced \TRLT{} under-charges it, so we report both.
Neither charges the NVLS stall, which we re-price separately, and both weight contexts equally, not by call frequency.
The supervised comparators are fit-once selectors; the tree models were tuned with the learners' budget on the H100
splits and with NVLS excluded (Appendix Table~\ref{tab:tuned-gbm-nvls}), but the GB200, label-budget and largest-scale
static rows use fixed settings.
We do not claim superiority over supervised methods in general, and we did not run active-learning selectors such as
ACCLAiM or FACT.
We ran no head-to-head against other online collective tuners (STAR-MPI, OMPICollTune,
AutoCCL), so the cost-model prior, our only differentiator from them, is untested against them.
Tuning used the selection-priced objective; re-tuning five families on the realized objective
did not change the picture (Appendix~\ref{app:realized-tuned}), but ETC and the supervised
models were not re-tuned on it.

\paragraph{Artifact availability.}
The NCCL tuner plugin is planned for release inside SPCX (we cannot give a date or terms); release of the evaluation
harness, policy code, and processed data is under consideration, not committed, so a reader cannot yet re-run the tables.

\section{Conclusion and future work}
\label{sec:conclusion}
On a forced-arm sweep of Spectrum-X H100 nodes under a scale shift, the argmin of NCCL's cost model is exact on most
contexts and badly wrong on a few, where it misplaces a protocol or algorithm crossover.
A per-context bandit seeded by that model cuts the contexts losing more than 10\% from 19 to 1 (p99 103\% to 9.9\%);
with hyperparameters frozen, on two GB200 fabrics it cuts p99 but not reliably the count (16 to 8, 10 to 13), and
scenario-level tests are null.
Against fit-once supervised selectors tuned with the same budget, online selection has a similar mean and, under scale
extrapolation, a narrow tail edge (7 against 1 contexts above 10\%, $p{=}0.06$); at seen scales the static fits match or
beat it.

Next are a shared contextual bandit and channel-count arms.
GB200 is evaluated offline only, and a live GB200 run is future work: we lacked the resources to complete a verified one
before the deadline.

\bibliography{main}
\bibliographystyle{mlsys2025}

\section*{AI Usage Disclosure}
\label{sec:ai-disclosure}
We used LLM-based coding agents in the Cursor IDE (Anthropic Claude Opus and OpenAI GPT models) throughout this work,
always directed and reviewed by the authors.
\emph{Code:} the agents wrote most of the offline emulation harness, the bandit policy implementations, the hyperparameter
search drivers, the data-processing, analysis, plotting and table-generation scripts, and their unit tests; the authors
specified the designs and reviewed the code.
The tuner and profiler plugin was developed by the authors.
\emph{Experiment design and analysis:} the agents proposed and ran ablations and robustness analyses (for example the
transient, split-robustness, NVLS-exclusion and largest-scale holdout studies) and summarized their outputs; the authors
chose which analyses to run, what to report and how to interpret it.
\emph{Data:} no data were generated by AI. Every latency measurement comes from NCCL runs on the hardware described in
the paper; the synthetic workloads are seeded generators written as code.
\emph{Literature:} the agents helped search for and summarize related work; the authors checked every cited reference
against its source.
\emph{Writing:} the agents drafted and edited prose, captions and the appendices. Numbers in the text were
copied from analysis-script output (generated tables or their CSVs); reviewer rounds checked prose against the tables
and found and corrected mismatches, so some may remain.
\emph{Internal review:} before submission, LLM agents acting as reviewers critiqued successive drafts in repeated rounds, and we revised the
paper in response to each.
The authors take full responsibility for all content, including the correctness of every claim and citation.

\ifblindbuild\clearpage\fi
\appendix
\section{Notation and acronyms}
\label{app:notation}
Table~\ref{tab:notation} collects the symbols and acronyms used in the paper; each is also
defined at first use.

\begin{table}[h]
  \centering
  \caption{Notation and acronyms (defined at first use in the body).}
  \label{tab:notation}
  \scriptsize
  \setlength{\tabcolsep}{3pt}
  \begin{tabular}{@{}>{\raggedright\arraybackslash}p{0.24\columnwidth}
                   >{\raggedright\arraybackslash}p{0.68\columnwidth}@{}}
    \toprule
    symbol / acronym & meaning \\
    \midrule
    \MAB{} & multi-armed bandit \\
    \AUTO{} & NCCL's static cost-model selector (baseline) \\
    \KPI{} & key performance indicator \\
    \SEM{} & standard error of the mean \\
    ETC & explore-then-commit \\
    FPL & follow-the-perturbed-leader \\
    KL-UCB & Kullback--Leibler upper confidence bound \\
    GBM+cost & gradient-boosted regressor with the cost-model prediction as a feature \\
    $\hat{\mu}(a)$ & cost model's predicted latency for arm~$a$ \\
    $\hat{\mu}_{\min}$ & best predicted latency, $\min\{\hat{\mu}(a):\hat{\mu}(a)\ \text{defined}\}$ \\
    $m$ & predicted runner-up margin $\hat{\mu}_{(2)}/\hat{\mu}_{(1)}-1$ \\
    $\bar{\mu}_a$ & reference-sweep mean latency of arm~$a$ \\
    $a^\star$ & measured-best (oracle) arm, $\arg\min_a \bar{\mu}_a$ \\
    $A_t$ & arm selected at iteration~$t$ \\
    $r(t)$ & pull reward $\hat{\mu}_{\min}/t$ (ratio to predicted best) \\
    $P$ & number of ranks participating in the collective \\
    $n$ & message size (bytes) \\
    $K$ & number of (filtered) arms in a context \\
    $T$ & horizon (iterations per context replay; $T{=}100$ main) \\
    \RLT{} & Relative Lost Time (per iteration) \\
    \CRLT{} & Cumulative Relative Lost Time \\
    \TRLT{} & Total Relative Lost Time (per context replay; headline accounting is realized latency) \\
    \bottomrule
  \end{tabular}
\end{table}

\section{Empirical motivation: figures and analysis}
\label{app:empirical-motivation}

This appendix gives the figures and analysis behind the failure-mode and design
motivation of Sections~\ref{sec:failure} and~\ref{sec:method}.

\subsection{Few recurring contexts}
\label{subsec:context-space}

\begin{figure}[!ht]
  \centering
  \includegraphics[width=\linewidth]{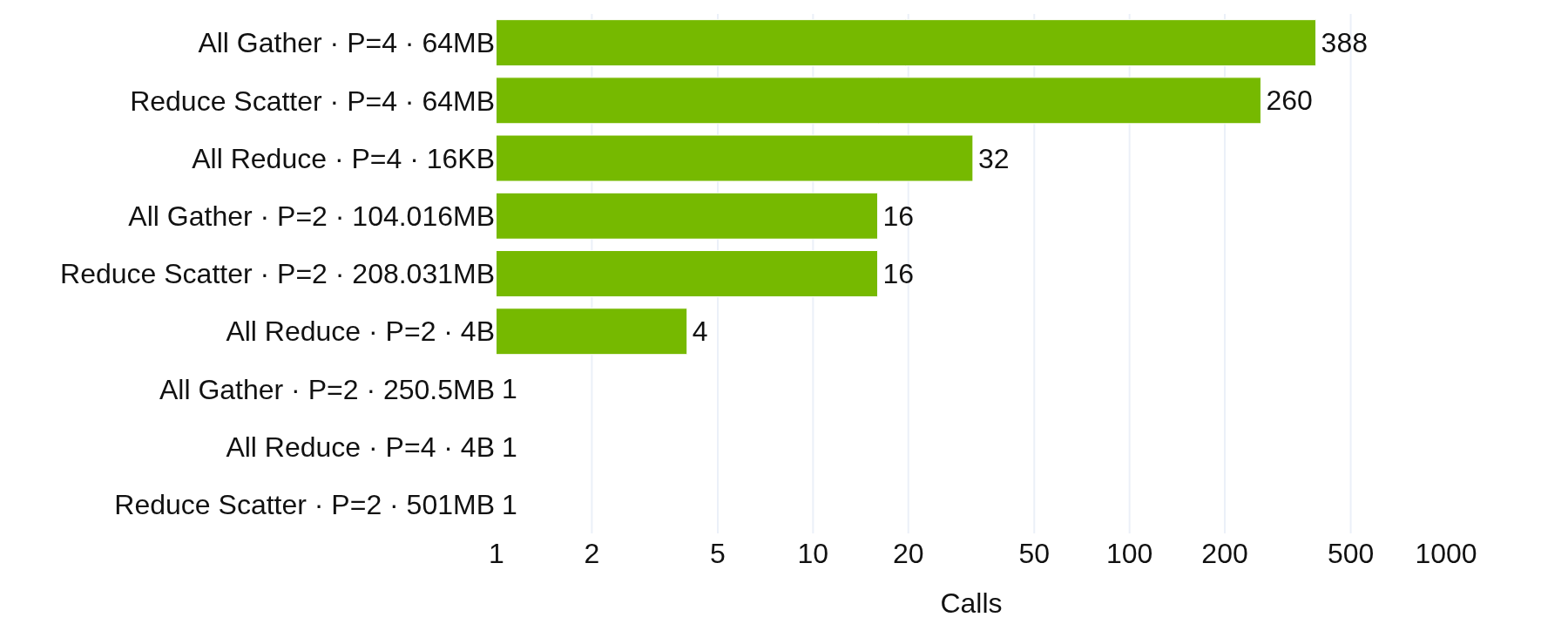}
  \caption{CCL-Bench Llama3-8B torchtitan \emph{training} trace~\cite{cclbench2026}
    (Perlmutter, 16~GPUs): collective-call counts
    per bandit context (collective type, trace process-group size~$g$, message
    size~$n$). \cclBenchCalls{} calls over \cclBenchContextsWord{} contexts
    ($\cclBenchMeanReuse{}\times$ \cclBenchMeanReuseDef{}); log-scale
    $x$-axis.}
  \label{fig:ccl-bench-contexts}
\end{figure}

Production LLM training jobs revisit a modest set of collective communication
contexts.
In a representative Llama3-8B torchtitan \emph{training} trace from
CCL-Bench~\cite{cclbench2026} (public Kineto workload on Perlmutter,
16~GPUs), \cclBenchCalls{} collective calls
collapse to only \cclBenchContextsWord{} distinct
(collective, $g$, message-size) contexts
($\cclBenchMeanReuse{}\times$ \cclBenchMeanReuseDef{}; $g$ is the trace
process-group size, e.g.\ four ranks on tensor-parallel groups, distinct from
the measurement-side rank count~$P$);
the two dominant contexts (64~MiB \texttt{all\_gather} and \texttt{reduce\_scatter} with
$g{=}4$) account for $\cclBenchTopTwoTrafficPct{}\%$ of calls
(Figure~\ref{fig:ccl-bench-contexts}).
Long jobs therefore revisit the same contexts often, which amortizes in-job exploration
without hundreds of learners.
This small, heavily reused context set is why we run one independent bandit per context
(Section~\ref{sec:method}): with so few contexts, cross-context generalization is unnecessary.

\subsection{Non-stationarity and steady-state arm evaluation}
\label{subsec:non-stationarity}

\begin{figure}[t]
  \centering
  \includegraphics[width=\linewidth]{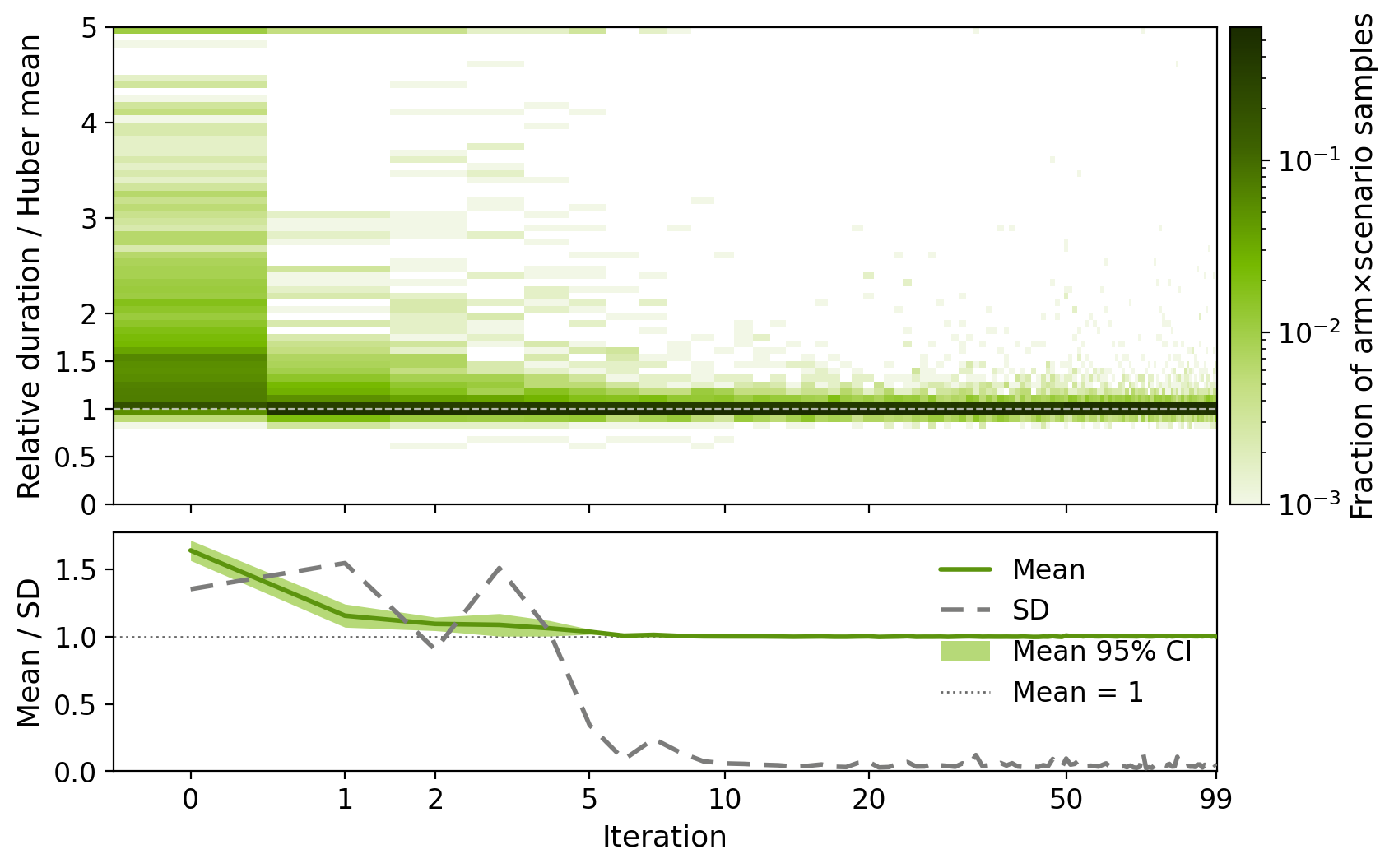}
  \caption{Pooled column-normalized heatmap of per-arm relative duration
    (latency divided by the arm's Huber location over iterations $0,\ldots,99$)
    vs.\ iteration on the H100 sweep grid.
    Early iterations concentrate probability mass above steady state
    (${\gg}1{\times}$ Huber mean), motivating the transient window
    excluded before ranking measured-best arms
    (Figure~\ref{fig:cohens-d-error-rate}).}
  \label{fig:arm-duration-heatmap}
\end{figure}

Defining the measured-fastest arm is itself non-trivial.
Figure~\ref{fig:arm-duration-heatmap} pools every arm's per-iteration
\emph{relative duration}, defined as latency divided by that arm's Huber location
(a robust mean that down-weights outliers~\cite{huber1964}) over the
full trace, and column-normalizes each iteration into a distribution, with color
encoding probability mass (log-scaled).
Early iterations concentrate mass well above $1\times$: iteration~0 (the perf
binary's warmup, dropped from the pools) and the next several pulls are
heavy-tailed outliers, after which the mass collapses onto the steady-state band
near $1\times$ within the first $10$ pulls (the lower panel traces the pooled mean
and standard deviation).
NCCL lazily initializes connections and related state on first use of each
$(\text{collective}, \text{nodes}, \text{algo-proto})$ combination, so this
transient is not only job-start remanence: it recurs when switching to an
algo-proto not exercised before in the same process (and may be absent when
an algo-proto was already warmed by a prior collective at that size).
That the transient is \emph{lazy initialization} is inferred from its dependence on the
iteration index within each arm's isolated job, not shown by a controlled experiment such as
connecting peers at init (\texttt{NCCL\_RUNTIME\_CONNECT=0}), which we did not run.
These outliers motivate robust (Huber) location estimates.
To locate the transient-to-steady boundary instead of judging it by eye, we fit a Bayesian
switchpoint model to the pooled relative-duration trace.
The model is one versus two Gaussian segments with a discrete-uniform prior on the change
point~$\tau$ (sampled with PyMC), and we compare the two by leave-one-out cross-validation (LOO) and WAIC (predictive model-comparison
criteria that estimate out-of-sample fit~\cite{vehtari2017,watanabe2010}),
with a fast BIC grid (Bayesian Information Criterion, a likelihood score
penalized for model complexity~\cite{schwarz1978}) as a companion check.
The two-segment model is decisively favored (LOO weight $1.00$), with an
elevated transient segment ($\mu_1{=}1.13\times$ the Huber mean) dropping
to a steady segment ($\mu_2{=}1.00\times$); the posterior concentrates the
change point at iteration $\tau{=}8$ with a posterior standard deviation of
only $0.2$ iterations. Its 90\% highest-density interval (HDI), the shortest
interval containing that share of the posterior mass~\cite{kruschke2015}, lies
within a single iteration.
We therefore conservatively begin the steady-state window one transient beyond
that estimate and rank algorithm--protocol arms by mean latency over iterations
$11,\ldots,99$ (iteration~0 dropped; iterations $1$--$10$ excluded as the
lazy-init transient).
The early transient therefore sets two evaluation choices used throughout:
robust (Huber) location estimates to blunt transient outliers, and ranking arms
by steady-state performance over iterations $11$--$99$ instead of by
final-iteration or naive-mean latency.
The same transient also constrains the \emph{online} learner, not only the offline
ground truth.
After first pulling an arm, a policy must either skip the initial lazy-init samples
(the cold-start window) or apply robust online updates (e.g.\ clipped-SGD-style
estimation~\cite{sankararaman2023}) that down-weight the early heavy-tailed outliers,
so that a handful of transient samples do not corrupt its arm-value estimates.
\subsection{Cost-model error analysis}
\label{subsec:cost-model-error}

\begin{figure}[t]
  \centering
  \includegraphics[width=\linewidth]{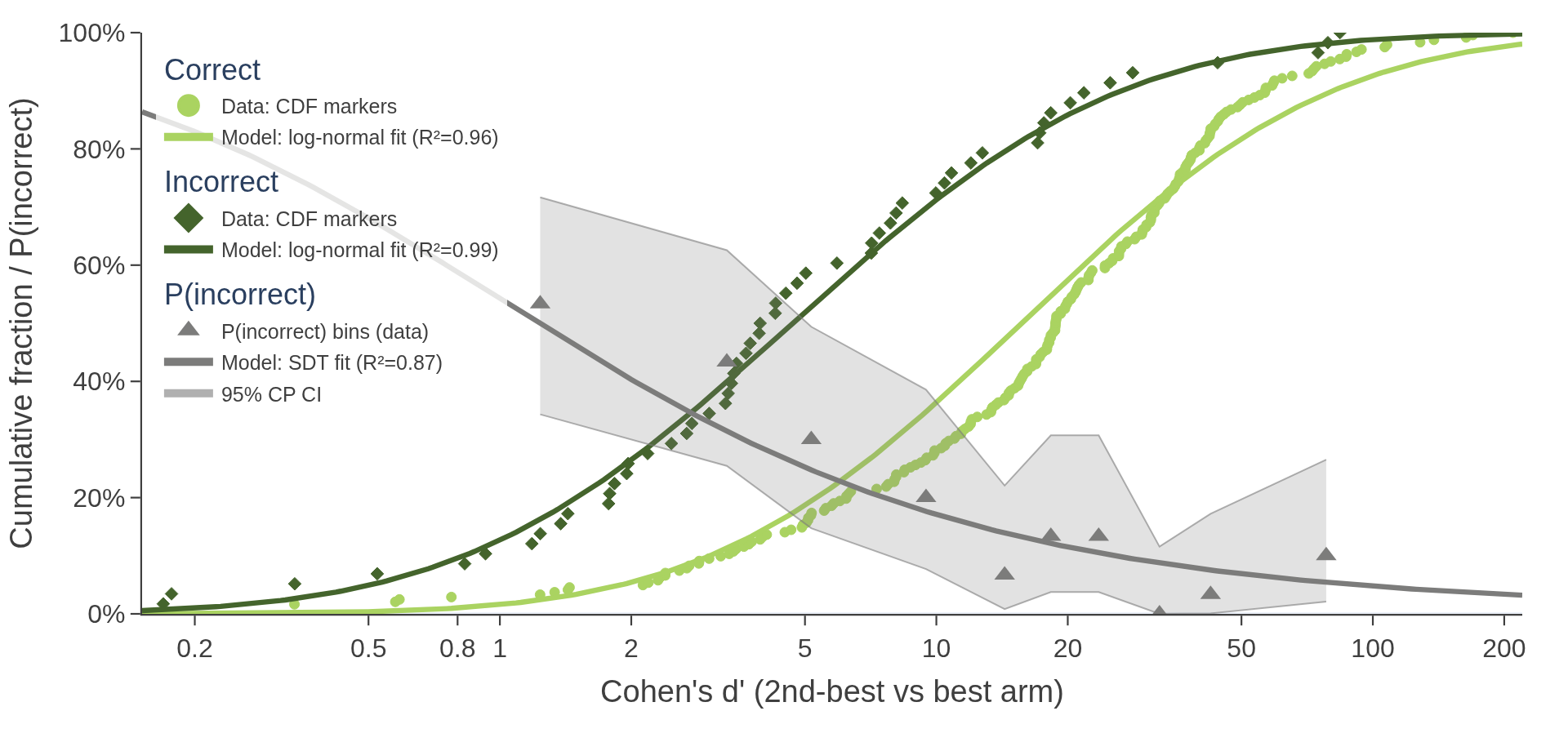}
  \caption{When the two best arms are close in latency (small Cohen's~$d'$), the cost
    model is most likely to mis-rank them, making static $\arg\min \hat{\mu}(a)$
    selection unreliable; online measurement can break such near-ties.
    H100 sweep grid ($n{=}300$ contexts with cost tables; measured-best ranked on
    iterations $11$--$99$).
    Light/dark green: empirical CDF of Cohen's~$d'$~\eqref{eq:cohens-d-prime}
    for cost-model-correct and incorrect contexts (markers) with log-normal
    fits in log-$d'$ ($R^2{=}0.96$ / $0.99$).
    Gray: empirical $P(\mathrm{incorrect})$ in equal-count deciles of~$d'$
    (triangles; $x$-axis at bin median~$d'$; shaded band: 95\%
    Clopper--Pearson~\cite{clopperpearson1934} CI) and Bayes posterior from the fitted class-conditional
    log-normal densities and empirical priors ($R^2{=}0.87$).}
  \label{fig:cohens-d-error-rate}
\end{figure}

NCCL's shipped selector plays $\arg\min_a \hat{\mu}(a)$ on exported cost tables;
we audit that fixed choice against the same open-loop, forced-arm sweep measurements used
in emulation, ranking the measured-best arm on the steady-state window
(iterations $11$--$99$) established above.
At each scenario (collective $\times$ node count $\times$ message size) let
$\mu_{\mathrm{best}}$ and $\mu_{\mathrm{2nd}}$ denote the means of the
fastest and second-fastest arms; Cohen's~$d'$ (a standardized effect
size~\cite{cohen1988}) measures their separation in pooled-standard-deviation units as
\begin{equation}
  \label{eq:cohens-d-prime}
  d' = \frac{\mu_{\mathrm{2nd}} - \mu_{\mathrm{best}}}{s_{\mathrm{pooled}}},
\end{equation}
where $s_{\mathrm{pooled}}$ is the standard two-sample pooled standard deviation
of the per-iteration latency samples for those two arms
($s_{\mathrm{pooled}} = \sqrt{((n_1{-}1)s_1^2 + (n_2{-}1)s_2^2)/(n_1{+}n_2{-}2)}$).
Across $n{=}300$ contexts with both measurements and cost predictions, the
model's predicted-best arm matches the measured fastest arm in roughly four of five
cases (the pooled error rate of Section~\ref{sec:failure} is $20.0\%$).
Figure~\ref{fig:cohens-d-error-rate} overlays empirical CDFs of Cohen's~$d'$ for
cost-model-correct and incorrect scenarios, log-normal fits to each
($R^2{=}0.96$ / $0.99$), the decile-binned $P(\mathrm{incorrect})$ (the
fraction where $\arg\min \hat{\mu}(a)$ differs from the measured-best
arm), and a Bayes posterior derived from the fitted class-conditional
densities ($R^2{=}0.87$ vs.\ bin medians).
Two fitted model families summarize this figure.
First, for each subgroup we fit a class-conditional log-normal to~$d'$ by
maximum likelihood, taking $\mu$ and $\sigma$ as the mean and standard deviation
of $\ln d'$ over that subgroup's scenarios; the fitted CDF is
$\Phi\!\big((\ln d' - \mu)/\sigma\big)$ (Gaussian in $\ln d'$, i.e.\ log-normal
in~$d'$), and the reported $R^2$ values compare each fit against its empirical
CDF evaluated at the observed~$d'$.
Second, the gray $P(\mathrm{inc}\mid d')$ curve is not fit to the binned
points separately; it is derived from the fitted log-normal densities
$f_c, f_i$ (correct/incorrect) and the empirical class priors
$\pi_c, \pi_i$ (correct/incorrect; empirically $\pi_c\approx0.8$,
$\pi_i\approx0.2$, the split above) via Bayes' rule,
\begin{equation}
  \label{eq:sdt-posterior}
  P(\mathrm{inc}\mid d') =
  \frac{\pi_i\, f_i(d')}
       {\pi_i\, f_i(d') + \pi_c\, f_c(d')},
\end{equation}
which is exactly the likelihood-ratio (signal-detection-theory) classifier for
the two classes; its $R^2{=}0.87$ is measured against the decile-binned
empirical $P(\mathrm{incorrect})$ at bin medians.
Smaller~$d'$ coincides with higher $P(\mathrm{incorrect})$: when the two
leading arms are close in latency, a small prediction error flips the ranking,
so a static analytic model's argmin is least reliable in this near-tie regime.
This explains the error \emph{rate}, not the tail: a near-tie flip costs little, and the
contexts that dominate the loss in Section~\ref{sec:failure} are not near-ties (their
measured gaps are large and survive within-pool resampling).
We do not claim the gap is irreducible for a static selector in general: a supervised fit with the cost-model prediction
as a feature matches the online mean on H100 (Section~\ref{sec:h100}).
The pooled-standard-deviation $d'$ also treats the 89 steady iterations as independent, which
overstates separation if they are autocorrelated.
Unlike the main-text runner-up margin
(Figure~\ref{fig:cost-model-margin-error-rate}), $d'$ uses measured means and
variances and is therefore a \emph{post-hoc} separation diagnostic, not a
decision-time signal. It complements the observable audit by explaining why
measured near-ties are fragile; that figure shows directly that the model's own
small predicted margin does not bound realized \TRLT{}.

\subsection{Cost-model arm filter threshold}
\label{subsec:arm-filter-threshold}

\begin{figure}[t]
  \centering
  \includegraphics[width=\linewidth]{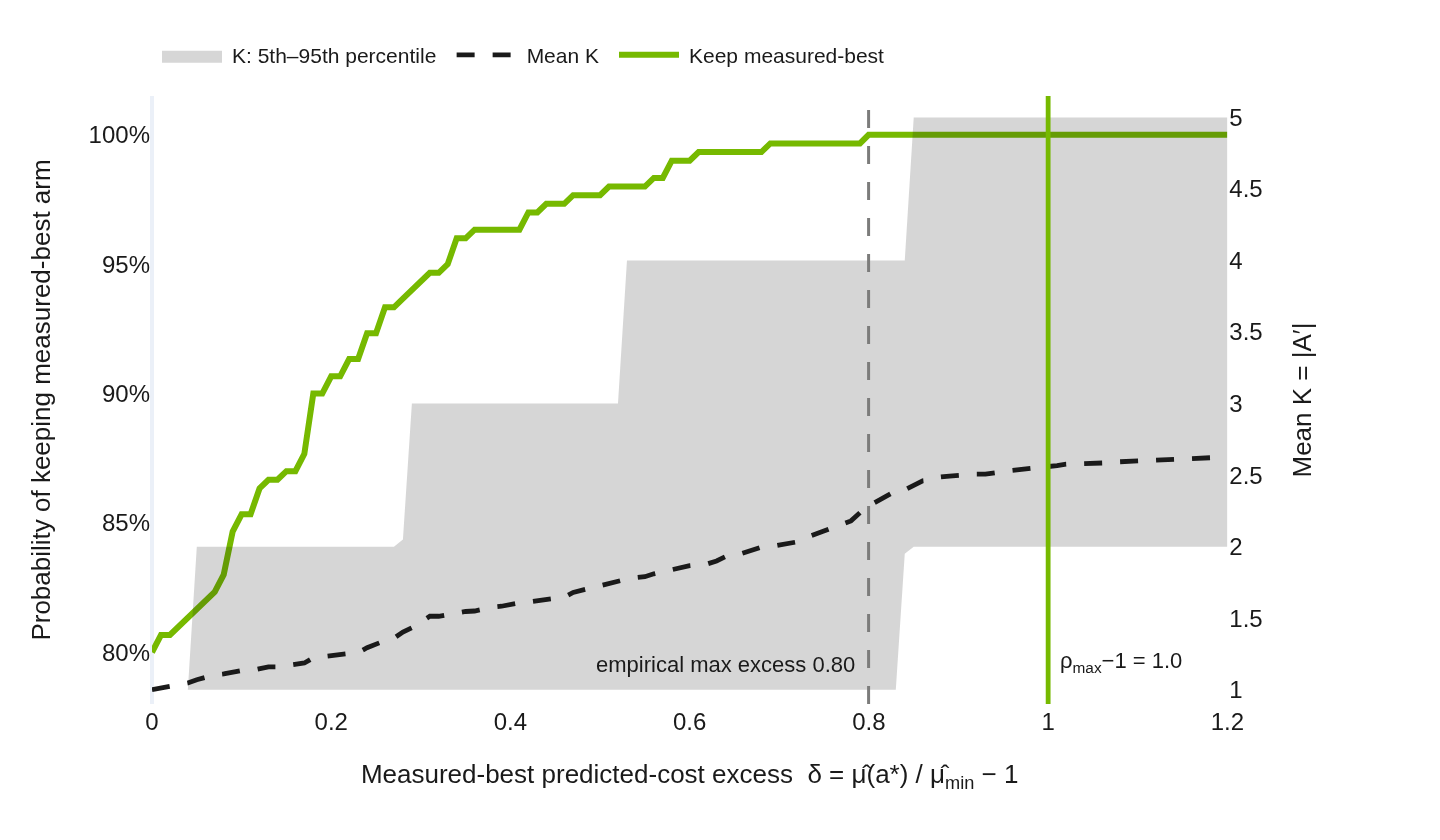}
  \caption{Choosing the predicted-cost filter threshold in excess-ratio units,
    $\delta=\hat{\mu}(a^\star)/\hat{\mu}_{\min}-1$
    (Equation~\eqref{eq:filter}): probability of retaining the measured-best
    arm (solid green, left) and mean offered-arm count $K{=}|\mathcal{A}'|$
    (dashed, right) with a 5th--95th percentile ribbon across contexts,
    vs.~$\delta$.
    H100 sweep grid ($n{=}300$ contexts with cost tables; measured-best on
    iterations~$11$--$99$).
    The measured-best arm's predicted-cost excess never exceeds $\delta=0.80$
    (vertical gray dashed line); the unchanged $\rho_{\max}{=}2$ filter appears
    as $\rho_{\max}-1=1.0$ (solid green) and keeps
    $P(\mathrm{keep\ best}){=}100\%$ with mean $K{\approx}2.56$ of ${\sim}4$
    arms (5th--95th percentile $[2,5]$). This retrospective measured-best
    excess ratio is not the observable runner-up margin in
    Figure~\ref{fig:cost-model-margin-error-rate}.}
  \label{fig:arm-filter-tradeoff}
\end{figure}

Filter~\eqref{eq:filter} prunes arms whose predicted cost exceeds
$\rho_{\max}$ times the model minimum.
Figure~\ref{fig:arm-filter-tradeoff} shows the safety--pruning tradeoff on the
H100 sweep grid in excess-ratio units
$\delta=\hat{\mu}(a^\star)/\hat{\mu}_{\min}-1$: the measured-best arm's
predicted cost has empirical maximum excess $\delta=0.80$, so
$P(\mathrm{keep\ best})$ reaches $100\%$ at that floor. The configured
$\rho_{\max}{=}2$ is unchanged and appears as $\rho_{\max}-1=1.0$, a round
value just above it that
retains a mean of ${\approx}2.56$ of ${\sim}4$ offered arms
(5th--95th percentile $[2,5]$ across contexts).
This retrospective ratio evaluates where the \emph{measured-best} arm falls in
the predicted ordering; it is distinct from the observable runner-up margin
$m=\hat{\mu}_{(2)}/\hat{\mu}_{\min}-1$ in
Figure~\ref{fig:cost-model-margin-error-rate}.

\section{Bandit policy specifications}
\label{app:policies}

All policies share the loop in Algorithm~\ref{alg:loop}: maintain per-arm
statistics, select $A_t$, observe reward $R_t$, update.
Let $\hat{\mu}^R_a$ denote the running mean reward on arm~$a$, $n_a$ its pull
count, and $N_t = \sum_a n_a$.
Unpulled arms ($n_a = 0$) are explored first where noted.
Selection rules below are implementation-agnostic; champion hyperparameters are
documented in Appendix~\ref{app:hyperopt} (Table~\ref{tab:policy-config}), and the
train/test holdout in Section~\ref{sec:protocol}.

\paragraph{Warm start seeds sufficient statistics per family.}
Warm start converts the cost-model reward guess $r(\hat{\mu}(a)) =
\hat{\mu}_{\min}/\hat{\mu}(a)$ into each policy's own initial statistics, weighted by
a per-family tuned pseudo-count $\nu$ (equivalently $n_0$; a number of \emph{virtual}
prior pulls). Mean-based policies (UCB1, UCB-V, UCB-tuned, Gaussian Thompson,
$\varepsilon$-greedy, KL-UCB, Softmax) initialize the running mean reward
$\hat{\mu}^R_a \gets r(\hat{\mu}(a))$ backed by $n_a \gets \nu$ virtual pulls
(the Gaussian-Thompson form is stated in its algorithm block below); loss-based policies (EXP3,
Tsallis-Inf, FPL) instead seed the cumulative loss $\hat{L}_a(0) = \nu\,(1 -
r(\hat{\mu}(a)))$. In every family $\nu{=}0$ recovers the cold start exactly, so a
champion tuned to $\nu{=}0$ (EXP3, FPL) is cold even when the warm flag is set, which
is why the ranking table's ``(warm)''/``(cold)'' suffix and
Table~\ref{tab:policy-config} agree per family. The three families whose
implementations expose no such seeding (EXP3++, ETC, Madan RR-ETC) are always cold.

\subsection{NCCL cost model (\AUTO{})}
Non-learning baseline: always play $A_t = \arg\min_{a \in \mathcal{A}'} \hat{\mu}(a)$
matching NCCL's built-in cost-model selector (minimum predicted elapsed
time)~\cite{nccl,williams2025nccltuning}.
The policy does not update from observations; it fixes the same argmin arm at
each context.
\subsection{Uniform random}
$A_t \sim \mathrm{Uniform}(\mathcal{A}')$; ignores rewards.
It is an unstructured lower reference.
\subsection{UCB1}
\begin{equation}
  \label{eq:ucb1}
  A_t = \arg\max_{a \in \mathcal{A}'}
  \left[
    \hat{\mu}^R_a + \alpha \sqrt{\frac{2 \ln N_t}{n_a}}
  \right],
\end{equation}
with exploration coefficient $\alpha > 0$~\cite{auer2002}.
Unpulled arms are chosen uniformly at random until each arm has at least one pull
(cold start only).
\subsection{UCB-V}
Uses empirical variance $\hat{\sigma}^2_a$ (Welford) and Bernstein-style bounds~\cite{audibert2009}:
\begin{equation}
  \label{eq:ucbv}
  A_t = \arg\max_{a \in \mathcal{A}'}
  \left[
    \hat{\mu}^R_a +
    \sqrt{\frac{2\hat{\sigma}^2_a \zeta \ln N_t}{n_a}}
    + \frac{c\,\zeta \ln N_t}{n_a}
  \right],
\end{equation}
with parameters $\zeta > 0$, $c > 0$.
Unpulled arms are chosen uniformly at random until each arm has at least one pull.
\subsection{UCB-tuned}
Variance-aware UCB of Auer et al.~\cite{auer2002}: with empirical variance
$V_a = \hat{\sigma}^2_a + \sqrt{2\ln N_t / n_a}$,
\begin{equation}
  \label{eq:ucb-tuned}
  A_t = \arg\max_{a \in \mathcal{A}'}
  \left[
    \hat{\mu}^R_a + \sqrt{\frac{\ln N_t}{n_a}\min\!\left(\tfrac{1}{4}, V_a\right)}
  \right].
\end{equation}
Unpulled arms are chosen uniformly at random until each arm has at least one pull.
\subsection{KL-UCB}
Bernoulli KL-UCB of Garivier and Capp{\'e}~\cite{garivier2011klucb}: each arm is
pulled once, then
\begin{multline}
  \label{eq:kl-ucb}
  A_t = \arg\max_{a \in \mathcal{A}'}
  \max\bigl\{
    q \in [\hat{\mu}^R_a, 1] : \\
    n_a\,\mathrm{KL}(\hat{\mu}^R_a, q) \le c\,\ln N_t
  \bigr\},
\end{multline}
with exploration constant $c > 0$ (bisection on $q$).
Empirical means are clamped to $[0,1]$ for the Bernoulli KL
(harness rewards occasionally exceed~$1$).
Warm start seeds per-arm mean/count from cost-model guesses when enabled.
\subsection{\texorpdfstring{$\varepsilon$}{epsilon}-greedy}
With exploration rate $\varepsilon_t$ (fixed or decayed as
$\varepsilon_t = \varepsilon_0 / \sqrt{t}$ when enabled),
explore with probability $\varepsilon_t$ ($A_t$ uniform on $\mathcal{A}'$), else
$A_t = \arg\max_a \hat{\mu}^R_a$~\cite{auer2002}.
\subsection{Explore-then-commit (ETC)}
For $m$ rounds per arm, pull each $a \in \mathcal{A}'$ exactly $m$ times, then
commit to $\arg\max_a \hat{\mu}^R_a$.
During exploration, each pull selects uniformly among arms with fewer than $m$
observations ($m$ exploration rounds).
\subsection{Madan et al. round-robin (deterministic ETC)}

Following Madan et al.~\cite{madan2026}, this policy uses fixed
round-robin exploration, IQR-trimmed commit, and an \AUTO{} safety gate.
Arms $\mathcal{A}'$ are sorted lexicographically once per context (same order as
the emulation environment).
Let $m$ be the number of exploration rounds and $B {=} m\,|\mathcal{A}'|$.
\paragraph{Exploration ($p < B$ inner updates).}
While the exploration counter $p < B$, play
$A_t = \mathcal{A}'[p \bmod |\mathcal{A}'|]$ (the index advances only on an
inner-policy update).
Each observed pull stores harness reward $R_t$; define per-pull cost
$c_t = -R_t$ (higher reward $\Leftrightarrow$ lower cost under
~\eqref{eq:ratio-reward}).
\paragraph{Commit.}
For each arm $a$, let $\mathcal{C}_a$ be the costs from its exploration pulls.
Compute the Tukey IQR-trimmed mean $\bar{c}_a$: drop samples outside
$[Q_1 - 1.5\,\mathrm{IQR},\, Q_3 + 1.5\,\mathrm{IQR}]$ (linear interpolation
for quartiles), then average the remainder.
Let $a^\star = \arg\min_a \bar{c}_a$ and $a_{\AUTO{}}$ be the cost-model arm
($\arg\min \hat{\mu}$).
With $\bar{c}_{\AUTO{}}$ and $\bar{c}_{a^\star}$, define relative improvement
$\iota = (\bar{c}_{\AUTO{}} - \bar{c}_{a^\star})/\bar{c}_{\AUTO{}}$ when
$\bar{c}_{\AUTO{}} > 0$; else $\iota = +\infty$ if $\bar{c}_{a^\star} < \bar{c}_{\AUTO{}}$,
and $\iota = 0$ otherwise.
Let $\gamma$ be the safety-gate margin.
If $\iota < \gamma$, commit to $a_{\AUTO{}}$; else commit to $a^\star$.
\paragraph{Exploitation.}
Play the committed arm on every subsequent pull.
\subsection{Follow the perturbed leader (FPL)}
We use the best-of-both-worlds FPL of Honda et al.~\cite{honda2023}, which
operates in loss space with a Fr\'echet-tailed perturbation.
Let $\ell_s = 1 - R_s$ and maintain an importance-weighted estimate
$\hat{L}_a$ of the cumulative loss of arm $a$.
Each round draw $r_a \sim \mathcal{F}_2$ i.i.d.\ from the Fr\'echet distribution
with shape $\alpha = 2$ (c.d.f.\ $F(x) = e^{-1/x^2}$, $x \ge 0$) and play
$A_t = \arg\min_a [\hat{L}_a - r_a / \eta_t]$ with learning rate
$\eta_t = \eta / \sqrt{t}$~\cite{honda2023}.
The unknown selection probability $w_{t,A_t}$ is estimated without bias by
geometric resampling~\cite{neu2013,honda2023}: resample perturbations until the
perturbed $\arg\min$ matches $A_t$; the number of draws $m$ estimates
$1/w_{t,A_t}$, and $\hat{L}_{A_t} \mathrel{+}= \ell_{t,A_t}\, m$.
The heavy Fr\'echet tail is what yields the $O(\sqrt{KT})$ adversarial and
$O(\log T)$ stochastic regret bounds; an exponential perturbation on
cumulative rewards~\cite{kalai2005} does not.
Warm start seeds $\hat{L}_a(0) = \nu\,(1 - r(\hat{\mu}(a)))$.
\subsection{EXP3}
Classic adversarial bandit with exponential weights~\cite{auer2002exp3}: loss
$\ell_t = 1 - R_t$ (optionally clipped at $0$), exploration rate
$\gamma \in [0,1]$, and learning rate $\eta > 0$.
Maintain weights $w_a > 0$; each round
\begin{equation}
  \label{eq:exp3}
  \pi_t(a) = (1-\gamma)\frac{w_a}{\sum_b w_b} + \frac{\gamma}{|\mathcal{A}'|},
  \qquad
  A_t \sim \pi_t,
\end{equation}
then update $w_{A_t} \leftarrow w_{A_t}\exp(-\eta\,\ell_t / \pi_t(A_t))$.
Warm start sets $w_a \propto \exp(\eta\, r(\hat{\mu}(a)))$ when enabled.
\subsection{EXP3++}
EXP3++ augments EXP3 with arm-specific exploration rates derived from empirical
loss-gap estimates~\cite{seldin2014exp3pp}.  It retains adversarial robustness
while reducing exploration when stable gaps emerge.  Our implementation
maintains importance-weighted cumulative losses, samples from the mixture of
exponential weights and per-arm exploration rates, and searches its
$\alpha$, $\beta$, and reward-clipping parameters.  Unlike EXP3, this
implementation does not use a cost-model warm start.
\subsection{Softmax (Boltzmann)}
Sample
\begin{equation}
  \label{eq:softmax}
  A_t \sim \pi_t, \qquad
  \pi_t(a) = \frac{\exp(\hat{\mu}^R_a / \tau)}{\sum_{b \in \mathcal{A}'} \exp(\hat{\mu}^R_b / \tau)},
\end{equation}
with temperature $\tau > 0$~\cite{kuleshov2014}.
With temperature decay, use $\tau_t = \tau/\sqrt{N_t}$ with $N_t$ the number
of softmax selections after cold-start exploration.
Unpulled arms are chosen uniformly at random until each arm has at least one pull
(cold start only).
\subsection{Gaussian Thompson sampling}
Maintain $(\hat{\mu}^R_a, \hat{\sigma}^2_a, n_a)$; prior variance $\sigma^2_0$
applies until $n_a < 2$.
Posterior scale $s_a = \sqrt{\hat{\sigma}^2_a / n_a}$; each step draw
$\tilde{\mu}_a \sim \mathcal{N}(\hat{\mu}^R_a, s_a^2)$ and play
$A_t = \arg\max_a \tilde{\mu}_a$~\cite{thompson1933}.
Under robust estimation, $\hat{\mu}^R_a$ and $\hat{\sigma}^2_a$ come from
clipped-SGD mean/variance instead of Welford.
\begin{algorithm}[h]
\caption{Gaussian Thompson sampling with warm start (sketch)}
\label{alg:gt}
\begin{algorithmic}[1]
\Require Arms $\mathcal{A}$, priors $\{r(\hat{\mu}(a))\}$, pseudo-count $\nu$, prior variance $\sigma^2_0$
\For{each $a \in \mathcal{A}$}
  \State $\hat{\mu}^R_a \gets r(\hat{\mu}(a))$;\quad $n_a \gets \nu$
\EndFor
\For{$t = 1, 2, \ldots$}
  \For{each $a \in \mathcal{A}$}
    \State Update $\hat{\sigma}^2_a$ via Welford; draw $\tilde{\mu}_a \sim \mathcal{N}(\hat{\mu}^R_a, \hat{\sigma}^2_a/n_a)$
  \EndFor
  \State $A_t \gets \arg\max_a \tilde{\mu}_a$; observe $R_t$; update statistics for $A_t$
\EndFor
\end{algorithmic}
\end{algorithm}

\subsection{Hierarchical bandit (EXP3 parent / searched child)}

Arms partition into groups $\mathcal{G}$ by their algorithm/protocol structure:
grouping by protocol or by algorithm.
A parent policy selects $G_t \in \mathcal{G}$; the child policy for $G_t$ selects
$A_t \in G_t$; both receive the same $R_t$~\cite{avin2025}.
Because each child learns online, the arm mixture within group $g$ changes with
time.  The parent's reward sequence for ``arm'' $g$ is therefore induced by
the evolving child policy and need not be i.i.d.\ or stationary even when each
physical NCCL arm is stationary.  We therefore fix the parent family to
adversarial EXP3 and let Optuna choose the child from Gaussian Thompson, UCB1,
and Softmax.  Adversarial robustness is the appropriate conservative model for
this endogenous reward schedule; it is not the only possible solution, since a
non-stationary stochastic parent could model the same coupling explicitly.
Warm start splits $r(\hat{\mu}(a))$ across levels when enabled; with
best-in-group parent aggregation, the parent group prior is the
maximum child $r(\hat{\mu}(a))$ in that group.
\begin{algorithm}[h]
\caption{Hierarchical step (benchmark champion: EXP3 parent, Softmax child,
algorithm grouping)}
\label{alg:hier}
\begin{algorithmic}[1]
\Require Parent policy $\pi_p$ over groups $\mathcal{G}$,
  child policies $\{\pi_c\}_{c \in \mathcal{G}}$ over arms in each group
\State $G_t \gets \pi_p.\textsc{Select}()$
\State $A_t \gets \pi_{G_t}.\textsc{Select}()$
\State Observe reward $R_t$ from arm $A_t$
\State $\pi_{G_t}.\textsc{Update}(A_t, R_t)$;\quad $\pi_p.\textsc{Update}(G_t, R_t)$
\end{algorithmic}
\end{algorithm}

\subsection{Tsallis-Inf}

Loss $\ell_t = 1 - R_t$ (optionally clipped at $0$).
Maintains importance-weighted cumulative losses and sampling weights $w_a$ on the
simplex via optimistic mirror descent with Tsallis entropy; step size
$\alpha \in (0,1)$~\cite{zimmert2021}.
Each step samples $A_t \sim w$; after observing $R_t$, update losses and recompute
$w$ by solving the dual OMD problem (fixed Tsallis order $1/2$ in the
implementation; $\alpha$ is the OMD step size)~\cite{zimmert2021}.
Warm start sets $w_a \propto r(\hat{\mu}(a))$ when enabled.

\begin{algorithm}[h]
\caption{End-to-end pipeline for one context $(n, P)$}
\label{alg:pipeline}
\begin{algorithmic}[1]
\Require Latency pools $\{\mathcal{D}_a\}$, cost model $\hat{\mu}$, threshold $\rho_{\max}$,
  policy $\pi$, horizon $T$
\State $\mathcal{A}' \gets \textsc{FilterArms}(\mathcal{A}, \hat{\mu}, \rho_{\max})$
  \Comment{Eq.~\eqref{eq:filter}}
\State Build prior rewards $r(\hat{\mu}(a))$ on $\mathcal{A}'$
\State Initialize $\pi$ with warm-start priors and $\nu$ if enabled
\For{$t = 1, \ldots, T$}
  \State $A_t \gets \pi.\textsc{Select}()$;\quad draw $X_t \sim \mathcal{D}_{A_t}$
  \State $R_t \gets \hat{\mu}_{\min} / X_t$;\quad $\pi.\textsc{Update}(A_t, R_t)$
\EndFor
\State Report $\mathrm{CRLT}$, $\mathrm{TRLT}$, $\mathrm{RLT}_i$, and $\mathrm{RLB}_i$
\end{algorithmic}
\end{algorithm}

\section{Extended results and sensitivity}
\label{app:extended-results}

This appendix holds the full policy ranking, the 32-node bandwidth slice, the
ablation suite (warm start, horizon, drift, update frequency, secondary knobs),
and supporting analyses behind Section~\ref{sec:results}.
None of these replaces the realized-latency comparison in Table~\ref{tab:realized}; the
tables below are selection-priced unless stated otherwise.

\subsection{Full policy ranking}
\label{app:full-ranking}
\begin{table}[!htb]
  \centering
  \caption{Grouped method comparison on the holdout harness (train: 4/8-node contexts excluding the replay holdout; test: all 16/32-node grid contexts plus eight test-only CCL-Bench replay contexts mapped to 4-node measurement pools, $P{=}4$; $T{=}100$ sequential replay; train ${n{=}}146$, test ${n{=}}154$ contexts). Train and held-out \textbf{test} mean \TRLT{} (\%) are reported as mean${\pm}$\SEM{} across contexts; test p99 is the context-level 99th-percentile point estimate with a 95\% bootstrap CI (contexts resampled with replacement; Section~\ref{sec:metrics}). Thus, ${\pm}$ values quantify uncertainty in a mean, whereas brackets quantify uncertainty in p99. Rows grouped as Fixed / Supervised / Bandits and sorted by test p99; \textbf{bold} marks the best per column (lower is better). Per-champion hyperparameters in Table~\ref{tab:policy-config}; full protocol in Appendix~\ref{app:hyperopt}.  \capkey{}}
  \label{tab:policy-ranking-full}
  \scriptsize
  \setlength{\tabcolsep}{2.5pt}
  \resizebox{\columnwidth}{!}{%
  \begin{tabular}{@{}l r@{\,}l r@{\,}l r@{\quad}l@{}}
    \toprule
    & \multicolumn{6}{c}{\TRLT{} (\%)} \\
    \cmidrule(l){2-7}
    policy & \multicolumn{2}{c}{train} & \multicolumn{2}{c}{test} & \multicolumn{2}{c}{test p99 [95\% CI]} \\
    \midrule
    \multicolumn{7}{@{}l}{\textit{Fixed}} \\
    NCCL \AUTO{} (cost model) & $4.04$ & ${\pm}0.99$ & $5.99$ & ${\pm}1.58$ & $103.25$ & $[46.48,\,134.61]$ \\
    Random & $39.73$ & ${\pm}2.95$ & $45.22$ & ${\pm}3.72$ & $190.30$ & $[147.02,\,195.35]$ \\
    \midrule
    \multicolumn{7}{@{}l}{\textit{Supervised}} \\
    Linear regression + cost & $2.94$ & ${\pm}0.48$ & $2.04$ & ${\pm}0.47$ & $28.68$ & $[17.02,\,42.58]$ \\
    Gradient boosting + cost & $\textbf{0.04}$ & ${\pm}0.03$ & $1.76$ & ${\pm}0.51$ & $36.89$ & $[17.77,\,40.37]$ \\
    Gradient boosting & $0.18$ & ${\pm}0.10$ & $4.70$ & ${\pm}1.35$ & $76.41$ & $[39.77,\,129.99]$ \\
    Random forest & $0.22$ & ${\pm}0.11$ & $6.79$ & ${\pm}1.86$ & $102.44$ & $[66.16,\,184.75]$ \\
    Random forest + cost & $0.29$ & ${\pm}0.14$ & $3.90$ & ${\pm}1.44$ & $103.25$ & $[27.35,\,134.61]$ \\
    Linear regression & $78.90$ & ${\pm}14.38$ & $66.31$ & ${\pm}12.39$ & $600.45$ & $[556.59,\,706.16]$ \\
    \midrule
    \multicolumn{7}{@{}l}{\textit{Bandits}} \\
    Softmax (warm) & $0.87$ & ${\pm}0.17$ & $\textbf{0.93}$ & ${\pm}0.11$ & $\textbf{6.59}$ & $[4.26,\,7.62]$ \\
    KL-UCB (warm) & $0.85$ & ${\pm}0.18$ & $1.18$ & ${\pm}0.24$ & $7.63$ & $[4.31,\,32.72]$ \\
    UCB1 (warm) & $0.86$ & ${\pm}0.18$ & $1.00$ & ${\pm}0.13$ & $7.97$ & $[5.58,\,9.39]$ \\
    ETC (cold) & $2.43$ & ${\pm}0.24$ & $2.70$ & ${\pm}0.49$ & $8.68$ & $[7.57,\,71.39]$ \\
    Madan RR-ETC (cold)\textsuperscript{\dag} & $2.42$ & ${\pm}0.24$ & $2.70$ & ${\pm}0.49$ & $8.68$ & $[7.57,\,71.39]$ \\
    UCB-tuned (warm) & $4.34$ & ${\pm}0.21$ & $4.38$ & ${\pm}0.20$ & $11.60$ & $[9.52,\,12.48]$ \\
    Gaussian Thompson (warm) & $1.25$ & ${\pm}0.25$ & $1.45$ & ${\pm}0.27$ & $16.94$ & $[7.59,\,22.20]$ \\
    Hierarchical EXP3/Softmax (warm) & $2.75$ & ${\pm}0.36$ & $3.23$ & ${\pm}0.36$ & $17.30$ & $[13.80,\,18.47]$ \\
    UCB-V (warm) & $6.58$ & ${\pm}0.38$ & $6.66$ & ${\pm}0.37$ & $18.81$ & $[16.51,\,19.87]$ \\
    $\varepsilon$-greedy (warm) & $1.07$ & ${\pm}0.27$ & $1.52$ & ${\pm}0.43$ & $22.15$ & $[10.36,\,51.66]$ \\
    Tsallis-Inf (warm) & $6.98$ & ${\pm}0.43$ & $7.59$ & ${\pm}0.46$ & $24.20$ & $[20.38,\,28.36]$ \\
    EXP3 (cold) & $7.52$ & ${\pm}0.49$ & $8.41$ & ${\pm}0.52$ & $24.23$ & $[21.95,\,26.32]$ \\
    FPL (cold) & $8.52$ & ${\pm}0.61$ & $9.91$ & ${\pm}0.72$ & $35.55$ & $[31.85,\,41.05]$ \\
    EXP3++ (cold) & $26.03$ & ${\pm}1.72$ & $27.74$ & ${\pm}1.85$ & $89.95$ & $[76.39,\,102.15]$ \\
    \bottomrule
    \multicolumn{7}{@{}l}{\scriptsize \textsuperscript{\dag}search-tuned to $m{=}1$, $1\%$ safety gate on our offline harness; \emph{not} the published live $m{=}10$ configuration.} \\
  \end{tabular}%
  }
\end{table}

\subsection{Fabric summary}
\label{app:fabric}
\begin{table}[t]
  \centering
  \caption{Fabric summary. H100 node entries use 8 GPUs per node; sweeps are isolated micro-benchmark jobs.}
  \label{tab:fabric}
  \begin{tabular}{@{}lr@{}}
    \toprule
    component & specification \\
    \midrule
    Node & Dell HGX H100 Gen-11; 8$\times$ H100 \\
    Intra-node & NVLink \\
    NIC & BlueField-3 SuperNIC \\
    Network & Spectrum-X Ethernet \\
    NCCL & 2.30.4 (CUDA 13.0) \\
    Scale & 4--32 nodes (32--256 GPUs) \\
    Sweep & Isolated jobs \\
    \bottomrule
  \end{tabular}
\end{table}

\subsection{32-node bandwidth slice}
\label{app:collective-32}
\begin{figure*}[t]
  \centering
  \includegraphics[width=\linewidth]{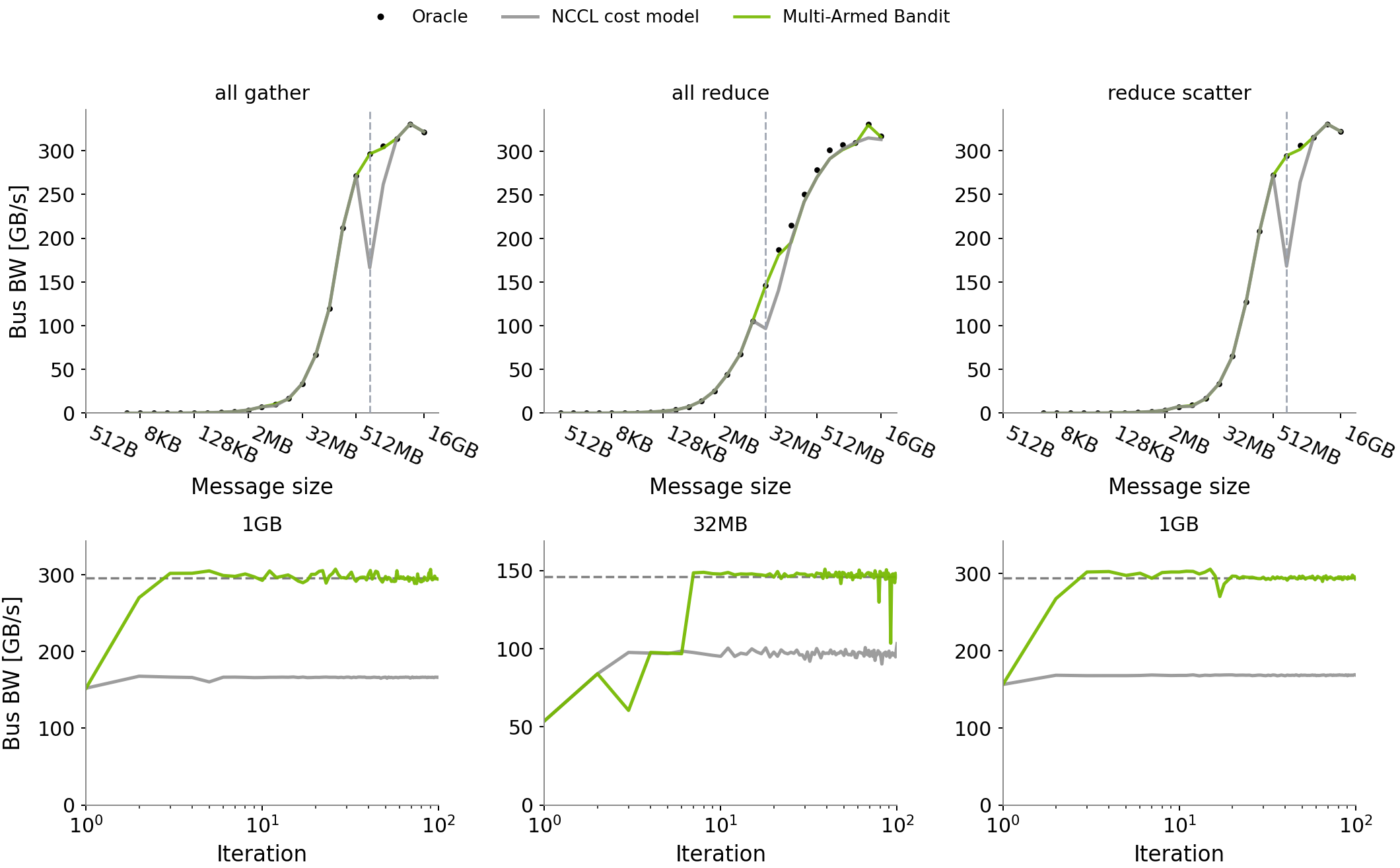}
  \caption{HGX H100 offline emulation at 32 nodes ($T{=}100$ sequential replay),
    using the held-out 32-node slice of the holdout harness; the three featured
    policies (oracle, \AUTO{}, and the deterministic UCB1 champion) are
    seed-independent under sequential replay.
    On this slice, the
    \CollectiveThirtyTwoBanditCaption{} reduces \TRLT{} from
    $\CollectiveThirtyTwoNcclMeanTrltPct\%$ mean /
    $\CollectiveThirtyTwoNcclPNineNineTrltPct\%$ p99 for NCCL \AUTO{} to
    $\CollectiveThirtyTwoBanditMeanTrltPct\%$ mean /
    $\CollectiveThirtyTwoBanditPNineNineTrltPct\%$ p99.
    Columns are the three collectives.
    \textbf{Top:} steady-state bus bandwidth vs.\ message size (black = oracle,
    gray = \AUTO{}, green = \CollectiveThirtyTwoBanditCaption{}); the green band
    is the central 95\% of observed bandwidth, not a seed-level CI.
    \textbf{Bottom:} sequential learning at the worst \AUTO{} message size
    (gray dashed marker above); dashed black = oracle.}
  \label{fig:collective-32}
\end{figure*}

\subsection{Ablations and sensitivity}
\label{sec:ablations}

Each ablation below changes exactly one design choice. Warm start and horizon run
on the same holdout harness as Table~\ref{tab:policy-ranking-full}
(Section~\ref{sec:protocol}); the drift and diluted-update experiments use targeted
sensitivity harnesses whose numbers are therefore not comparable with held-out test
\TRLT{}, and are noted as such where they appear.

\subsubsection{Cost-model prior and learning horizon}
\label{sec:abl-prior-horizon}

\paragraph{Warm start.}
(Realized-latency version: Table~\ref{tab:warm-cold}.)
Table~\ref{tab:warm-start-ablation} toggles cost-model warm start off vs.\ on for
every warm-capable champion at otherwise frozen hyperparameters (top-level and, for
the hierarchical policy, nested per level); the three families whose implementations
do not support warm start (EXP3++, Madan RR-ETC, ETC) are omitted.
Warm start seeds each arm's sufficient statistics with a per-family tuned
pseudo-count $n_0$ of virtual observations at the cost-model reward guess
$\hat{\mu}_{\min}/\hat{\mu}(a)$ (Section~\ref{sec:reward}), so a champion tuned to
$n_0{=}0$ is effectively cold even with the flag on.
The search varied warm/cold jointly per bandit, so this table isolates the
warm-start effect at fixed knobs.
Warm start helps most families and matters most where learning is coupled: cold
Hierarchical EXP3\,/\,Softmax reaches $44.0\%$ test \TRLT{}, which warm start
collapses to $3.23\%$. One plausible
explanation is short-horizon coupling: with uninformative priors at $T{=}100$ the
EXP3 parent can move across algorithm groups before the Softmax children stabilize.
The remaining families are nearly prior-insensitive on this scale-OOD test
(UCB-tuned, UCB-V, Tsallis-Inf, EXP3, and FPL each move by ${<}0.6$~pp), and two do
not improve at all: UCB-V worsens slightly ($+0.21$~pp) and FPL is
unchanged.\footnote{The FPL champion is tuned cold with pseudo-count $n_0{=}0$
(Table~\ref{tab:policy-config}); toggling the warm flag on while $n_0{=}0$ leaves the
seeded cumulative losses at zero, a literal no-op, so its paired $\Delta$ is
exactly $0.00$.}
\begin{table}[t]
  \centering
  \caption{Warm-start ablation on the holdout harness, \emph{selection-priced} \TRLT{} (each step priced at the chosen arm's steady mean; the realized-latency version is Table~\ref{tab:warm-cold}): search champion hyperparameters per family, only warm start toggled off vs.\ on. Held-out \textbf{test} \TRLT{} (\%) at $T{=}100$; columns report \mbox{mean${\pm}$\SEM{}} across the test context pool (${n{=}}154$ contexts). \underline{Underlined} = setting Table~\ref{tab:policy-ranking-full} used for that champion. $\Delta$ is paired true$-$false per context (not independent-column propagation). \capkey{} \capbold{}}
  \label{tab:warm-start-ablation}
  \scriptsize
  \setlength{\tabcolsep}{3pt}
  \resizebox{\columnwidth}{!}{%
  \begin{tabular}{@{}lrrr@{}}
    \toprule
    & \multicolumn{3}{c}{\TRLT{} (\%)} \\
    \cmidrule(l){2-4}
    & \multicolumn{2}{c}{warm start} & \\
    \cmidrule(lr){2-3}
    policy & false & true & $\Delta$ (true$-$false) \\
    \midrule
    Softmax & $5.45{\pm}0.87$ & \underline{$\textbf{0.93}{\pm}0.11$} & $-4.53{\pm}0.88$ \\
    UCB1 & $7.68{\pm}1.46$ & \underline{$1.00{\pm}0.13$} & $-6.68{\pm}1.46$ \\
    KL-UCB & $3.71{\pm}0.69$ & \underline{$1.18{\pm}0.24$} & $-2.52{\pm}0.68$ \\
    Gaussian Thompson & $\textbf{2.87}{\pm}0.50$ & \underline{$1.45{\pm}0.27$} & $-1.42{\pm}0.49$ \\
    $\varepsilon$-greedy & $5.43{\pm}1.25$ & \underline{$1.52{\pm}0.43$} & $-3.91{\pm}1.35$ \\
    Hierarchical EXP3/Softmax & $44.03{\pm}3.85$ & \underline{$3.23{\pm}0.36$} & $-40.80{\pm}3.85$ \\
    UCB-tuned & $4.94{\pm}0.21$ & \underline{$4.38{\pm}0.20$} & $-0.56{\pm}0.09$ \\
    UCB-V & $6.44{\pm}0.32$ & \underline{$6.66{\pm}0.37$} & $0.21{\pm}0.10$ \\
    Tsallis-Inf & $7.71{\pm}0.45$ & \underline{$7.59{\pm}0.46$} & $-0.12{\pm}0.07$ \\
    EXP3 & \underline{$8.41{\pm}0.52$} & $7.89{\pm}0.50$ & $-0.52{\pm}0.10$ \\
    FPL & \underline{$9.91{\pm}0.72$} & $9.91{\pm}0.72$ & $0.00{\pm}0.00$ \\
    \bottomrule
  \end{tabular}
  }
\end{table}

\paragraph{Horizon.}
Held-out test \TRLT{} improves monotonically through $T{=}100$ for every learning
champion, computed post-hoc from the same benchmark run at
$T{\in}\{10,20,50,100\}$ (Appendix~\ref{app:extended-results},
Table~\ref{tab:horizon-rtl}). Sequential replay preserves lazy-init overhead on
first arm pulls, so short horizons are pessimistic; deterministic fixed selectors
such as \AUTO{} are horizon-invariant, choosing the same arm at every pull.

\subsubsection{Non-stationarity: discounted forgetting}
\label{sec:drift}

\paragraph{Regime construction.}
Our main harness is piecewise stationary: within a job the best arm is fixed and a
policy need only identify it once. Production is not always so benign: a routing or
congestion change, compute--communication overlap, or a co-scheduled job can make a
previously poor arm the new winner \emph{mid-run}~\cite{madan2026,xu2025}.
Adapting to such change is what a bandit should add over a static cost model,
so we test it explicitly. The space of non-stationary
environments is unbounded (what changes, how, how much, how often); non-stationary
bandit theory collapses it into \emph{switching} and \emph{slowly-varying}
families~\cite{slivkins2019}, and we take one worst-case instance of each, built
from real measurements instead of synthetic magnitudes.
On the held-out \emph{test} contexts (154 jobs, 10 seeds, $T{=}100$) we identify
each context's measured best arm $a^\star$ and worst arm $a^\circ$ from its real
steady-state pools (iterations $\geq11$; Section~\ref{sec:protocol}), and define a
regime by a time-varying map from each offered arm to the \emph{pool it samples
from}: unswapped, every arm draws its own pool; swapped, $a^\star$ and $a^\circ$
exchange pools, so the former winner becomes a ``trap'' and the former worst the new
winner. Rewards are bootstrap draws from the mapped pool, converted to the harness
reward $r{=}\hat{\mu}_{\min}/t$ (Section~\ref{sec:reward}).
\textbf{Stationary} never swaps; \textbf{abrupt} swaps for all steps $t\geq50$;
\textbf{gradual} crossfades by swapping with probability $1{-}p_{\mathrm{old}}(t)$,
where $p_{\mathrm{old}}$ ramps linearly $1\!\to\!0$ over steps 35--65. The regime
trajectory is drawn from a policy-independent RNG stream, so all policies face the
\emph{identical} realized drift within a (context, seed).
One caveat: a best$\leftrightarrow$worst swap makes the new optimum the previously
\emph{least}-sampled arm, which favors optimism-under-uncertainty exploration, so
this is a targeted stress test, not a claim of adversarial-worst-case
robustness.

\paragraph{Dynamic oracle and leakage-free forgetting rate.}
Because the optimal arm now moves, \emph{static} regret (against the single best
\emph{fixed} arm in hindsight) is meaningless: we score against a \emph{dynamic
oracle} (a time-varying, per-step-optimal comparator) that plays whichever arm
currently draws the best pool and report \emph{dynamic} \CRLT{}/\TRLT{} against
that time-varying denominator (the stationary regime reduces to ordinary \TRLT{}). The mechanism we ablate is
exponential \emph{forgetting}: discount each arm's running sufficient statistics by
$\gamma\in(0,1]$ per update, capping the effective sample size at $1/(1-\gamma)$
($\gamma{=}1$ recovers plain means, which react to a switch only once fresh samples
dilute the stale ones). This is not a new mechanism: it is Discounted
UCB~\cite{kocsis2006discounted,garivier2011switching}, whose sliding-window sibling
SW-UCB replaces the geometric weights with a hard window of the last $\tau$ samples;
both attain $\tilde{O}(\sqrt{\Gamma T})$ dynamic regret over $\Gamma$ breakpoints, and
both require $\gamma$ (or $\tau$) to be matched to an unknown switching rate, which
we select below. Applied to our Softmax champion, whose estimator is
an ordinary sample mean, $\gamma$ yields the discounted mean of that literature
verbatim. Applied to our UCB1 champion it does not: that champion's mean is the
clipped-SGD estimator~\cite{sankararaman2023}, so $\gamma$ discounts only its
\emph{counts} (both the confidence radius' $n_a$ and the estimator's effective sample
size), leaving a hybrid, not textbook D-UCB. We select $\gamma$ for each of the
two champions on the \textbf{train} partition only before freezing
it for test; the train/test \emph{context} split, not the seeds, prevents leakage.
The three regimes' \TRLT{} scales differ by more than an order of magnitude, so a
plain average would be swamped by the drift numbers and ignore the stationary cost.
We therefore sweep $\gamma$ over an 18-point grid and minimize the mean \emph{competitive
ratio}: each regime's mean dynamic \TRLT{} divided by the best any grid point
achieves for that regime, which puts every regime on a ``fraction of its own
best'' scale. The sweep selects Softmax $\gamma{=}0.85$ (aggressive forgetting) and UCB1
$\gamma{=}0.995$ (essentially plain).

\begin{table}[t]
  \centering
  \scriptsize
  \setlength{\tabcolsep}{4pt}
  \caption{Non-stationarity ablation: mean dynamic \TRLT{} (\%\,$\pm$\,\SEM{}, \textbf{lower is better}); best per drift column in \textbf{bold}. \emph{Stationary} is the sequential CV holdout replay---the \emph{same} harness as Table~\ref{tab:policy-ranking-full} ($T{=}100$, 154 held-out test jobs, NCCL lazy-init cold start included), so the plain rows match Table~\ref{tab:policy-ranking-full} and the plain$\to$discounted change is the true stationary cost of forgetting. \emph{Abrupt} (best$\leftrightarrow$worst pool swap at step~50) and \emph{gradual} (crossfade over steps 35--65) are the bootstrap steady-pool drift replay against a dynamic oracle ($T{=}100$, $154\times10$ runs). Discounting ($\gamma$ selected leakage-free on the drift train partition) cuts drift \TRLT{} several-fold while adding ${<}0.2$~pp on the stationary metric. Per-regime p99 and the purpose-built adversarial baselines (EXP3, EXP3++, Tsallis-Inf, FPL) are in Appendix~\ref{app:extended-results} (Table~\ref{tab:drift-ablation-full}).}
  \label{tab:drift-ablation}
  \begin{tabular}{@{}lccc@{}}
    \toprule
    Policy & stationary & abrupt & gradual \\
    \midrule
    NCCL \AUTO{} (static) & 5.99\,\tiny$\pm$1.58 & 45.01\,\tiny$\pm$3.93 & 44.60\,\tiny$\pm$3.89 \\
    Softmax (plain) & \textbf{0.93}\,\tiny$\pm$0.11 & 31.57\,\tiny$\pm$2.63 & 31.99\,\tiny$\pm$2.65 \\
    Softmax ($\gamma{=}0.85$) & 1.07\,\tiny$\pm$0.12 & 5.96\,\tiny$\pm$0.29 & 13.11\,\tiny$\pm$1.17 \\
    UCB1 (plain) & 1.00\,\tiny$\pm$0.13 & 6.60\,\tiny$\pm$0.61 & 10.93\,\tiny$\pm$0.87 \\
    UCB1 ($\gamma{=}0.995$) & 1.16\,\tiny$\pm$0.13 & \textbf{5.88}\,\tiny$\pm$0.54 & \textbf{10.68}\,\tiny$\pm$0.87 \\
    \bottomrule
  \end{tabular}
\end{table}

\begin{figure*}[t]
  \centering
  \includegraphics[width=\linewidth]{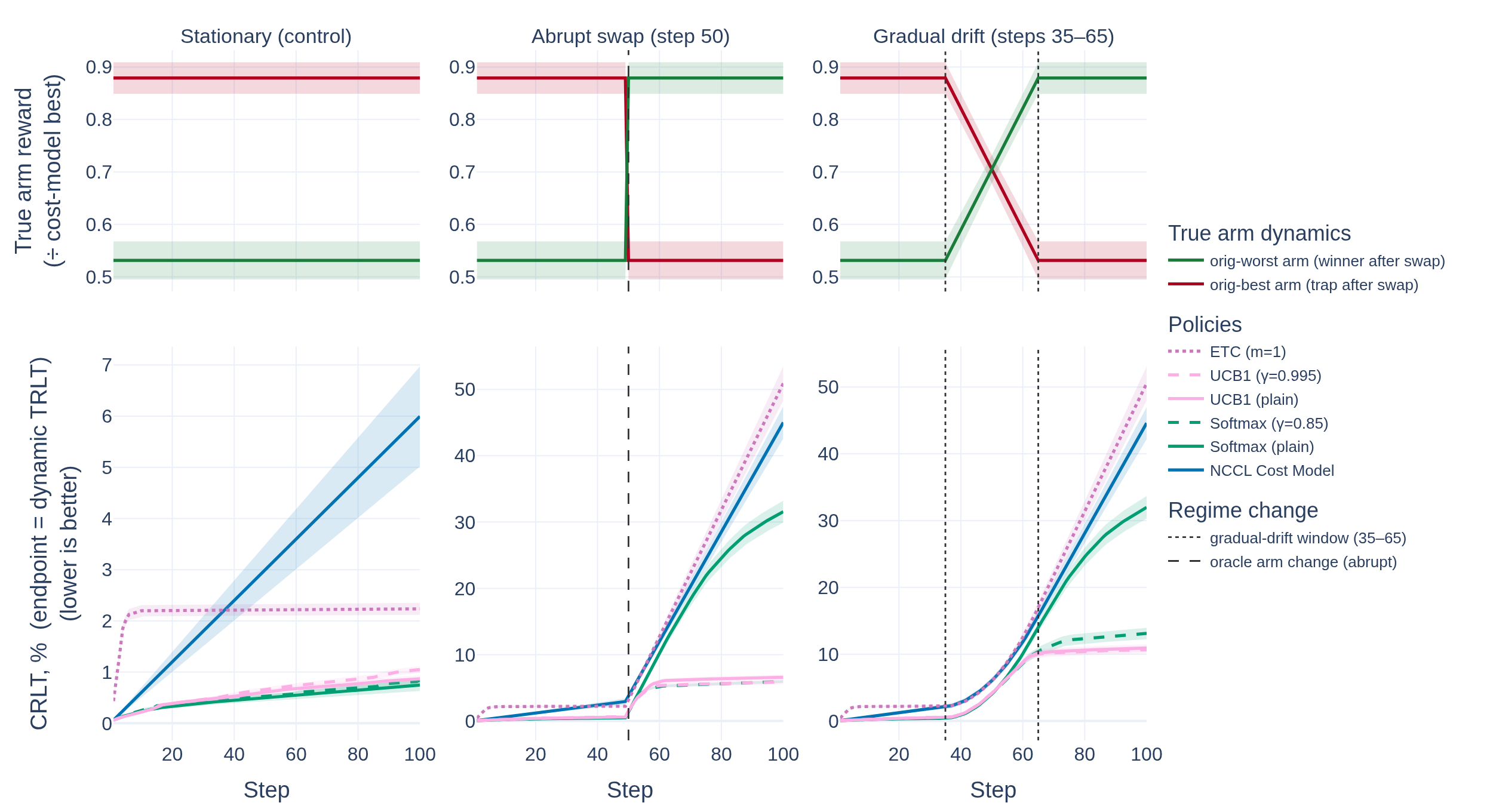}
  \caption{Non-stationarity ablation on the held-out test contexts (154 jobs,
    10 seeds, $T{=}100$). Columns are three regimes: stationary control, an abrupt
    best$\leftrightarrow$worst arm-pool swap at step~50, and a gradual crossfade
    over steps 35--65 (gray dashed = abrupt change point; gray dotted = gradual
    window). \textbf{Top:} true policy-independent reward of the two flipping arms
    ($\div$ cost-model best), showing the underlying regime change.
    \textbf{Bottom:} mean dynamic \CRLT{} (\%) vs.\ step against a dynamic oracle
    (\textbf{lower is better}); each curve's endpoint equals its dynamic \TRLT{},
    shaded bands are 95\% CI across contexts $\times$ seeds. Discounting
    (Softmax $\gamma{=}0.85$, UCB1 $\gamma{=}0.995$, selected leakage-free on the
    train partition by a three-regime competitive-ratio grid sweep) cuts drift
    \TRLT{} several-fold below plain Softmax and far below the static NCCL \AUTO{}
    cost model, while leaving the stationary control essentially unchanged. These
    panels are the bootstrap steady-pool replay; the stationary column of
    Table~\ref{tab:drift-ablation} instead re-measures on the sequential
    Table~\ref{tab:policy-ranking-full} harness for direct comparability.
    The tuned explore-then-commit champion (ETC, $m{=}1$, dotted; Madan et al.'s
    plugin is an instance of this family) is shown as the non-adaptive commit-once
    foil: cheap in the stationary control but the worst policy under both drift
    regimes, exceeding even the static cost model, because it commits and cannot
    re-rank.
    Purpose-built adversarial bandits (EXP3, EXP3++, Tsallis-Inf, FPL) are omitted
    for clarity and discussed in Appendix~\ref{app:extended-results}.}
  \label{fig:drift-ablation}
\end{figure*}

\paragraph{Forgetting helps under drift at negligible stationary cost.}
Under the abrupt swap, discounting cuts Softmax's dynamic \TRLT{} from $31.6\%$
(plain) to $5.96\%$ ($5.3\times$), and under gradual drift from $32.0\%$ to
$13.1\%$ ($2.4\times$); the static \AUTO{} cost model pays about $45\%$ in both
because it never re-selects (Table~\ref{tab:drift-ablation}). The stationary column
re-measures each policy on the \emph{sequential} harness of
Table~\ref{tab:policy-ranking-full}, so its plain rows reproduce those numbers exactly
and the cost of forgetting is directly readable: ${+}0.14$~pp for Softmax
($0.93{\to}1.07\%$), whose p99 even \emph{improves} ($6.59{\to}5.34\%$), and
${+}0.16$~pp for UCB1 ($1.00{\to}1.16\%$). UCB1 needs less help because its
clipped-SGD robust mean already tracks a moving target part of the way: its step
size is $\lambda/n_a^{p}$ with a tuned $p{\approx}0.47$, which is $5\times$ the
sample average's $1/n_a$ at $100$ pulls, so a fresh observation still moves the
estimate long after a sample mean has stopped responding. \emph{Plain}
UCB1 is therefore strong under drift ($6.60\%$ abrupt, $10.9\%$
gradual) and extra forgetting adds little ($5.88\%$/$10.7\%$), which is why its
selected $\gamma$ sits near $1$. The step size still decays to zero, so the
estimate keeps responding for longer but has no bounded memory; that difference shows
only over horizons far longer than the $T{=}100$ evaluated here (see the caveat below).
Figure~\ref{fig:drift-ablation} shows the mechanism: the two flipping arms'
true rewards cross at the change point, and each policy's \CRLT{} departs from the
oracle only around that crossing, with faster forgetting closing the gap sooner.
The four purpose-built adversarial champions (EXP3, EXP3++, Tsallis-Inf,
FPL) are the \emph{weakest} learners in every regime here, at or above static
\AUTO{} under the abrupt swap, because this environment is stochastic and only
piecewise non-stationary, not adversarial
(Appendix~\ref{app:extended-results}).

\paragraph{Commit-once explore-then-commit cannot recover.}
The explore-then-commit (ETC) family, of which Madan et al.'s round-robin plugin
(Section~\ref{sec:related}) is an instance, shows the linear-regret argument
in practice. We evaluate our \emph{tuned} ETC champion ($m{=}1$, selected on the same
CV harness as the bandit champions, so the comparison is matched). In the
stationary control it is cheap ($2.2\%$ dynamic \TRLT{}, near the adaptive
champions and well below the static \AUTO{} cost model's $6.0\%$), because it
commits almost immediately to the cost-model-filtered best arm. But having
committed it has no mechanism to re-rank: under the abrupt swap it stays on the
now-``trap'' arm (post-change oracle-selection rate $1.2\%$), and its dynamic
\TRLT{} is the \emph{worst} of all policies ($50.9\%$ abrupt, $50.5\%$ gradual,
exceeding even the static cost model's $45.0\%$), while the discounted adaptive
champions stay near $6\%$ (Figure~\ref{fig:drift-ablation}). This is the empirical
counterpart of the commit-once failure of Section~\ref{sec:related}: after a
permanent commit, dynamic regret grows linearly in the horizon, and since the
stationary cost is negligible, no exploration budget fixes it.

\paragraph{Alternatives we do not evaluate, and why.}
Geometric discounting is the cheapest of several standard responses to a moving
optimum. \emph{Change-detection} methods run a statistical test alongside the bandit
and reset the affected statistics on an alarm (CUSUM-UCB~\cite{liu2018cusum} and
M-UCB~\cite{cao2019mucb}), while \emph{parameter-free} methods such as
AdSwitch~\cite{auer2019adswitch} and the MASTER black-box
reduction~\cite{wei2021master} remove the need to know the switching rate at all.
Both families are a poor match for our horizon, though not for our problem. A
detector must distinguish a real shift from sampling noise, and M-UCB's own tuning
rule prescribes a window of several hundred to over a thousand samples \emph{per arm}
for reward changes of the magnitude our swap induces, alongside forced uniform
sampling at a rate that saturates when the horizon is short. At $T{=}100$
collectives per context the prescribed window exceeds the entire run. These methods
become the natural choice at horizons of $10^4$ collectives and beyond, which is
also where the parameter-free guarantees start to pay for their constants.

\paragraph{Caveat: these conclusions are horizon-bound.}
Our regimes contain a single change within $T{=}100$, which is the deployment
horizon of Table~\ref{tab:policy-ranking-full} but far short of a long-lived
communicator. Two asymptotic effects are invisible at this scale. First, any
estimator whose step size decays (a sample mean, and more slowly the clipped-SGD
mean) eventually stops responding on the arm it is \emph{currently playing}: after
$n_a$ pulls a fresh observation moves $\hat{\mu}_a$ by $O(1/n_a)$, so the arm with
the most data adapts the slowest, and a degraded incumbent is never re-ranked.
Second, UCB1's radius revisits a neglected arm only when
$\alpha\sqrt{2\ln N_t/n_a}$ regrows past the gap, which caps that arm's lifetime
pulls at $O(\ln N_t)$ and spaces retests by a constant \emph{factor} in $N_t$.
Bounded-memory estimators---D-UCB, SW-UCB, or a constant-gain filter---address both,
and the near-unity $\gamma$ selected for UCB1 here should therefore be read as
``discounting does not pay for itself within $100$ steps'', not as evidence that a
long-running tuner can keep a sample average.
We replayed the constant-gain filter on this same swap and on the stationary, NVLS-excluded split
(Appendix~\ref{app:recommended}); like everything here, that evidence stops at $T{=}100$ with one change.

Within that horizon, forgetting is a cheap, leakage-free safeguard: it recovers most
of the mid-run-switch loss while leaving the stationary common case essentially
untouched.

\subsubsection{Update frequency}
\label{sec:update-freq}

\paragraph{Diluted updates.}
Every accepted update triggers a cross-rank synchronization of sufficient statistics
plus $O(K)$ policy arithmetic, a fixed cost independent of message size
(about $0.5$--$0.9$\,ms per sync on H100-A, Appendix Table~\ref{tab:live-overhead-size}). That is
disproportionate for small payloads: on our H100 grid the measured-best arm
completes in $39$--$458$~µs for all-reduce and $46$--$996$~µs for all-gather and reduce-scatter up to $32$~MiB, comparable to
a single sync round, so a practical deployment must amortize learning over many fast
collectives. We therefore decouple \emph{selection} from \emph{learning}: the policy
still selects an arm on every collective, leaving stochastic exploration intact, but
the harness accumulates per-arm rewards and observed collective time and, once the
accumulated time reaches an interval $\tau$, applies one update per observed arm
using that arm's \emph{mean} reward over the window ($\tau{=}0$ recovers
per-collective updates). Diluting does not simply discard information: aggregating
$m$ observations into one mean-reward update trades update \emph{frequency} for
update \emph{quality}, since the batch mean has variance $\sigma^2/m$ rather than
$\sigma^2$. The policy therefore sees fewer but cleaner updates, most valuable at
small message sizes, where latency is noisiest relative to its mean
(Section~\ref{sec:failure}) and many measurements accumulate per flush.

\paragraph{It still works at a low update rate.}
Table~\ref{tab:update-freq} sets $\tau{=}1$~ms and replays the featured UCB1
champion across the full grid (all collectives $\times$ node counts $\times$ message
sizes) for $T{=}200$ sequential collectives, wrapping the steady-state tail so the
NCCL lazy-init transient is replayed only on first exposure. Diluting cuts the median
update count from $200$ to $51$ yet leaves mean \TRLT{} unchanged
($0.69{\to}0.68\%$) and even improves p99 ($9.20{\to}6.56\%$), both far below NCCL
\AUTO{} ($5.04\%$ mean). The effect concentrates at small, sub-millisecond sizes and
is a no-op at large ones, whose single-collective latency already exceeds $\tau$.
Under this time-based rule, which batches mainly the small sizes, learning survives a lower update rate.
This is a single-seed sensitivity run, not the holdout harness, and the headline results do not batch.
\begin{table}[t]
  \centering
  \caption{Update-frequency ablation: featured UCB1 champion across the full grid (all collectives $\times$ node counts $\times$ message sizes; $300$ contexts, $T{=}200$ sequential with the steady-state tail wrapped so the NCCL lazy-init transient is replayed only on first exposure). Selection runs on every collective; only the \texttt{update()} cadence changes. ``updates'' is the median batched updates per context; \TRLT{} columns report \mbox{mean${\pm}$\SEM{}} and p99 with 95\% bootstrap CI across contexts. Diluting to $\tau{=}1$~ms batches updates at small (sub-ms) sizes and is a no-op at large sizes where a single collective already exceeds $\tau$. Single-seed sensitivity run, not the holdout harness. \capTRLTany{} \capbold{}}
  \label{tab:update-freq}
  \setlength{\tabcolsep}{3pt}
  \resizebox{\columnwidth}{!}{%
  \begin{tabular}{@{}lrrr@{}}
    \toprule
    & & \multicolumn{2}{c}{\TRLT{} (\%)} \\
    \cmidrule(l){3-4}
    policy & updates & mean & p99 [95\% CI] \\
    \midrule
    NCCL \AUTO{} (cost model) & --- & $5.04{\pm}0.94$ & $75.01_{[50.93,\,131.74]}$ \\
    UCB1, update every collective & 200 & $0.69{\pm}0.10$ & $9.20_{[5.53,\,14.59]}$ \\
    UCB1, update every ${\ge}1$~ms & 51 & $\textbf{0.68}{\pm}0.09$ & $\textbf{6.56}_{[3.61,\,9.26]}$ \\
    \bottomrule
  \end{tabular}%
  }
\end{table}

\subsubsection{Secondary design checks}
\label{sec:abl-secondary}

Three further knobs move the held-out test result little, and we report them in full
in Appendix~\ref{app:extended-results}. The cost-model arm filter is neutral for the
featured UCB1 champion ($1.00\%$ on vs.\ $1.01\%$ off at $T{=}100$) even though
disabling it nearly doubles the average offered arm count ($2.56{\to}4.01$), so
pruning pessimistic arms costs nothing and keeps selection cheap
(Table~\ref{tab:filter-ablation}). Clipped-SGD robust mean estimation does not help
and usually hurts (Softmax $0.93{\to}1.36\%$, UCB-tuned $4.38{\to}6.18\%$, UCB-V
$6.66{\to}8.23\%$), because at $K{\le}5$ arms and $T{=}100$ the bias from clipping
outweighs its variance reduction (Table~\ref{tab:robust-ablation}). The featured
UCB1 champion nonetheless carries robust estimation \emph{on} because Optuna
selected it on the train objective; on test this is a within-noise $+0.06$~pp
change ($0.94{\to}1.00\%$, paired SEM $0.06$), so on the stationary test it is a harmless tuning artifact
and no evidence that robustness helps. Under drift it may matter: Appendix~\ref{sec:drift} attributes plain
UCB1's drift robustness to the clipped-SGD step size, but we did not replay a sample-mean UCB1 under drift, so that
attribution is a mechanism argument, not a measurement. The adaptive
observed-max wrapper is essentially a no-op, since rewards $r{=}\hat{\mu}_{\min}/t$
rarely exceed their early running maximum: most families move by $0.00$~pp and only
the coupled hierarchy is materially worse ($3.23{\to}3.81\%$;
Table~\ref{tab:aom-ablation}). The champions' gains therefore come from cost-model
warm start and exploration control; reward renormalization adds almost nothing.

\subsection{Population vs.\ conditional \TRLT{}}
\label{sec:worst-case}

Uniform job pooling (Section~\ref{sec:limitations}) weights all $300$
multi-arm contexts equally (12 scenarios $\times$ up to 26 message sizes; the
full sweep grid holds $306$ measured cells, but 6 large-size cells offer only a
single measured arm and admit no selection, Section~\ref{sec:protocol}).
NCCL \AUTO{} attains
$\mathrm{RLT}{=}0$ on roughly $80\%$ of contexts (matching its static
\PBest{}), so its $5.99\%$ held-out test mean is driven by the remaining roughly
$20\%$ where the cost model mis-ranks arms.
Table~\ref{tab:conditional-rtl} restricts to the contexts where \AUTO{} has positive
normalized regret; this is an exploratory stratification with all partitions pooled,
not a held-out measurement, which is why the headline stays with
Table~\ref{tab:policy-ranking-full}.
On those failing contexts \AUTO{} averages $26.1\%$ \TRLT{} at $T{=}100$ against
$2.4\%$ for Softmax, $2.8\%$ for UCB1, and $4.8\%$ for Gaussian Thompson, a
$9.2$--$11.0\times$ conditional gap, wider than the $6.4\times$ population-mean
ratio for Softmax ($6.0\times$ for the featured UCB1 champion); these ratios are
selection-priced, unlike the realized headline. Learning policies
still pay a small exploration tax on easy contexts ($0.2\%$ mean \TRLT{} where
\AUTO{} is already oracle-equivalent).
Both views matter: population \TRLT{} is deployable overhead under uniform context
reuse, while conditional \TRLT{} isolates recovery where static selection fails.

\begin{table}[t]
  \centering
  \caption{Conditional \TRLT{} (\%) on the contexts where NCCL \AUTO{} mis-ranks arms (measured-best $\neq$ predicted-best; exploratory stratification, all holdout partitions pooled). Rows are the cost model and the four warm-start champions discussed in the main text; the full policy set is in the benchmark CSV. \capTRLTany{} \capbold{}}
  \label{tab:conditional-rtl}
  \begin{tabular}{lr}
    \toprule
    policy & conditional \TRLT{} (\%) \\
    \midrule
    NCCL \AUTO{} & 26.08 \\
    Softmax (warm) & \textbf{2.38} \\
    UCB1 (warm) & 2.82 \\
    Gaussian Thompson (warm) & 4.76 \\
    Hierarchical EXP3/Softmax (warm) & 7.76 \\
    \bottomrule
  \end{tabular}
\end{table}

\subsection{Learning horizon}
\label{app:horizon}

Table~\ref{tab:horizon-rtl} reports held-out \textbf{test} \TRLT{} at
$T{\in}\{10,20,50,100\}$ for each search champion, computed post-hoc from the same
benchmark run (Section~\ref{sec:abl-prior-horizon}). Every learning champion improves
monotonically with $T$, so the $T{=}100$ ranking is not an artifact of a horizon
sweet spot; sequential replay preserves lazy-init overhead on first arm pulls, which
is why short horizons look pessimistic. The two fixed selectors bracket the range:
\AUTO{} is horizon-invariant at $5.99\%$, while uniform random stays high across
horizons ($46.8\%$ at $T{=}10$, $45.2\%$ at $T{=}100$) because it keeps sampling bad arms.
\begin{table}[t]
  \centering
  \caption{Held-out \textbf{test} \TRLT{} (\%) vs.\ horizon on the holdout benchmark harness (search champion per family; Table~\ref{tab:policy-ranking-full}). Deterministic fixed selectors (e.g.\ NCCL \AUTO{}) are horizon-invariant: the same arm is chosen every pull. \capkey{} \capbold{}}
  \label{tab:horizon-rtl}
  \setlength{\tabcolsep}{3pt}
  \resizebox{\columnwidth}{!}{%
  \begin{tabular}{@{}lrrrr@{}}
    \toprule
    & \multicolumn{4}{c}{epoch length $T$} \\
    \cmidrule(l){2-5}
    policy & 10 & 20 & 50 & 100 \\
    \midrule
    NCCL \AUTO{} (cost model) & 5.99 & 5.99 & 5.99 & 5.99 \\
    Random & 46.83 & 44.52 & 44.94 & 45.22 \\
    \multicolumn{5}{@{}c@{}}{\dashfill}\\
    Softmax (warm) & 5.76 & 3.22 & 1.52 & \textbf{0.93} \\
    UCB1 (warm) & \textbf{3.73} & \textbf{2.73} & \textbf{1.42} & 1.00 \\
    KL-UCB (warm) & 4.96 & 3.01 & 1.74 & 1.18 \\
    Gaussian Thompson (warm) & 5.95 & 3.88 & 2.10 & 1.45 \\
    $\varepsilon$-greedy (warm) & 4.58 & 2.92 & 1.88 & 1.52 \\
    Madan RR-ETC (cold) & 22.16 & 11.35 & 4.86 & 2.70 \\
    ETC (cold) & 22.16 & 11.35 & 4.86 & 2.70 \\
    Hierarchical EXP3/Softmax (warm) & 12.48 & 9.98 & 5.29 & 3.23 \\
    UCB-tuned (warm) & 11.91 & 9.21 & 6.12 & 4.38 \\
    UCB-V (warm) & 20.89 & 16.15 & 10.20 & 6.66 \\
    Tsallis-Inf (warm) & 28.41 & 20.47 & 12.17 & 7.59 \\
    EXP3 (cold) & 28.57 & 19.98 & 12.44 & 8.41 \\
    FPL (cold) & 29.05 & 22.24 & 14.55 & 9.91 \\
    EXP3++ (cold) & 44.33 & 41.28 & 33.85 & 27.74 \\
    \bottomrule
  \end{tabular}%
  }
\end{table}

\subsection{Non-stationarity: full table and adversarial baselines}
\label{app:drift-full}

Table~\ref{tab:drift-ablation-full} gives the per-regime p99 alongside the means
summarized in Table~\ref{tab:drift-ablation}, under the construction and
$\gamma$-selection protocol of Section~\ref{sec:drift}.
\begin{table*}[t]
  \centering
  \small
  \caption{Non-stationarity ablation, full detail (compact version: Table~\ref{tab:drift-ablation}): dynamic \TRLT{} (\%, \textbf{lower is better}). \emph{Stationary} is the sequential CV holdout replay---the \emph{same} harness as Table~\ref{tab:policy-ranking-full} ($T{=}100$, 154 held-out test jobs, NCCL lazy-init cold start included), so the plain rows match Table~\ref{tab:policy-ranking-full} and the plain$\to$discounted change is the true stationary cost of forgetting. \emph{Abrupt} (best$\leftrightarrow$worst pool swap at step~50) and \emph{gradual} (crossfade over steps 35--65) are the bootstrap steady-pool drift replay against a dynamic oracle ($T{=}100$, $154\times10$ runs). Values are job-pool mean\,$\pm$\,\SEM{} and p99; best per drift column in \textbf{bold}. Discounting ($\gamma$ selected leakage-free on the drift train partition) cuts drift \TRLT{} several-fold while adding ${<}0.2$~pp on the stationary metric. Purpose-built adversarial baselines (EXP3, EXP3++, Tsallis-Inf, FPL) are discussed in Appendix~\ref{app:extended-results}.}
  \label{tab:drift-ablation-full}
  \begin{tabular}{l cc cc cc}
    \toprule
    & \multicolumn{2}{c}{Stationary (seq., cf.\ Table~\ref{tab:policy-ranking-full})} & \multicolumn{2}{c}{Abrupt} & \multicolumn{2}{c}{Gradual} \\
    \cmidrule(lr){2-3}\cmidrule(lr){4-5}\cmidrule(lr){6-7}
    Policy & mean & p99 & mean & p99 & mean & p99 \\
    \midrule
    NCCL \AUTO{} (static) & 5.99\,\tiny$\pm$1.58 & 103.25 & 45.01\,\tiny$\pm$3.93 & 197.46 & 44.60\,\tiny$\pm$3.89 & 195.36 \\
    Softmax (plain) & \textbf{0.93}\,\tiny$\pm$0.11 & 6.59 & 31.57\,\tiny$\pm$2.63 & 128.39 & 31.99\,\tiny$\pm$2.65 & 133.57 \\
    Softmax ($\gamma{=}0.85$) & 1.07\,\tiny$\pm$0.12 & \textbf{5.34} & 5.96\,\tiny$\pm$0.29 & \textbf{15.70} & 13.11\,\tiny$\pm$1.17 & 58.12 \\
    UCB1 (plain) & 1.00\,\tiny$\pm$0.13 & 7.97 & 6.60\,\tiny$\pm$0.61 & 33.71 & 10.93\,\tiny$\pm$0.87 & 44.31 \\
    UCB1 ($\gamma{=}0.995$) & 1.16\,\tiny$\pm$0.13 & 7.59 & \textbf{5.88}\,\tiny$\pm$0.54 & 30.32 & \textbf{10.68}\,\tiny$\pm$0.87 & \textbf{43.15} \\
    \bottomrule
  \end{tabular}
\end{table*}

\paragraph{Purpose-built adversarial bandits underperform here.}
One might expect algorithms designed for adversarial or non-stationary rewards to
dominate this benchmark. They do not. When we evaluate the tuned holdout champions of
EXP3~\cite{auer2002exp3}, EXP3++~\cite{seldin2014exp3pp},
Tsallis-Inf~\cite{zimmert2021}, and FPL~\cite{honda2023} on the same replay, all four
are the \emph{weakest} learners in every regime. Under the abrupt swap their mean
dynamic \TRLT{} is $30$--$45\%$ (EXP3 $40.5\%$, EXP3++ $45.2\%$, Tsallis-Inf
$33.1\%$, FPL $30.1\%$), at or above the static \AUTO{} cost model ($45.0\%$) and
$5$--$8\times$ worse than the discounted stochastic champions ($5.9\%$); on the
stationary control they lose $6.3$--$26.6\%$ against roughly $1\%$ for the champions
on the same harness.
This is not because the environment is easy. It is \emph{stochastic and
only piecewise non-stationary}, not truly adversarial, and four properties of these
policies cost them here: (i)~their minimax regret
guarantees carry loose finite-time constants that have not amortized at $T{=}100$;
(ii)~they enforce a uniform exploration floor that wastes pulls with only $K{\le}5$
arms; (iii)~they take no cost-model warm start, so they start cold where the
stochastic champions start from the \AUTO{} prior; and (iv)~their loss scaling is
tuned for adversarial $[0,1]$ losses rather than our cost-normalized rewards. On
switching stochastic environments, discounted stochastic estimators are
near-optimal~\cite{garivier2011switching}
and a robust worst-case policy pays for insurance it never uses. We therefore keep
them out of Figure~\ref{fig:drift-ablation} for clarity.

\subsection{Secondary configuration checks}
\label{app:secondary-checks}

\paragraph{Arm filter (matched UCB1 champion; holdout harness).}
Table~\ref{tab:filter-ablation} compares filter-on (the holdout benchmark UCB1
champion, $\rho_{\max}{=}2$) against a matched filter-off rerun with the same
hyperparameters and holdout split. Held-out test \TRLT{} at pulls 50/100 is
unchanged within display precision ($1.42/1.00\%$ on vs.\ $1.41/1.01\%$ off) while
the average offered-arm count rises from $2.56$ to $4.01$ (max $5$ to $7$): the
filter buys a smaller action space at no measured accuracy cost on this policy.
\begin{table}[t]
  \centering
  \caption{Cost-model arm filter ablation on the UCB1 champion (holdout harness: filter-on from Table~\ref{tab:policy-ranking-full} benchmark vs.\ matched filter-off rerun). T50/T100 are held-out \textbf{test} \TRLT{} (\%) at pulls 50/100; K columns summarize offered-arm counts after filtering. \capkey{} \capbold{}}
  \label{tab:filter-ablation}
  \resizebox{\columnwidth}{!}{%
  \begin{tabular}{@{}lrrrrrr@{}}
    \toprule
    config & T50 & T100 & Kmin & Kmed & Kmax & Kavg \\
    \midrule
    On ($\rho\_{\max}{=}2$) & 1.42 & \textbf{1.00} & 2 & 2.00 & 5 & 2.56 \\
    Off & \textbf{1.41} & 1.01 & 2 & 3.00 & 7 & 4.01 \\
    \bottomrule
  \end{tabular}%
  }
\end{table}

\paragraph{Robust estimation (holdout harness; search champions).}
Table~\ref{tab:robust-ablation} toggles clipped-SGD robust mean estimation off vs.\
on for every robust-capable champion, applying defaults when enabling it on a
champion that ran without it. Robust estimation does \emph{not} help on the held-out
test suite and usually hurts: it raises test \TRLT{} for Softmax
($0.93{\to}1.36\%$), UCB-tuned ($4.38{\to}6.18\%$), UCB-V ($6.66{\to}8.23\%$), and
Gaussian Thompson ($1.45{\to}1.49\%$), and is roughly neutral for the UCB1 champion
($0.94{\to}1.00\%$). These variants are empirical clipped-SGD estimators rather than
exact ones; at $K{\le}5$ arms and $T{=}100$ the extra bias from clipping outweighs
its variance reduction, so plain sample means are preferable here.
\begin{table}[tbp]
  \centering
  \caption{Robust-estimation ablation on the holdout harness: search champion hyperparameters per robust-capable family, only robust (clipped-SGD) mean estimation toggled off vs.\ on (defaults used when enabling on a champion that ran without it). Held-out \textbf{test} \TRLT{} (\%) at $T{=}100$; columns report \mbox{mean${\pm}$\SEM{}} across the test context pool (${n{=}}154$ contexts). \underline{Underlined} = setting Table~\ref{tab:policy-ranking-full} used for that champion. $\Delta$ is paired true$-$false per context. \capkey{} \capbold{}}
  \label{tab:robust-ablation}
  \scriptsize
  \setlength{\tabcolsep}{3pt}
  \resizebox{\columnwidth}{!}{%
  \begin{tabular}{@{}lrrr@{}}
    \toprule
    & \multicolumn{3}{c}{\TRLT{} (\%)} \\
    \cmidrule(l){2-4}
    & \multicolumn{2}{c}{robust estimation} & \\
    \cmidrule(lr){2-3}
    policy & false & true & $\Delta$ (true$-$false) \\
    \midrule
    UCB1 & $0.94{\pm}0.13$ & \underline{$\textbf{1.00}{\pm}0.13$} & $0.06{\pm}0.06$ \\
    Softmax & \underline{$\textbf{0.93}{\pm}0.11$} & $1.36{\pm}0.28$ & $0.43{\pm}0.23$ \\
    Gaussian Thompson & \underline{$1.45{\pm}0.27$} & $1.49{\pm}0.31$ & $0.05{\pm}0.17$ \\
    UCB-tuned & \underline{$4.38{\pm}0.20$} & $6.18{\pm}0.35$ & $1.80{\pm}0.20$ \\
    UCB-V & \underline{$6.66{\pm}0.37$} & $8.23{\pm}0.52$ & $1.57{\pm}0.17$ \\
    \bottomrule
  \end{tabular}
  }
\end{table}

\paragraph{Adaptive observed-max (reward renormalization; holdout harness).}
Table~\ref{tab:aom-ablation} toggles the adaptive observed-max wrapper off vs.\ on
for every learning champion. The wrapper divides cost-model-normalized pull rewards
by their running maximum, rescaling inner statistics when a new maximum appears. On
this harness it is essentially a \emph{no-op}: rewards $r{=}\hat{\mu}_{\min}/t$ rarely
exceed their early running maximum, so the divisor stays near its initial value and
test \TRLT{} is unchanged to two decimals for most families (Softmax, UCB1, KL-UCB,
$\varepsilon$-greedy, Gaussian Thompson, Madan RR-ETC, ETC). It is mildly harmful
where it does fire: Hierarchical $3.23{\to}3.81\%$, FPL $9.91{\to}10.45\%$,
EXP3++ $27.74{\to}28.68\%$.
\begin{table}[tbp]
  \centering
  \caption{Adaptive-observed-max (reward renormalization) ablation on the holdout harness: search champion hyperparameters per learning family, only adaptive observed-max toggled off vs.\ on. The wrapper divides cost-model-normalized pull rewards by their running maximum, rescaling inner statistics when a new maximum appears. Held-out \textbf{test} \TRLT{} (\%) at $T{=}100$; columns report \mbox{mean${\pm}$\SEM{}} across the test context pool (${n{=}}154$ contexts). \underline{Underlined} = setting Table~\ref{tab:policy-ranking-full} used for that champion. $\Delta$ is paired true$-$false per context. \capkey{} \capbold{}}
  \label{tab:aom-ablation}
  \scriptsize
  \setlength{\tabcolsep}{3pt}
  \resizebox{\columnwidth}{!}{%
  \begin{tabular}{@{}lrrr@{}}
    \toprule
    & \multicolumn{3}{c}{\TRLT{} (\%)} \\
    \cmidrule(l){2-4}
    & \multicolumn{2}{c}{adaptive observed-max} & \\
    \cmidrule(lr){2-3}
    policy & false & true & $\Delta$ (true$-$false) \\
    \midrule
    Softmax & \underline{$\textbf{0.93}{\pm}0.11$} & $\textbf{0.93}{\pm}0.11$ & $0.00{\pm}0.00$ \\
    UCB1 & \underline{$1.00{\pm}0.13$} & $1.00{\pm}0.13$ & $0.00{\pm}0.00$ \\
    KL-UCB & \underline{$1.18{\pm}0.24$} & $1.18{\pm}0.24$ & $0.00{\pm}0.00$ \\
    Gaussian Thompson & $1.45{\pm}0.27$ & \underline{$1.45{\pm}0.27$} & $0.00{\pm}0.00$ \\
    $\varepsilon$-greedy & $1.52{\pm}0.43$ & \underline{$1.52{\pm}0.43$} & $0.00{\pm}0.00$ \\
    Madan RR-ETC & \underline{$2.70{\pm}0.49$} & $2.70{\pm}0.49$ & $0.00{\pm}0.00$ \\
    ETC & $2.70{\pm}0.49$ & \underline{$2.70{\pm}0.49$} & $0.00{\pm}0.00$ \\
    Hierarchical EXP3/Softmax & \underline{$3.23{\pm}0.36$} & $3.81{\pm}0.46$ & $0.58{\pm}0.26$ \\
    UCB-tuned & \underline{$4.38{\pm}0.20$} & $4.62{\pm}0.19$ & $0.23{\pm}0.05$ \\
    UCB-V & \underline{$6.66{\pm}0.37$} & $6.87{\pm}0.37$ & $0.21{\pm}0.06$ \\
    Tsallis-Inf & \underline{$7.59{\pm}0.46$} & $7.59{\pm}0.47$ & $0.01{\pm}0.03$ \\
    EXP3 & \underline{$8.41{\pm}0.52$} & $8.58{\pm}0.52$ & $0.17{\pm}0.03$ \\
    FPL & \underline{$9.91{\pm}0.72$} & $10.45{\pm}0.70$ & $0.53{\pm}0.11$ \\
    EXP3++ & \underline{$27.74{\pm}1.85$} & $28.68{\pm}1.93$ & $0.94{\pm}0.23$ \\
    \bottomrule
  \end{tabular}
  }
\end{table}

\subsection{Extended ranking notes}
\label{app:ranking-notes}

\paragraph{Hierarchical parent choice.}
The tuned hierarchy uses an EXP3 parent over algorithm groups and a Softmax child
within each selected group ($3.23\%$ held-out test mean \TRLT{}, $17.30\%$ p99).
The child policy changes as it learns, so the reward the parent observes for a fixed
group is generated by a changing within-group action distribution rather than a
stationary arm distribution. We therefore use an adversarial parent, whose update
does not assume stationary group rewards. This is a robustness choice, not a
necessity: a non-stationary stochastic meta-bandit would also serve, but a
stationary UCB or Thompson interpretation at the parent level is misspecified for
the coupled learner.

\paragraph{ETC and Madan RR-ETC coincide at $m{=}1$.}
Both search champions tuned to a single exploration round, and
Table~\ref{tab:policy-ranking-full} therefore renders them identically ($2.70\%$ test
mean, $8.68\%$ p99; their train \TRLT{} is bit-identical and their test values
differ only beyond two-decimal display, $2.699$ vs.\ $2.698$). When each arm
contributes one learning update before commit, the $1\%$
safety gate rarely overrides \AUTO{} at these short horizons
($|\mathcal{A}'|{\in}\{2,\ldots,5\}$ on most contexts), so both commit to the same
empirical best. They remain distinct policies, not a duplicated row.

\section{Hyperparameter tuning}
\label{app:hyperopt}

\begin{figure}[t]
  \centering
  \includegraphics[width=\linewidth]{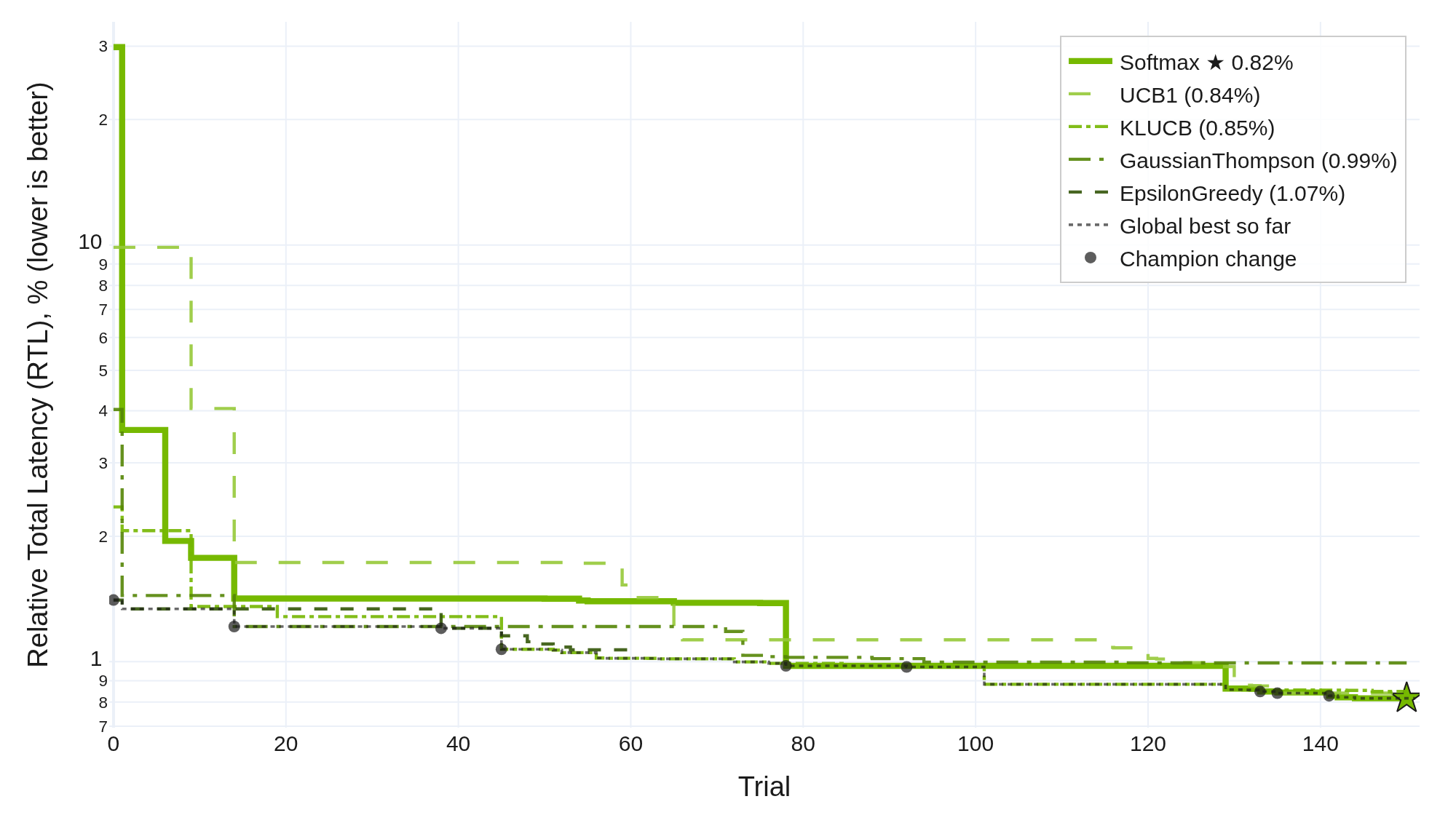}
  \caption{Per-bandit hyperparameter search progress on the holdout tuning objective
    (train \TRLTintro{}, \textbf{lower is better}).
    Lines show the best-so-far configuration for the five strongest families
    among the available trial histories (including every family that held the
    global record); distinct dash patterns make traces readable without color.
    The dotted near-black staircase is the global best-so-far across bandits.}
  \label{fig:optuna-progress}
\end{figure}

This appendix records the Optuna search procedure and frozen champion settings
behind Table~\ref{tab:policy-ranking-full}.
The train/test context split, sequential replay harness, and reporting protocol
are defined in Section~\ref{sec:protocol}; headline rankings use held-out
\textbf{test} realized \TRLT{}.
Train columns and Figure~\ref{fig:optuna-progress} are search diagnostics only.

\paragraph{Search objective and harness.}
Each Optuna trial runs the same emulation job as the main benchmark
(Section~\ref{sec:protocol}): one independent bandit per context, $T{=}100$
sequential replay, base seed~42, pull reward $R_t = \hat{\mu}_{\min}/X_t$
from~\eqref{eq:ratio-reward}, cost-model filter~\eqref{eq:filter} with
$\rho_{\max}{=}2$, and \TRLT{} denominators from arm means over iterations
$\ge 11$.
The objective is mean \textbf{train}-partition \TRLT{} only ($146$ contexts);
test is never used for selection.

\paragraph{Optuna search.}
We run 14 independent Optuna~\cite{optuna2019} studies (one per learning family
in Table~\ref{tab:policy-config}); \AUTO{} and Random are baselines without search.
Each study runs $151$ trials on the train partition with the TPE sampler:
$1$ educated-guess seed (trial~0) $+ 50$ random startup trials $+ 100$ TPE trials,
sampler seed~$42$.
We use TPE~\cite{optuna2019} because it natively models \emph{hierarchical,
conditional} search spaces: parameters that only exist when a parent choice is
active. Our spaces are strongly conditional: the robust-SGD clip threshold and
learning rate exist only when robust estimation is on, the warm-start
pseudo-count only when warm start is on, and the hierarchical family has
per-level (parent/child) sub-spaces. TPE handles these tree-structured
dependencies directly (unlike samplers that assume a fixed-dimension continuous
cube, e.g.\ CMA-ES or a plain Gaussian-process surrogate), which is why we prefer
it here.
Table~\ref{tab:policy-config} summarizes, per family, the trial index at which the
champion was found, its search-objective improvement over the educated-guess
default, and the hyperparameters tuning changed from that default.
The winning trial per bandit is frozen and replayed once with no further tuning.

\paragraph{Wrappers and warm start (champion replay).}
Each champion is assembled as
inner policy $\rightarrow$ optional skip-first-$k$ rewards
$\rightarrow$ optional adaptive observed-max.
Adaptive observed-max divides cost-model-normalized pulls by the running maximum
observed ratio and rescales inner statistics when a new maximum appears.
Warm start seeds $r(\hat{\mu}(a))$ with pseudo-count $\nu$; hierarchical warm
start honors either the root or the nested per-level settings.
Robust estimation uses clipped-SGD means with a clip threshold and learning rate.

\begin{table*}[tbp]
  \centering
  \caption{Per-family hyperparameter-search summary (Appendix~\ref{app:hyperopt}), merging reproducibility and champion configuration. Each of the 14 learning families runs an independent Optuna study on the train partition of the holdout harness (1 educated-guess seed at trial~0 $+ 50$ random startup $+ 100$ TPE trials; sampler seed 42; objective mean \textbf{train} \TRLT{}). \emph{best trial} is the trial index at which the champion was found. \emph{train \TRLT{} default$\to$best} shows the search-objective improvement over the trial-0 educated guess (with speedup factor). The trials ran 99 steps per context and Table~\ref{tab:policy-ranking-full} 100, with identical choices over the shared steps, so the values differ slightly (UCB1 $0.84\%$ here, $0.86\%$ there). \emph{search changes} lists only the knobs tuning moved from that default. \AUTO{} and Random are baselines without search. EXP3++'s champion is its trial-0 educated-guess default (best trial~$0$; $22.2{\to}22.19$ over $151$ trials): no sampled configuration improved on it, which we read as an ineffective search space for that family on this scale-OOD harness rather than a strong default---consistent with EXP3++ being the weakest learner in the full policy ranking. \emph{Abbreviations:} pc $=$ warm-start pseudo-count $\nu$ (virtual prior pulls); pv $=$ Gaussian-Thompson prior variance; ign-$k$ $=$ skip-first-$k$ learning updates per arm; AOM $=$ adaptive observed-max reward wrapper; warm $=$ cost-model warm start; lr and $\eta_r$ are the \emph{distinct} robust clipped-SGD learning rate and step-size decay exponent (e.g.\ UCB1: lr${\to}0.468$ vs.\ $\eta_r{\to}0.467$); clip $=$ robust clip threshold $\lambda$; clipR $=$ clip loss/reward to $[0,1]$; gate $=$ Madan safety-gate margin; $\varepsilon$dec $=$ $\varepsilon$ decay; decay $=$ Softmax temperature decay; $m$ $=$ ETC exploration rounds. A ``${\to}$value'' entry with no left-hand side is a conditional knob that was inactive at the trial-0 default (its governing flag was off) and that tuning switched on at the stated value.}
  \label{tab:policy-config}
  \scriptsize
  \setlength{\tabcolsep}{4pt}
  \resizebox{\textwidth}{!}{%
  \begin{tabular}{@{}lrrl@{}}
    \toprule
    policy & best trial & train \TRLT{} default$\to$best (\%) & search changes (init$\to$best) \\
    \midrule
    Softmax & 144 & 29.8$\to$0.82 (37$\times$) & $\tau$ 1$\to$0.089; decay F$\to$T; pc $\to$1.24; warm F$\to$T \\
    UCB1 & 135 & 9.9$\to$0.84 (12$\times$) & robust F$\to$T; clip $\to$0.389; lr $\to$0.468; $\eta_r$ $\to$0.467; $\alpha$ 0.05$\to$0.0928; pc $\to$2.89; warm F$\to$T \\
    KL-UCB & 146 & 2.4$\to$0.85 (3$\times$) & c 0.217$\to$0.0373; pc $\to$1.56; warm F$\to$T \\
    Gaussian Thompson & 123 & 4.0$\to$0.99 (4$\times$) & ign-$k$ 0$\to$1; pv 0.002$\to$0.000301; pc $\to$0.491; warm F$\to$T; AOM F$\to$T \\
    $\varepsilon$-greedy & 53 & 1.4$\to$1.07 (1$\times$) & pc $\to$0.892; warm F$\to$T; AOM F$\to$T \\
    Madan RR-ETC & 48 & 11.1$\to$2.44 (5$\times$) & ign-$k$ 0$\to$1; $m$ 10$\to$1; gate 0.05$\to$0.0102; pc $\to$0 \\
    ETC & 58 & 10.0$\to$2.44 (4$\times$) & ign-$k$ 0$\to$1; AOM F$\to$T; pc $\to$0 \\
    Hierarchical EXP3/Softmax & 144 & 12.4$\to$2.86 (4$\times$) & ign-$k$ 0$\to$3 \\
    UCB-tuned & 143 & 5.1$\to$4.36 (1$\times$) & pc $\to$2; warm F$\to$T \\
    Tsallis-Inf & 122 & 7.4$\to$5.35 (1$\times$) & $\alpha$ 0.85$\to$0.883; clipR F$\to$T; loss\_estimator iw$\to$rv; warm F$\to$T \\
    EXP3 & 90 & 14.8$\to$5.48 (3$\times$) & $\gamma$ 0.1$\to$0.0523; $\eta$ 0.1$\to$1.51; pc $\to$0 \\
    UCB-V & 138 & 21.2$\to$6.60 (3$\times$) & $\zeta$ 1$\to$0.501; c 3$\to$1.02; pc $\to$1.73; warm F$\to$T \\
    FPL & 146 & 10.3$\to$8.27 (1$\times$) & $\eta$ 1$\to$4.85; pc $\to$0 \\
    EXP3++ & 0 & 22.2$\to$22.19 & warm $\to$F; pc $\to$0 \\
    \bottomrule
  \end{tabular}%
  }
\end{table*}

\section{Deployment: plugin architecture and measured overhead}
\label{app:deployment}

\paragraph{Scope.}
This appendix documents the NCCL tuner-plugin architecture used for the live
closed-loop study in Section~\ref{sec:live} (illustrated for Spectrum-X; the same
feedback hooks apply on other fabrics). The interface mechanisms below are absent
from upstream NCCL; we implemented them in the SPCX environment plugin, which is
what let the bandit run in-loop on real hardware. The measured profiling and
cross-rank synchronization overhead is reported in Table~\ref{tab:live-overhead}.

\begin{figure}[t]
  \centering
  \resizebox{\linewidth}{!}{\providecolor{spcxNcclDraw}{HTML}{2E7D32}
\providecolor{spcxNcclFill}{HTML}{E6F4EA}
\providecolor{spcxPluginDraw}{HTML}{1565C0}
\providecolor{spcxPluginFill}{HTML}{E8F0FE}
\providecolor{spcxBanditDraw}{HTML}{E68A00}
\providecolor{spcxBanditFill}{HTML}{FFE7B3}
\providecolor{spcxStep}{HTML}{C62828}
\providecolor{spcxLbl}{gray}{0.35}
\begin{tikzpicture}[
  every node/.style={transform shape=false},
  collbox/.style={draw=spcxNcclDraw, fill=white, rounded corners=2pt,
                  font=\scriptsize\bfseries, text=spcxNcclDraw, align=center,
                  minimum width=2.2cm, minimum height=1.5cm},
  cmbox/.style={draw=spcxNcclDraw, fill=white, rounded corners=2pt,
                align=center, inner sep=3pt},
  barbox/.style={draw=spcxPluginDraw, fill=spcxPluginDraw, text=white,
                 rounded corners=2pt, font=\scriptsize\bfseries, align=center,
                 minimum height=0.7cm},
  banditbox/.style={draw=spcxBanditDraw, fill=spcxBanditFill, rounded corners=3pt,
                    font=\scriptsize\bfseries, align=center, minimum height=0.9cm},
  contbox/.style={rounded corners=4pt},
  arr/.style={-{Stealth[length=2.2mm]}, thick},
  biarr/.style={{Stealth[length=2mm]}-{Stealth[length=2mm]}, thick},
  lbl/.style={font=\tiny, text=spcxLbl, align=center},
  stepnum/.style={circle, draw=spcxStep, fill=white, text=spcxStep,
                  font=\scriptsize\bfseries, inner sep=0pt, minimum size=3.4mm}
]
  \begin{scope}[on background layer]
    \node[contbox, draw=spcxNcclDraw, fill=spcxNcclFill,
          minimum width=8.6cm, minimum height=2.4cm] (core) at (4.5,4.2) {};
    \node[contbox, draw=spcxPluginDraw, fill=spcxPluginFill,
          minimum width=9.0cm, minimum height=4.6cm] (plugin) at (4.5,-1.3) {};
  \end{scope}

  \node[cmbox] (cm) at (1.7,4.2) {%
    {\scriptsize\textbf{Cost Model}}\\[1pt]
    {\tiny\setlength{\tabcolsep}{3pt}\renewcommand{\arraystretch}{0.9}%
     \begin{tabular}{@{}lcc@{}}
       \textbf{AR 1K} & Ring & Tree\\
       LL     & \textcolor{spcxStep}{\textbf{3.2}} & 5.0\\
       LL128  & 5.2  & 6.2\\
       Simple & 10.0 & 17.3\\
     \end{tabular}}%
  };
  \node[collbox] (coll) at (6.0,4.2) {Collective};

  \node[barbox, minimum width=4.8cm] (tuner)    at (3.2,0.0) {Tuner};
  \node[barbox, minimum width=1.8cm] (profiler) at (6.8,0.0) {Profiler};
  \node[banditbox, minimum width=8.6cm] (bandit) at (4.5,-2.6)
    {Per-Context Multi-Arm Bandit\\[1pt]{\tiny (one instance per (collective, size, scale); arms = Algo / Proto)}};

  \node[anchor=north west, font=\scriptsize\bfseries, text=spcxNcclDraw]
    at ($(core.north west)+(3pt,-3pt)$) {NCCL Core};
  \node[anchor=south west, font=\scriptsize\bfseries, text=spcxPluginDraw]
    at ($(plugin.south west)+(3pt,3pt)$) {SPCX NCCL Plugin};

  \draw[arr, spcxNcclDraw] (coll.west) -- (cm.east);
  \node[stepnum] at ($(coll.west)!0.5!(cm.east)$) {1};

  \draw[arr, spcxPluginDraw] (cm.south) -- ([xshift=-1.5cm]tuner.north);
  \node[stepnum] at (1.7,1.9) {2};
  \node[lbl, anchor=west, text width=1.5cm] at (1.95,1.9)
    {cost table + msg size + job dims};

  \draw[biarr, spcxBanditDraw] ([xshift=0cm]tuner.south) -- ([xshift=-1.3cm]bandit.north);
  \node[stepnum] at (3.2,-1.25) {3};
  \node[lbl, anchor=west] at (3.45,-1.25) {online\\parameter selection};

  \draw[arr, spcxPluginDraw] ([xshift=2.0cm]tuner.north) -- ([xshift=-0.8cm]coll.south);
  \node[stepnum] at (5.2,1.9) {4};
  \node[lbl, anchor=east] at (4.95,1.9) {provide\\params};

  \draw[arr, spcxNcclDraw] ([xshift=0.8cm]coll.south) -- (profiler.north);
  \node[stepnum] at (6.8,1.9) {5};
  \node[lbl, anchor=west] at (7.05,1.9) {collective\\events};

  \draw[arr, spcxBanditDraw] (profiler.south) -- ([xshift=2.3cm]bandit.north);
  \node[stepnum] at (6.8,-1.25) {6};
  \node[lbl, anchor=west] at (7.05,-1.25) {reward};

  \node[banditbox, minimum width=2.6cm, minimum height=0.7cm,
        densely dashed, fill opacity=0.4, text opacity=1] (peerbandit) at (1.5,-1.05)
    {Bandit {\tiny\mdseries(rank $r{+}1$)}\\[1pt]%
     \textcolor{spcxStep}{\tiny new online-sync API}};
  \draw[biarr, spcxStep] (peerbandit.south) -- ([xshift=-3.0cm]bandit.north);
  \node[stepnum] at (1.5,-1.78) {7};
  \node[lbl, anchor=west, text width=1.3cm] at (1.75,-1.78)
    {synchronize state across ranks};
\end{tikzpicture}}%
  \caption{Tuner-plugin architecture. Green: NCCL core (cost model, collective
    execution). Blue: SPCX plugin (tuner, profiler); orange: the per-context bandit.
    (1)~the collective triggers the cost model; (2)~NCCL invokes the tuner with the cost
    table, message size, and job dimensions; (3)~the bandit selects an algorithm--protocol
    pair; (4)~parameters return to NCCL; (5)--(6)~the profiler returns the measured
    latency as the reward; (7)~bandit state is synchronized across ranks.}
  \label{fig:spcx-arch}
\end{figure}
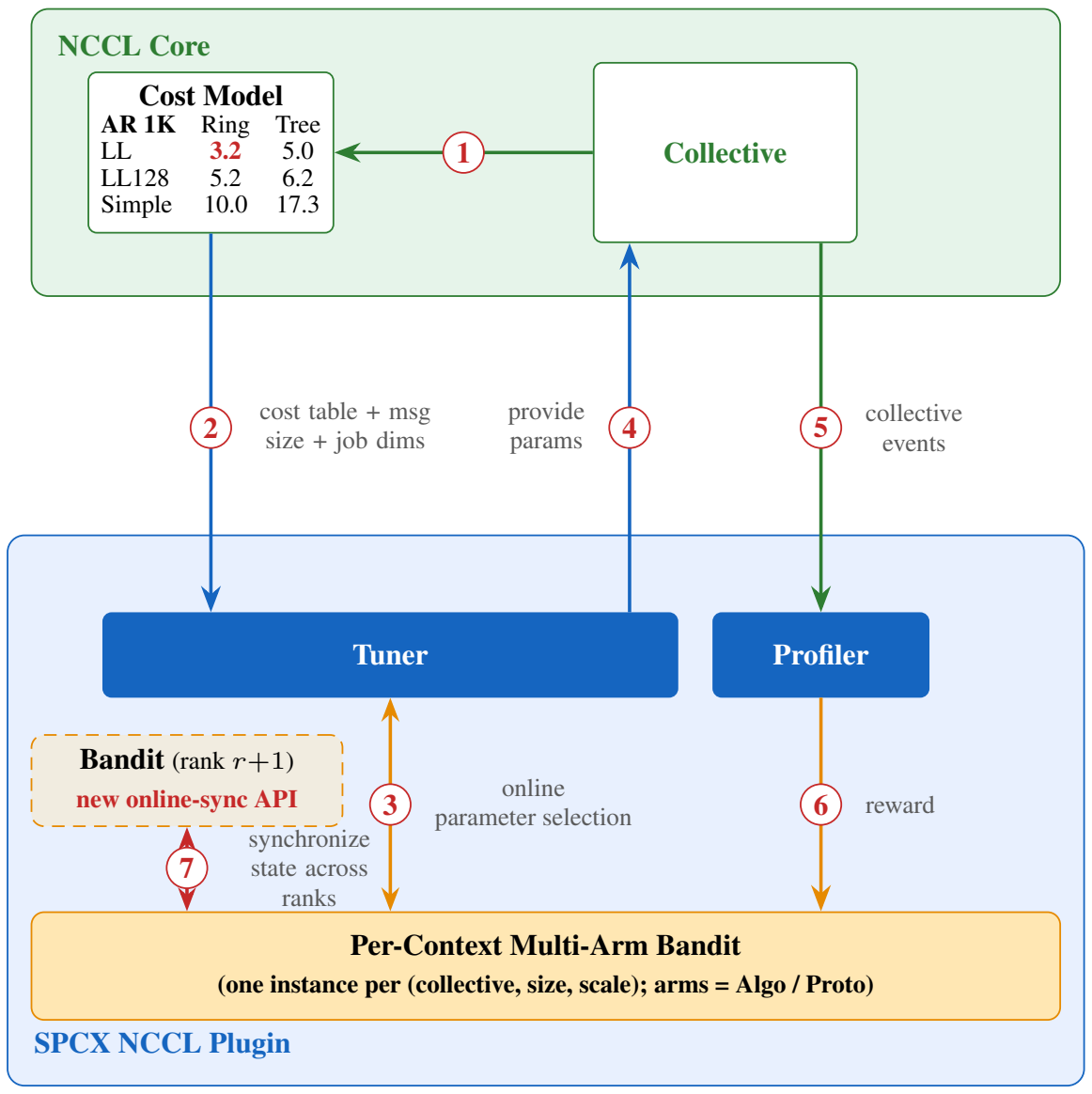

\paragraph{Architecture.}
Figure~\ref{fig:spcx-arch} shows the deployed plugin stack.
NCCL's cost model supplies per-arm predicted latencies to the \emph{tuner} callback;
a \emph{profiler} observes collective completion times and returns rewards to a
\emph{multi-arm bandit} over algorithm--protocol pairs (one instance per context).
The bandit selects parameters for the next collective invocation; a new API
synchronizes bandit state across ranks when an update is accepted.
This mirrors the offline loop (Algorithm~\ref{alg:loop}) but executes inside the
training job rather than via offline replay.
On each \texttt{pluginGetCollInfo} invocation, NCCL supplies collective type,
communicator size~$P$, and message size~$n$; the plugin looks up (or lazily
creates) the bandit for that context key and writes algo/proto costs into the
returned table, matching our offline lazy per-$(\text{collective}, P, n)$
instances rather than one global policy across all sizes.

\paragraph{Interface inventory (upstream vs.\ implemented).}
The deployment relies on one upstream mechanism and three we added. \emph{Upstream:}
the per-call cost-table override in \texttt{getCollInfo} (tuner interface v3, NCCL~2.22 and later~\cite{nccl222blog}), which
changes the next collective's algo/proto. \emph{Implemented in the SPCX plugin:}
(i)~a profiler$\rightarrow$tuner \emph{reward-feedback channel} returning each
collective's measured latency, aggregated across ranks (each rank times only its
own completion); (ii)~\emph{durable per-context bandit state} across
revisits (the callback otherwise has no memory of past latencies); and (iii)~a
\emph{cross-rank state-synchronization API} so all ranks share one policy and
agree on the selected arm. Upstream NCCL ships none of (i)--(iii) as a supported path
between its tuner and profiler plugins, so we built them in the SPCX plugin environment,
and the live study of Section~\ref{sec:live} exercises all three.
Upstream NCCL does document a profiler plugin interface, and one library can export
both plugin types, so (i) in particular may be reproducible without modifying NCCL.
We have not established which of (i)--(iii) strictly require NCCL changes.
In outline, and without a formal specification: (i)~the channel delivers, per collective, the measured
latency in $\mu$s together with the communicator, collective type, message size, and the algorithm--protocol pair
that was played, to the tuner before its next call for that context; (ii)~the state is a per-context table keyed by
(collective, ranks, message size) holding per-arm counts, reward means (and the second moments some policies use),
created on first sight and kept for the life of the communicator; (iii)~in the intended design, the synchronization is an all-reduce of the
per-arm reward sums and counts accumulated since the last flush, after which every rank applies the same update.
Sums and counts reproduce a sample mean exactly; the clipped-SGD and constant-gain estimators depend on the order of
individual samples, so with them every rank must apply the same flushed samples in the same order (an all-gather) or
update once per flush from the batch mean; which the live UCB1 build did is not recorded.
The sync rule differs between what we evaluated and what ran: the offline evaluation updates after every collective and
charges no sync cost; the live overhead runs synchronized every 4, 16 or 64 collectives; and the rule the live learners
used is not recorded in the run metadata we hold.
The run metadata we hold does not record which tuner interface version the plugin was built against.
The plugin is planned for release (Section~\ref{sec:limitations}); this outline is not a substitute for its interface.
Per-context state is $O(K)$ words (counts, means, last arm), so thousands of contexts are
kilobytes to megabytes, but we have not measured footprint or the cost of synchronizing
many contexts.

\paragraph{Rank synchronization and action consistency (sketch).}
NCCL requires \emph{identical} algo/proto choices on every participating rank,
which needs two distinct synchronizations. First, \emph{inputs must agree}: each
rank times only its own completion and cannot observe when the collective
\emph{globally} finished, so the reward must be a cross-rank aggregate. We take
the all-reduced \emph{maximum} completion time as a proxy for the
application-visible collective latency (the last rank to finish); the mean or a
leader's local time are possible alternatives we leave to future work. The
per-arm sufficient statistics must likewise be replicated; without this each
rank's estimates diverge and even a pure $\arg\max$ selection can pick different
arms, so state synchronization is required for \emph{all} policies, deterministic
or stochastic. Second, \emph{the decision must be reproducible} given identical
statistics: a \emph{stochastic} policy (Softmax, Gaussian Thompson, EXP3/EXP3++,
Tsallis-Inf, or a hierarchical child) draws from the shared distribution~$\pi$
independently on each rank and can still \emph{sample different arms}.
That is an NCCL correctness violation, not a performance issue, so a stochastic policy
additionally needs shared randomness (a communicator-shared RNG seed synchronized with~$\pi$, or a leader
rank that samples and broadcasts the arm). A policy that is a deterministic
function of the shared statistics avoids this second step; UCB1 and
$\varepsilon$-greedy qualify only once their cold-start/tie-break exploration
(``unpulled arms chosen uniformly at random'') is likewise derandomized (a fixed
order) or seeded.
In the intended design (we cannot verify that the build that ran followed it), the plugin replicates \emph{bandit sufficient statistics} rather
than raw latency traces: per-arm pull counts, empirical reward means (and
variances where used), and the last committed arm. It all-reduces a fixed-size
vector whenever the flush rule accepts an update for a context $(\text{collective}, P, n)$, so every rank enters the next selection with the same $\pi$ (and,
for stochastic or seeded policies, the same drawn arm).
The decision of \emph{whether} to update and sync on a given collective must
itself be a collectively-agreed predicate (leader-gated, or derived from
communicator-wide agreed context keys), so that divergent lazy context creation or
update gating across ranks cannot deadlock the sync all-reduce, with some ranks
entering it while others skip, before the \AUTO{} fallback can trigger.
NCCL's current tuner interface falls back to its default selection for a collective when \texttt{getCollInfo} returns an error, so a
per-call fallback to \AUTO{} is available when a rank detects disagreeing context keys (the same safety pattern as Madan et
al.'s commit gate).
A failed synchronization all-reduce, by contrast, is a communicator error that a tuner cannot recover
from per call, so the predicate has to prevent it, not recover from it.

\paragraph{Measured sync cost.}
By design the per-update sync is a small collective: each update all-reduces an $O(K)$-word
state vector (${\ll}1$~KiB for $K{\le}5$ arms, independent of~$n$), and $C$
concurrently active contexts batch into one $O(C\!\cdot\!K)$-word collective. We
measure its cost directly with plugin runs that profile and synchronize state but leave \AUTO{}'s selection untouched, so any
bandwidth gap versus stock NCCL is pure overhead
(Table~\ref{tab:live-overhead}). Profiling alone (sync off) costs $2.1\%$ of
average bus bandwidth; adding cross-rank sync every $4$/$16$/$64$ collectives
costs $7.5$/$3.5$/$2.3\%$. \emph{Frequency} dominates: the most
frequent sync (every $4$) is costliest, and batching to a longer period
amortizes it.
Each figure is one run; the same policy run in the two spacing modes, separate jobs, differs by a median $0.4$--$1.5\%$ at
${\ge}64$\,MiB (Table~\ref{tab:live-overhead-size}).
The average hides where the cost lands.
Profiling alone adds $0.6\%$ per collective below 1\,MiB and $5.7\%$ at 64--512\,MiB, so it is not a fixed per-call cost,
and we do not know its mechanism.
The sync cost does behave like a fixed per-call cost: each period implies about $0.5$--$0.9$\,ms per sync below 1\,MiB,
which the near-zero small-size bandwidth hides from the average. The diluted-update ablation (Section~\ref{sec:update-freq}) shows
learning survives such batching essentially unchanged. On H100-A this overhead
($2.1$--$7.5\%$) is between about a tenth and two fifths of the $18$--$24\%$ average-bandwidth gain of the adaptive learners.
That average is over 26 sizes and dominated by the large ones: per size, every adaptive learner is slower than stock at all
11 sizes below 1\,MiB (median $+11$ to $+14\%$ out-of-place time, against $+0.3\%$ for profiling alone on that measure and
$+0.6\%$ on the in/out mean above), and gains only at the large sizes of
\AUTO{}'s error.
On H100-A the NVLS stall falls in the warm-up pass of \texttt{nccl-tests} in every run, \AUTO{}'s included, so these averages
exclude it.
H100-B has no plugin-off run, but its forced-arm companion sweep on the same nodes has one stock run (a separate job).
Against it the 26-size average bus bandwidth of every H100-B learner is $0.94$--$0.99\times$ stock (removing the stalled size
changes this by at most $0.02$), an average the large sizes dominate; Table~\ref{tab:stock-band} splits it by size band.
Every learner is more than $5\%$ faster than stock at each 1--8\,MiB size except $\varepsilon$-greedy at 1\,MiB, and more than
$5\%$ slower at every size below 1\,MiB.
Below 8\,MiB the learners play the cost-model argmin on ${\ge}99.8\%$ of timed collectives.
At 1--8\,MiB the forced Tree/LL128 run, that argmin, is $1.26$--$1.51\times$ faster than stock, and the learners are $0.88$--$0.94\times$ that forced run, so the band's gain is a
gap between the stock run and the forced arms, not a selection gain.
Below 1\,MiB the sync explains only part of the loss.
The median slowdown ($0.59$--$0.65\times$, i.e.\ $+54$--$69\%$ time) is far above H100-A's $+13$--$15\%$.
Against the forced sweep's run of the same arm the learners are still $0.76$--$0.86\times$, while that forced run is itself
$0.97$--$1.70\times$ stock's time.
ETC, whose H100-B re-exploration setting is not recorded, shows the loss too, unlike H100-A's commit-once ETC.
In the back-to-back H100-A runs with \AUTO{}'s selection fixed (the mode H100-B ran), a sync every 64 collectives costs $+4\%$
host time below 1\,MiB over stock and $+2\%$ over profiling alone, against $+13\%$ every 16 (Table~\ref{tab:live-overhead-size}).
In both campaigns every arm switch of UCB1, Softmax, Thompson and $\varepsilon$-greedy falls on one phase of a 16-call
counter per size, and every one of Greedy's on another phase; ETC's are spread.

\begin{table}[t]
  \centering
  \scriptsize
  \caption{H100-B learners against the companion sweep's stock run (a separate job), per size band: median over sizes of
    stock time over learner time (above 1: learner faster), per-size range in parentheses; \texttt{nccl-tests} out-of-place
    timed pass, host clock. Each run's NVLS-stall size is removed (ETC's lies below 1\,MiB, the others' at 1--8\,MiB),
    leaving 10--11, 3--4 and 11 sizes. Softmax's $0.11$ at 64\,MiB is a second first-touch gap (${\approx}0.64$\,s, Ring/LL128)
    in its timed pass. One run per policy. Bold: best (highest) per column.}
  \label{tab:stock-band}
  \begin{tabular}{@{}lrrr@{}}
\toprule
Policy & $<$1\,MiB & 1--8\,MiB & 16\,MiB--16\,GiB \\
\midrule
Greedy & 0.64 (0.54--0.89) & 1.27 (1.23--1.42) & 0.98 (0.93--1.03) \\
ETC ($m{=}1$) & \textbf{0.65} (0.61--0.90) & \textbf{1.31} (1.20--1.45) & 0.97 (0.82--1.07) \\
UCB1 & \textbf{0.65} (0.56--0.88) & 1.29 (1.27--1.41) & 0.97 (0.86--1.02) \\
Softmax & 0.61 (0.42--0.87) & 1.22 (1.19--1.40) & 0.97 (0.11--1.03) \\
Thompson & 0.59 (0.33--0.87) & 1.29 (1.26--1.40) & 0.98 (0.91--1.01) \\
$\varepsilon$-greedy & 0.64 (0.45--0.89) & 1.21 (1.00--1.24) & \textbf{0.99} (0.90--1.03) \\
\bottomrule
\end{tabular}

\end{table}

\begin{table}[t]
  \centering
  \caption{\textbf{Measured profiling + cross-rank sync overhead} at fixed \AUTO{} selection (H100-A: 8 nodes, 64 ranks, back-to-back). All rows make NCCL's default selection, so bandwidth lost relative to stock NCCL is the cost of running the plugin: profiling every collective and all-reducing bandit state every \emph{state sync} period. Bus bandwidth is the \texttt{nccl-tests} average over 26 sizes. \AUTO{} plugin (sync off) prices profiling alone; the sync/$k$ rows add cross-rank state exchange every $k$ collectives. Frequency, not payload (${<}1$~KiB), dominates: sync/4 is costliest and the overhead shrinks as the period grows, so batching updates amortizes it (Section~\ref{sec:lessons}). On this fabric the average overhead is smaller than the average-bandwidth gain of the adaptive learners (Table~\ref{tab:live-ranking}), which comes from the large sizes; below 1~MiB the learners are slower than stock, and the overhead is not size-independent (Table~\ref{tab:live-overhead-size}). Each row is one run. Bold: best per column (highest bandwidth, lowest overhead).}
  \label{tab:live-overhead}
  \resizebox{\columnwidth}{!}{%
  \begin{tabular}{@{}lrrr@{}}
    \toprule
    configuration & state sync & bus bw (GB/s) & overhead (\%) \\
    \midrule
    Stock NCCL & --- & 73.0 & --- (ref.) \\
    NCCL \AUTO{} & off & \textbf{71.5} & \textbf{2.1} \\
    \AUTO{} + sync/4 & 4 & 67.5 & 7.5 \\
    \AUTO{} + sync/16 & 16 & 70.5 & 3.5 \\
    \AUTO{} + sync/64 & 64 & 71.3 & 2.3 \\
    \bottomrule
  \end{tabular}}
\end{table}

\begin{table}[t]
  \centering
  \scriptsize
  \setlength{\tabcolsep}{2.5pt}
  \caption{Per-size plugin overhead on H100-A (back-to-back, \AUTO{}'s selection fixed): median per-collective time
    overhead against stock (\%) per size bin, with the bin's share of the 26-size average-bus-bandwidth drop in parentheses;
    ``1\,ms flush'' is the number of collectives per sync that the 1\,ms rule would imply at stock latency (median over the
    bin). The last row compares each run with the same policy's run in the other spacing mode (a separate job) at
    ${\ge}64$\,MiB, as a single-run noise yardstick. One run per column. Bold: best per row among the policies.}
  \label{tab:live-overhead-size}
  \resizebox{\columnwidth}{!}{\begin{tabular}{@{}lrrrrrr@{}}
\toprule
Size bin & $n$ & no sync & sync/4 & sync/16 & sync/64 & 1\,ms flush: coll./sync \\
\midrule
$<$1 MiB & 11 & +0.6 (0) & +61.6 (1) & +13.4 (1) & \textbf{+4.3} (1) & 3 \\
1--32 MiB & 6 & +1.4 (5) & +41.8 (32) & +10.1 (19) & \textbf{+3.8} (13) & 2 \\
64--512 MiB & 4 & +5.7 (80) & +13.0 (40) & +7.6 (58) & \textbf{+6.0} (69) & 1 \\
$\ge$1 GiB & 5 & +0.4 (14) & +2.9 (27) & +1.2 (22) & \textbf{+0.7} (17) & 1 \\
\midrule
avg.\ bus bw drop & 26 & 2.1 & 7.5 & 3.5 & \textbf{2.3} & \\
run-to-run, $\ge$64\,MiB & 9 & 0.4 & 1.5 & 1.3 & 0.5 & \\
\bottomrule
\end{tabular}
}
\end{table}

\begin{figure}[!htbp]
  \centering
  \includegraphics[width=\linewidth]{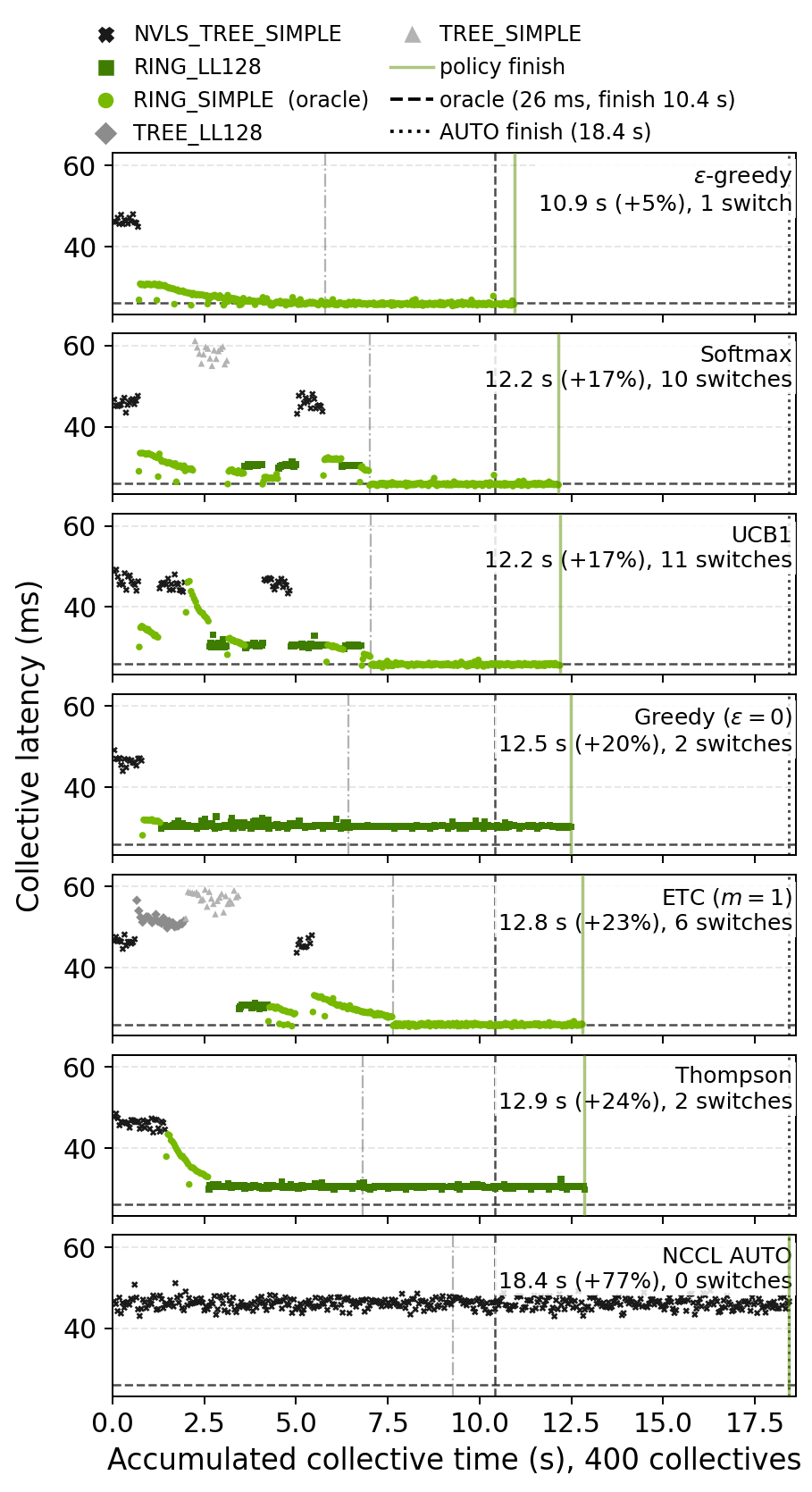}
  \caption{H100-A live arm choices at the largest \AUTO{} gap (AllReduce, 4\,GiB,
    back-to-back): one facet per policy, fastest first, each of the 400 collectives placed at
    accumulated collective time, with colour and marker giving the selected arm. Solid green:
    policy finish; dashed:
    oracle latency and finish; dotted: \AUTO{}'s finish. Every learner leaves \AUTO{}'s arm
    within 31 collectives; reaching the oracle depends on \emph{when} it samples
    \texttt{RING\_SIMPLE}, whose latency drifts early (probe means 28.9--40.4\,ms vs.\ 25.9\,ms
    settled) while \texttt{RING\_LL128} holds 30.5\,ms. Greedy ($\varepsilon$-greedy with
    $\varepsilon{=}0$) samples it at collective 17, measures 31.6\,ms, and never returns;
    Thompson likewise: its stationary posterior on \texttt{RING\_SIMPLE} concentrates on the
    drift-inflated 36.6\,ms, so although every arm keeps non-zero probability, a revisit is
    ${\sim}10^{-25}$ likely per decision and a longer run would not rescue it. ETC ($m{=}1$) reaches the oracle only because it samples
    \texttt{RING\_SIMPLE} fifth (28.9\,ms); sampled first, it would likely have committed to
    \texttt{RING\_LL128}. Finish times include the drift, so they exceed \TRLT{}. One run per
    policy: this illustrates the mechanism, not a ranking.}
  \label{fig:live-learning-trace}
\end{figure}

\begin{table*}[t]
  \centering
  \scriptsize
  \caption{Live decomposition across the two closed-loop H100 campaigns. Selection quality is
    priced by a profiler against a reconstructed oracle with setup excluded; plugin
    overhead is measured with \AUTO{}'s selection fixed (profiling alone, then with state syncs at fixed periods; one run
    per row); the setup toll is the first-use \texttt{NVLS\_TREE\_SIMPLE}
    stall; ``--'' marks fields not measured. No learner's rank consistency was verified live.}
  \label{tab:live-decomp}
  \begin{tabular}{p{0.16\linewidth}p{0.14\linewidth}p{0.52\linewidth}c}
\toprule
Component & Campaign & Finding & Measured \\
\midrule
selection quality & H100-A (64 ranks) & best learner epsilon-greedy: 2.86\% mean / 8.64\% p99 selection TRLT & \checkmark \\
 & H100-B (64 ranks) & best learner greedy (epsilon=0): 0.59\% mean / 1.98\% p99 selection TRLT & \checkmark \\
plugin / sync overhead & H100-A (64 ranks) & 2.1\% average bus bandwidth lost vs stock with profiling only (AUTO selection, state sync off; one run per row); sync every 4 collectives is costliest and the ladder falls toward stock as the sync period grows & \checkmark \\
 & H100-B (64 ranks) & not measured: this campaign has no plugin runs with AUTO selection fixed & -- \\
one-time setup toll & H100-A (64 ranks) & 6.6 s median (6.4--7.8 s over 10 runs) NVLS\_TREE\_SIMPLE first-use stall (6.1\% median of run wall); $\sim$123,717 median steady collectives to amortize at 1\% & \checkmark \\
 & H100-B (64 ranks) & 5.8 s median (5.7--5.9 s over 6 runs) NVLS\_TREE\_SIMPLE first-use stall (6.2\% median of run wall); $\sim$172,261 median steady collectives to amortize at 1\% & \checkmark \\
\bottomrule
\end{tabular}

\end{table*}

Table~\ref{tab:live-ranking} lists the H100-A policies, including the timed-pass bandwidth
column behind the ETC comparison above; the text quotes its values.
The H100-A row of Table~\ref{tab:live-decomp} is computed from the same profiler data by a separate pipeline; the two
differ by at most $0.1$\,pp and we did not reconcile them.

\begin{table}[t]
  \centering
  \caption{\textbf{Live closed-loop tuner on real hardware} (8 nodes, 64 ranks, AllReduce; H100-A campaign). The bandit ran \emph{inside} NCCL via the SPCX env plugin and selected the algo/proto arm online, with a per-collective profiler returning measured latency. UCB1, Softmax and Thompson ran the offline champions' hyperparameters, warm-started from the cost model; $\epsilon$-greedy ran an outdated constant ($\epsilon{=}0.033$, decaying) instead of its tuned value $\epsilon{=}0$, and no policy used the constant-gain mean we recommend as the default; Greedy ($\epsilon{=}0$ from the cost-model prior) and ETC (one sample per arm) are implemented in the plugin itself. Selection quality is mean/worst-size \TRLT{} over the $10$ scorable sizes (32~MiB--16~GiB, where every arm is well sampled), priced from pooled per-arm profiler latency; ``mean rank'' is the average placing over those sizes (1~=~best). ``Bus bw'' is the wall-clock average over all 26 sizes of the 100 timed iterations (after 100 warm-up iterations) from the \texttt{nccl-tests} \texttt{out} tables, so it includes whatever exploration falls in the timed pass, and ``vs stock'' is the ratio to stock NCCL (no plugin). Bold marks the best value per column. One run per policy (no seed variance); the oracle is reconstructed by pooling every arm observation across runs of the same spacing, including the scored runs. \TRLT{} charges exploration over all 400 collectives, so ETC ($m{=}1$) can lead on timed-pass bandwidth yet trail on \TRLT{}. No learner's rank consistency was verified (data checking off), and the stochastic learners may not have been rank-consistent, so the ranking is indicative only. \capTRLTany{}}
  \label{tab:live-ranking}
  \setlength{\tabcolsep}{4pt}
  \resizebox{\columnwidth}{!}{%
  \begin{tabular}{@{}lrrrrr@{}}
    \toprule
    & \multicolumn{2}{c}{\TRLT{} (\%)} & & \multicolumn{2}{c}{bus bw} \\
    \cmidrule(lr){2-3} \cmidrule(l){5-6}
    policy & mean & worst & rank & GB/s & vs stock \\
    \midrule
    $\epsilon$-greedy & \textbf{2.85} & \textbf{9.1} & \textbf{1.6} & 90.7 & 1.24$\times$ \\
    Greedy & 6.43 & 19.0 & 2.8 & 86.1 & 1.18$\times$ \\
    UCB1 & 8.28 & 12.5 & 4.0 & 89.3 & 1.22$\times$ \\
    Thompson & 8.49 & 20.4 & 3.9 & 88.0 & 1.21$\times$ \\
    Softmax & 9.00 & 15.6 & 4.0 & 90.0 & 1.23$\times$ \\
    ETC ($m{=}1$) & 21.96 & 49.4 & 5.7 & \textbf{93.5} & \textbf{1.28}$\times$ \\
    NCCL \AUTO{} & 31.35 & 77.3 & 5.1 & 71.5 & 0.98$\times$ \\
    \bottomrule
  \end{tabular}%
  }
\end{table}

\paragraph{Live campaign setup.}
\label{app:live-setup}
H100-A uses 2 queue pairs per connection and 10 for long-haul links; H100-B uses 1 for both.
The run metadata we hold record no long-haul emulation for the primary sweep.
The variable \texttt{NCCL\_IB\_LH\_LATENCY} is not in NCCL's public documentation, and we do not know its unit or where it
injects delay; H100-A ships no cost table, so we cannot check whether its model saw the emulated latency.
H100-B has no runs with \AUTO{}'s selection fixed, so its small-size loss is attributed to the sync only in part, by analogy
with H100-A; diluting the sync is a straightforward mitigation, but no learner ran live with it.
The first-use \texttt{NVLS\_TREE\_SIMPLE} stall ranges over $6.4$--$7.8$\,s on H100-A (10 runs; $6.1\%$ of run wall time)
and $5.7$--$5.9$\,s on H100-B (6 runs); amortizing it to 1\% takes about 660\,s of collectives, while the offline harness
charges at most the first sample's excess (${\le}9.8$\,ms).

\paragraph{How long live inflation lasts.}
The replay models the lazy-initialization transient as the first ten samples of each arm (Section~\ref{sec:protocol}).
On H100-A (back-to-back runs, 262 (policy, size, arm) cells with at least 100 pulls, each pull divided by the cell's
settled median over pulls beyond 50), the 90th percentile of that ratio is $1.28$ at the first pull, $1.09$--$1.12$ over pulls
3--10, and at most $1.07$ from pull 11 to 40, and the share of cells more than $5\%$ above settled falls from $32\%$ to $14\%$ by
pull 11.
So the live inflation lasts about as long as the replay's transient window; the cells are arms a policy kept playing, so arms
abandoned early (\texttt{RING\_SIMPLE} under Greedy in Figure~\ref{fig:live-learning-trace}) are not in this sample.

\paragraph{Relation to Madan et al.}
The tuned ETC ($m{=}1$) row beats \AUTO{} on held-out test \TRLT{}
(Table~\ref{tab:policy-ranking-full}) in the same direction as the latency reductions Madan et
al.\ report on live A100 clusters~\cite{madan2026}, which is not a validation of their
results (their report is not peer reviewed and our configuration differs).
In our live study the ETC ($m{=}1$) plugin reaches
$93.5$~GB/s ($1.28\times$ stock) on H100, the best timed-pass bandwidth of any
policy (Table~\ref{tab:live-ranking}). Because that average hides exploration
cost in the warm-up pass, this live result does not separate commit-once from adaptive policies.
On realized offline latency ETC has a $63.0\%$ p99 against UCB1's $9.9\%$ (Table~\ref{tab:realized}).
The residual differences reflect hardware generation (H100 vs.\ A100) and exploration budget:
their $10 \times |\mathcal{A}'|$ pulls inflate early \TRLT{} on longer horizons.
On our selection-quality metric the same commit-once policy indeed trails the adaptive bandits
($22.0\%$ vs.\ $\epsilon$-greedy's $2.9\%$) because \TRLT{} charges the
exploration that the timed-pass bandwidth hides.

\paragraph{Metric choice.}
We recommend \TRLT{} for systems comparisons, \CRLT{} for learning curves, and
\RLT{} for per-step gaps.
Classical cumulative reward-space regret remains a useful diagnostic but can
mis-rank policies when the cost-model prior skews the objective
(Section~\ref{sec:results}).

\paragraph{First-touch NVLS toll: sensitivity of the realized ranking.}
\label{app:nvls-toll}
In the live profiler traces the stall is the single largest inter-collective gap per run (at least 1\,s), attributed to the arm played right after it; we did not separate it from other arms' lazy initialization with a controlled experiment.
The live campaigns (Section~\ref{sec:live}) show a stall on the first use
of \texttt{NVLS\_TREE\_SIMPLE} (median $6.6$\,s, range $6.4$--$7.8$\,s, on H100-A; $5.8$\,s, $5.7$--$5.9$\,s, on H100-B), which the pinned offline sweep never charges.
Table~\ref{tab:nvls-reprice} re-prices the realized \TRLT{} by adding a one-time toll to the
first NVLS collective of every held-out context in which a policy (or the oracle) uses an
NVLS arm.
This is an \emph{upper bound}: in production the toll is paid once per communicator, not once
per context, so a job that touches NVLS in many contexts pays it once.
The fraction of contexts in which a policy touches an NVLS arm at all is $11.7\%$ for \AUTO{},
$3.9\%$ for GBM+cost, $16.2\%$ for UCB1, $21.4\%$ for Softmax and $22.1\%$ for Random, so exploration pays the toll more often than a static
choice.
At the measured toll the ordering of Table~\ref{tab:realized} inverts for any policy that
explores.
Arms with large one-time setup costs should therefore be excluded from exploration, or
pre-warmed at communicator creation, not priced inside a $T{=}100$ horizon.
Table~\ref{tab:nvls-policy} gives the two less pessimistic accountings
(Section~\ref{sec:lessons}): the benchmark rerun with \texttt{NVLSTree\_Simple} removed from
every policy's arm set (the cost-model minimum, \AUTO{}'s argmin, the filter, and the oracle
all see the reduced set; the frozen champions and the supervised fits are otherwise
unchanged), and a pooled, traffic-weighted \TRLT{} in which the $5.8$\,s toll is paid once
per communicator size in which a policy plays NVLS, against a pre-warmed oracle, with $V$
visits per context.
The oracle is pre-warmed and never pays.
Charging it the same tolls would add a common term to every denominator and
cannot reorder policies that touch NVLS in both communicator sizes, but could change their order relative to the supervised
models without the cost feature, which never play NVLS (not recomputed).
At $V{=}10^2$, the $T{=}100$ replay, the pooled collective time is only $89$\,s, so a $5.8$\,s
toll paid in two communicators moves every policy that touches NVLS by about $13$\,pp;
at $V{=}10^4$ it moves them by about $0.1$\,pp.

\begin{table}[tbp]
  \centering
  \small
  \caption{Realized \TRLT{} (held-out test mean, \%) if the first NVLS collective of each
    context costs a one-time toll (upper bound; see text). ``none'' reproduces
    Table~\ref{tab:realized}. \capkey{} \capbold{}}
  \label{tab:nvls-reprice}
  \begin{tabular}{lrrrr}
\toprule
Policy & none & 1\,s & 3\,s & 5.8\,s \\
\midrule
NCCL \AUTO{} & 6.1 & 98.9 & 285.1 & 545.8 \\
Supervised GBM+cost, fixed & 1.8 & \textbf{11.8} & \textbf{32.2} & \textbf{60.9} \\
Supervised GBM+cost, tuned & \textbf{1.4} & 12.2 & 34.4 & 65.5 \\
ETC ($m{=}1$, $k{=}1$) & 4.7 & 595.6 & 1779.2 & 3436.2 \\
Softmax & 1.8 & 411.5 & 1231.9 & 2380.6 \\
UCB1 & 1.6 & 209.7 & 626.6 & 1210.3 \\
KL-UCB & 2.1 & 380.0 & 1136.9 & 2196.6 \\
Random & 46.8 & 637.7 & 1821.3 & 3478.3 \\
\bottomrule
\end{tabular}

\end{table}

\begin{table}[tbp]
  \centering
  \scriptsize
  \setlength{\tabcolsep}{2.5pt}
  \caption{NVLS handling (held-out test, realized \TRLT{}, \%). \emph{NVLS offered}: the main
    benchmark. \emph{NVLS excluded}: \texttt{NVLSTree\_Simple} removed from every policy's arm
    set and from the oracle. \emph{Pooled}: traffic-weighted \TRLT{} over all test contexts
    with a $5.8$\,s toll paid once per communicator size in which the policy plays NVLS,
    against a pre-warmed oracle, at $V$ visits per context. \capkey{} \capbold{}}
  \label{tab:nvls-policy}
  \resizebox{\columnwidth}{!}{\begin{tabular}{lrrrrrrr}
\toprule
& \multicolumn{2}{c}{NVLS offered} & \multicolumn{2}{c}{NVLS excluded} & \multicolumn{3}{c}{Pooled, 5.8\,s once per comm.} \\
\cmidrule(lr){2-3}\cmidrule(lr){4-5}\cmidrule(lr){6-8}
Policy & mean & p99 & mean & p99 & $V{=}10^2$ & $10^3$ & $10^4$ \\
\midrule
NCCL \AUTO{} & 6.08 & 103.3 & 6.27 & 111.4 & 15.82 & 4.14 & 2.98 \\
Supervised GBM+cost & 1.78 & 32.2 & 1.79 & 32.2 & 14.08 & 2.41 & 1.24 \\
ETC ($m{=}1$) & 4.74 & 63.0 & 3.26 & 13.8 & 13.76 & 2.08 & 0.92 \\
Softmax & 1.75 & 11.9 & \textbf{1.22} & \textbf{8.6} & 13.50 & 1.82 & 0.66 \\
UCB1 & \textbf{1.61} & \textbf{9.9} & 1.23 & 11.2 & 13.60 & 1.93 & 0.76 \\
KL-UCB & 2.11 & 22.7 & 1.90 & 27.5 & \textbf{13.47} & \textbf{1.80} & \textbf{0.63} \\
\bottomrule
\end{tabular}
}
\end{table}

\section{Contextual bandits}
\label{app:contextual}
The per-context bank of Section~\ref{sec:method} shares nothing across contexts, which raises
the question of whether a single contextual bandit that shares a model across contexts would do
better.
We tested this with one native LinUCB policy over the same offered arms and the same
cost-model prior, and the same $T{=}100$-per-context horizon, on a balanced uniform-random
workload ($100$ visits to each of the $154$ held-out contexts, shuffled; five reshuffles).
This is the friendliest workload for a shared model, because it maximizes context diversity per
observation, and the least favorable to the bank; real traces revisit few contexts
(Section~\ref{sec:trace-replay}).
The features are a per-collective quadratic block in $\log_2$ message size and $\log_2$ node
count plus the cost model's predicted reward for the arm being scored and its product with
$\log_2$ size (``cost'' basis); the prior enters as one pseudo-observation per
context--arm, injected on first sight of a context, so the policy is deployable.
Without the cost-model feature LinUCB falls behind both banks, and a physics-motivated
Hockney basis is worst, because the observed quantity is a latency \emph{ratio} on a geometric
size grid.

Table~\ref{tab:contextual} gives the selection-priced, per-context \TRLT{} (the same unit as
elsewhere).
LinUCB with the cost feature reaches $0.61\pm0.04\%$ mean and $5.31\%$ p99, below the warm bank
($1.00\%$ / $7.3$--$8.0\%$) and the supervised GBM+cost selector ($1.76\%$ / $36.9\%$); cold
banks and \AUTO{} are far worse.
We do not promote this to a headline: it is selection-priced, one workload family, tuned with
its own search, and not yet scored with realized accounting or a real trace.
It does not change the thesis, since the contextual policy also depends on the cost-model
prior; it indicates that the per-context bank is a conservative, easily deployed design,
not the only one.

\begin{table}[tbp]
  \centering
  \small
  \caption{Per-context \TRLT{} (\%) on the balanced uniform workload over the $154$ held-out
    contexts (selection-priced; mean $\pm$ SD over five workload reshuffles where the policy is
    stochastic in the interleaving). \capkey{} Bank: one independent bandit per context; LinUCB: one linear contextual
    bandit shared across contexts. \capbold{}}
  \label{tab:contextual}
  \begin{tabular}{lrr}
    \toprule
    Policy & mean & p99 \\
    \midrule
    NCCL \AUTO{} & 5.99 & 103.3 \\
    Random & 45.73 & 194.0 \\
    Supervised GBM+cost & 1.76 & 36.9 \\
    Bank UCB1, cold & 7.68 & 102.8 \\
    Bank Softmax, cold & 5.26$\pm$0.49 & 56.7 \\
    Bank UCB1, warm & 1.00 & 8.0 \\
    Bank Softmax, warm & 1.00$\pm$0.04 & 7.3 \\
    LinUCB (cost basis), warm & \textbf{0.61}$\pm$0.04 & \textbf{5.3} \\
    \bottomrule
  \end{tabular}
\end{table}

\section{Workload weighting and cross-architecture detail}
\label{app:workload}
\begin{figure}[tbp]
  \centering
  \includegraphics[width=\linewidth]{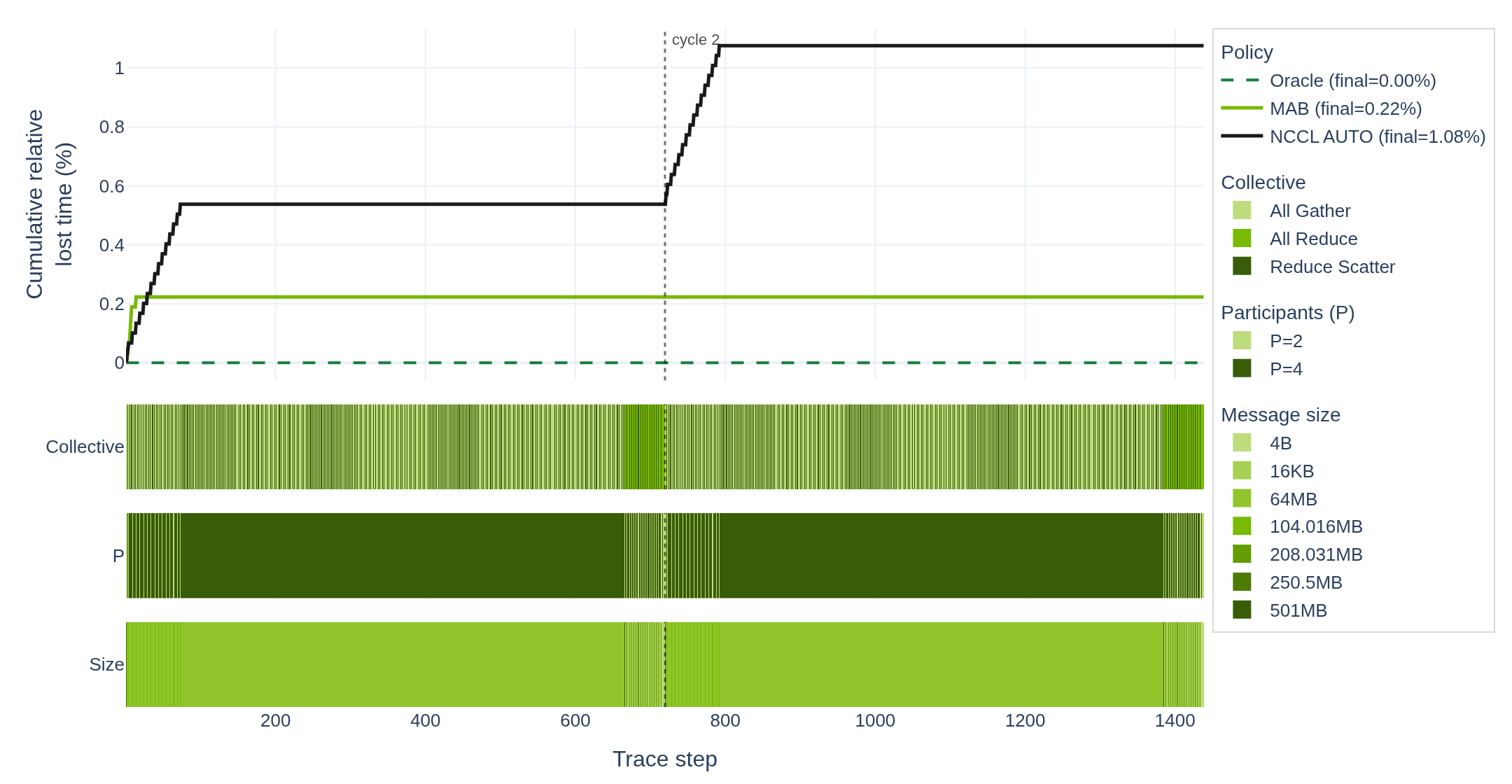}
  \caption{CCL-Bench Llama3 trace replay (two cycles). Warm UCB1 flattens after one-time
    exploration while \AUTO{} repeats the same loss.}
  \label{fig:ccl-bench-replay}
\end{figure}

\paragraph{Trace replay.}
Table~\ref{tab:ccl-bench-replay} and Figure~\ref{fig:ccl-bench-replay} give the Llama3 replay
behind Section~\ref{sec:trace-replay}: two cycles of the trace, warm UCB1 state carried
across cycles, sequential sampling that wraps past the lazy-init region so that repeated
cycles add no cold-start cost.
The statistic is cumulative relative lost time over the whole replay, which weights each
context by its call count; the per-context mean used in the main text weights every context
equally.
Both are computed from arm means (selection pricing).

\begin{table}[t]
  \centering
  \caption{CCL-Bench Llama3 trace replay: path \TRLT{} (\% $\Sigma\mu^\star$) over two cycles of the $\cclBenchCalls{}$-step ordered trace ($1438$ steps, bandit state carried across cycles). Oracle attains 0\%; the warm-started UCB1 champion and NCCL \AUTO{} are deterministic under fixed sequential measurement pools (ten epoch replicates are identical; no \SEM{}). \AUTO{} loses a fixed fraction each cycle; UCB1 amortizes exploration and trends to the oracle as the workload repeats. \capTRLTany{} \capbold{}}
  \label{tab:ccl-bench-replay}
  \begin{tabular}{@{}lr@{}}
    \toprule
    policy & TRLT \\
    \midrule
    UCB1 & \textbf{0.22} \\
    NCCL cost model & 1.08 \\
    \bottomrule
  \end{tabular}
\end{table}

\begin{table}[t]
  \centering
  \scriptsize
  \caption{Datasets and their roles. The H100 sweep and the two live setups are node sets of one Spectrum-X H100 system,
    so the live campaigns are not independent systems.}
  \label{tab:datasets}
  \begin{tabular}{p{0.22\linewidth}p{0.40\linewidth}p{0.24\linewidth}}
    \toprule
    Dataset & Platform / coverage & Role \\
    \midrule
    H100 sweep & HGX H100, Spectrum-X Ethernet, NCCL 2.30.4; 3 collectives $\times$ 4/8/16/32 nodes $\times$ 26 sizes & primary ground truth \\
    GB200 sweep & GB200, 2 rank placements (MNNVL off: one rank per physical node; on: whole allocation); $\sim$310 contexts/slice & cross-architecture \\
    CCL-Bench & Llama3 trace, nine contexts & workload weighting \\
    H100-A live & 8$\times$HGX H100 (64 GPUs), Spectrum-X Ethernet, NCCL 2.30.7 (development build), 2 queue pairs per connection (10 long-haul), long-haul latency emulation; 22 runs, AllReduce & live \\
    H100-B live & 8$\times$HGX H100 (64 GPUs), Spectrum-X Ethernet over RoCE, NCCL 2.32.0a4, 1 queue pair per connection, long-haul latency emulation; 6 runs, AllReduce, on the nodes and configuration of a forced-arm sweep with one stock run & live, stock control \\
    \bottomrule
  \end{tabular}
\end{table}

\begin{table}[t]
  \centering
  \caption{Collective tuning approaches. Day: 0 = built-in selector, 1 = offline pre-production tuning, 2 = online production adaptation. Offline tuning covers profilers and ML tuners (e.g., FACT/ACCLAiM). CM = cost model role; bench = included in our offline benchmark. RR-ETC = round-robin explore-then-commit; $\dagger$ rows are emulated, not live plugin deployments.}
  \label{tab:tuning-comparison}
  \begin{tabular}{@{}lrrrr@{}}
    \toprule
    approach & day & online & CM & bench \\
    \midrule
    NCCL \AUTO{} & 0 & No & Selector & Yes \\
    Offline tuning & 1 & No & No & No \\
    STAR-MPI~\cite{faraj2006starmpi} & 2 & Yes & No & No \\
    AutoCCL~\cite{xu2025} & 2 & Early iters & No & No \\
    Madan RR-ETC & 2 & Yes & Gate & Yes$^\dagger$ \\
    This work & 2 & Live + replay & Prior/filter & Yes$^\dagger$ \\
    \bottomrule
  \end{tabular}
\end{table}

\paragraph{GB200 train and test.}
The frozen H100 hyperparameters were not re-tuned on GB200.
On the GB200 train partition (4/8 nodes) the supervised GBM+cost selector fits its own
training contexts to $0.16\%$ (MNNVL off) and $0.04\%$ (MNNVL on), but on the held-out
16/32-node slice it reaches $4.41\%$ and $3.36\%$ mean (p99 $75.3\%$ and $59.4\%$, selection-priced), worse than \AUTO{} on the mean on both
fabrics: a selector fitted on small topologies does not carry to the large ones.
The forced-arm pinning confound of the GB200 sweeps (Section~\ref{sec:gb200}) bounds but does not
remove the comparison.

\section{Robustness analyses}
\label{app:robustness}
This appendix collects the statistics behind the claims of Sections~\ref{sec:failure}
and~\ref{sec:results} that go beyond the headline tables: where the tail lives, tail-specific
tests and scenario-level intervals, robustness to the split and to tuning, independently tuned
cold champions, a margin-gated hybrid, and the full realized ranking.
Every number is computed from the per-context results of the replays.

\subsection{Tail anatomy}
\label{app:tail-anatomy}
Table~\ref{tab:tail-regions} lists the 19 held-out contexts on which \AUTO{} loses more than
$10\%$ realized \TRLT{}, grouped by (\AUTO{}'s arm, measured-best arm).
In all three groups the forced-arm measurements place a protocol/algorithm crossover at a different size than the
cost model, with predicted margins $m$ of $0.06$--$0.57$.
GBM+cost removes the NVLS group and leaves the other two (up to $40\%$ and $26\%$), whereas UCB1 keeps every group at or
below $13.5\%$.
Table~\ref{tab:scenario-breakdown} splits the loss by (collective, nodes) scenario; the
4-node rows are the eight Llama3 trace contexts.
Each of the six large-scale scenarios contributes three contexts above $10\%$ and between
$12\%$ and $22\%$ of \AUTO{}'s total test loss, so the tail is not one scenario.
Splitting by NVLS involvement (\AUTO{}'s arm or the best arm is NVLS: 19 contexts) shows
\AUTO{} at $14.8\%$ mean there and $4.8\%$ on the other 135, with 6 and 13 contexts above
$10\%$ respectively; UCB1 is $4.5\%$ and $1.2\%$.

Within-pool sampling noise does not explain these contexts.
We resampled each arm's 89 steady iterations independently (iid and 5-iteration blocks,
2000 draws).
The expected exact share of \AUTO{} is $78.1\%$ ($78.2\%$ blocked) against the point estimate $78.6\%$, its mean selection-priced loss has a 95\% interval
$[5.92, 6.08]\%$ and its p99 $[102.3, 104.1]\%$, and 18 of 18 (17 of 18 blocked) contexts
above $10\%$ keep a lower bound above $10\%$.
The noise analysis uses selection-priced loss, where 18 contexts exceed $10\%$ (realized accounting: 19).
Because each arm's pool is one job, this says nothing about between-job variation.

The largest group, Ring/Simple chosen where Ring/LL128 is faster for all-gather and reduce-scatter,
is consistent across separate jobs (Table~\ref{tab:tail-busbw}).
Simple reaches about $320$\,GB/s, $80\%$ of the per-node NIC aggregate, at 4 and 8 nodes, so it is
not throttled everywhere. Its bandwidth ramp moves to larger messages as the ring grows, as
expected when each rank's share of the message shrinks, and the crossover with LL128 moves with it.
The crossover already appears at 8 nodes, a training scale, where Simple takes $2.8\times$ LL128's time at 128\,MiB.
A placement or congestion difference would have to reproduce this ordering in each of the eight
jobs per protocol; repeated jobs remain the direct test.

\begin{table}[t]
\centering
\small
\setlength{\tabcolsep}{3pt}
\caption{Steady bus bandwidth (GB/s per GPU, nccl-tests convention) of Ring/Simple and Ring/LL128 for all-gather (AG) and reduce-scatter (RS) on the primary H100 sweep, as Simple\,/\,LL128 per node count. Each node has eight 400\,Gb/s NICs, about 400\,GB/s aggregate. Simple plateaus near 320\,GB/s (80\%) at 4 and 8 nodes, and its ramp moves to larger messages as the ring grows: at the same per-rank share (size$/$ranks${=}4$\,MiB) it gives 126--207\,GB/s at every scale. LL128 leads by 2.3--4.1$\times$ at 16--32 nodes and 128--512\,MiB, and by 1.04--1.17$\times$ at 2\,GiB. Each value is the mean of one job's steady iterations; the eight jobs per protocol (4 scales $\times$ 2 collectives) are separate.}
\label{tab:tail-busbw}
\begin{tabular}{llrrrr}
\toprule
Coll. & Size & 4 nodes & 8 nodes & 16 nodes & 32 nodes \\
\midrule
AG & 128\,MiB & 207\,/\,267 & 90\,/\,250 & 51\,/\,209 & 32\,/\,120 \\
AG & 256\,MiB & 289\,/\,288 & 163\,/\,282 & 68\,/\,259 & 62\,/\,212 \\
AG & 512\,MiB & 314\,/\,288 & 258\,/\,296 & 126\,/\,292 & 92\,/\,271 \\
AG & 1\,GiB & 321\,/\,300 & 309\,/\,293 & 217\,/\,303 & 166\,/\,296 \\
AG & 2\,GiB & 322\,/\,307 & 327\,/\,303 & 284\,/\,297 & 262\,/\,305 \\
\midrule
RS & 128\,MiB & 203\,/\,266 & 91\,/\,250 & 51\,/\,204 & 33\,/\,127 \\
RS & 256\,MiB & 289\,/\,288 & 164\,/\,283 & 68\,/\,262 & 63\,/\,208 \\
RS & 512\,MiB & 315\,/\,287 & 259\,/\,297 & 125\,/\,293 & 93\,/\,272 \\
RS & 1\,GiB & 321\,/\,298 & 303\,/\,293 & 216\,/\,303 & 168\,/\,294 \\
RS & 2\,GiB & 322\,/\,306 & 327\,/\,302 & 286\,/\,297 & 264\,/\,306 \\
\bottomrule
\end{tabular}
\end{table}

\paragraph{Forcing against stock.}
The primary sweep has no stock run. H100-B has one (all-reduce, 8 nodes, a separate job on the same eight nodes and environment as
its forced runs except the pin; stock's per-size choice was not logged), which we compare with the forced run of the exported cost-model argmin. The same nodes also ran the six live learners, which select arms through the tuner's cost table, the interface through which a plugin applies the cost model's choice (whether stock without a plugin runs that arm the same way is not shown); every learner's first call at each of the 26 sizes was the exported argmin, so the cost table NCCL handed the plugin at run time has the exported argmin. Where a learner played one arm on all 100 timed calls of a size (no setup stall), its time prices that arm under tuner selection, plugin cost included; learners tend to keep arms that did well, so this biases tuner-selected prices low, and these are separate jobs.
Where that argmin is Ring/Simple (4--16\,GiB), stock is within $2.3\%$ of the forced run, and tuner-selected Ring/Simple
costs $1.02$--$1.04\times$ its forced run (2--8\,GiB, 1--3 learners per size): forced, tuner-selected and stock Ring/Simple agree within $4\%$ where their sizes overlap (4--8\,GiB, all-reduce, 8 nodes); the nine Ring/Simple tail contexts lose $16$--$134\%$.
At 16--256\,MiB, where the argmin is NVLSTree/Simple, stock is $11$--$39\%$ faster than the forced argmin and within
$-13\%$ to $+10\%$ of the forced best arm. Tuner-selected NVLSTree/Simple costs $0.70$--$0.98\times$ its forced run (32\,MiB--1\,GiB, 2--3 learners per size), consistent with forcing slowing NVLS; the six
NVLS tail contexts ($12$--$65\%$) may overstate the loss of the cost model's choice, by an amount the selection bias (which would make the NVLS ratio look lower than forcing alone causes) and job variation leave unquantified.
Tuner-selected Tree/LL128 costs $1.05$--$1.18\times$ its forced run (256\,KiB--16\,MiB, median over learners; the excess is
largest below 1\,MiB, where the plugin's per-call cost is) and Ring/LL128 $1.02$--$1.05\times$ (64--128\,MiB), so forcing
makes no non-NVLS arm we can check slower than tuner selection does.
Stock is the outlier: at 1--8\,MiB it is $26$--$51\%$ slower than Tree/LL128 forced and $18$--$34\%$ slower than Tree/LL128
selected through the tuner, and below 1\,MiB it is faster than both, for a reason we cannot identify (stock's choice and
channel count were not logged). So on H100-B the \AUTO{} row's forced price is the cost model's choice as a tuner plugin applies it, which
differs from stock's own run in both directions; the four Ring/LL128 tail contexts ($18$--$26\%$) sit in the
band where stock was slower, not faster.
These checks cover \AUTO{}'s tail arms more than the tail's oracle arms: LL128 is checked only as Ring/LL128 at 64--128\,MiB on
all-reduce, whereas the Ring/Simple tail's oracle is Ring/LL128 on all-gather and reduce-scatter at 512\,MiB--2\,GiB.
Only if the Ring/Simple bound transferred, keeping only that group, would \AUTO{} have 9 contexts above $10\%$ (lowest $13.3\%$ after a $2.3\%$ discount; 7 after
a $13\%$ discount) to UCB1's 0, whose one context above $10\%$ is NVLS; all 5 scenarios that differ favour UCB1 ($p{=}0.0625$,
the smallest five non-zero scenarios allow; the bounds were chosen post hoc).
One of the nine is the unshifted 4-node trace context; on scale-shifted contexts the count is 8 to 0 over four scenarios.
A single bound need not transfer, so we also give the discount at which each context falls to $10\%$: about $53\%$ for the
two 512\,MiB contexts at 16 nodes, $37$--$38\%$ for the 1\,GiB pair at 32 nodes, $21\%$ for the 1\,GiB pair at 16 nodes,
$14\%$ for the trace context, and $5$--$6\%$ for the 2\,GiB pair at 32 nodes. These bound the ratio of \AUTO{}'s to the
oracle arm's forced latency, so they cover distortion of either arm; where H100-B's argmin is its forced-best arm, stock
ranged from $41\%$ faster than the forced run (32\,KiB) to $51\%$ slower (2\,MiB), more than every break-even but the two 512\,MiB contexts.
The NVLS-excluded rerun (\AUTO{} 20 contexts above $10\%$, UCB1 2; Section~\ref{sec:lessons}) changes the arm set, not the
pricing, so it does not address forcing.
All of this is one node set, one collective and one stock job, on another NCCL version under latency emulation; it is
consistent with forcing slowing NVLSTree/Simple and not the other arms checked, finds stock without a plugin differing from the cost model's choice by more than most tail gaps, and does not bound either effect on the primary sweep.

\begin{table}[t]
  \centering
  \scriptsize
  \setlength{\tabcolsep}{3pt}
  \caption{The 19 held-out H100 contexts (\capsplit) on which \AUTO{} has realized \TRLT{} above 10\%, grouped by the
    arm \AUTO{} plays and the measured-best arm. \capTRLT{} $n$: contexts in the group; $m$: \AUTO{}'s predicted
    runner-up margin $\hat\mu_{(2)}/\hat\mu_{(1)}-1$; last three columns: the largest realized \TRLT{} in the group for
    \AUTO{}, UCB1 and the supervised GBM+cost selector. AG/AR/RS: all-gather/all-reduce/reduce-scatter. Bold: best per
    row among the three policies.}
  \label{tab:tail-regions}
  \resizebox{\columnwidth}{!}{\begin{tabular}{@{}llrlrrrr@{}}
\toprule
AUTO picks & best arm & $n$ & collectives, sizes & $m$ & AUTO & UCB1 & GBM+c \\
& & & & & \multicolumn{3}{c}{max realized \TRLT{} in the group (\%)} \\
\cmidrule(lr){6-8}
\midrule
Ring\_Simple & Ring\_LL128 & 9 & AG/RS, 128\,MiB--2\,GiB & 0.08--0.18 & 134 & \textbf{2.9} & 40 \\
NVLSTree\_Simple & Tree\_LL128 & 6 & AR, 16\,MiB--128\,MiB & 0.06--0.57 & 65 & 13.5 & \textbf{0} \\
Ring\_LL128 & Ring\_LL & 4 & AG/RS, 4\,MiB--8\,MiB & 0.20--0.21 & 26 & \textbf{9.1} & 26 \\
\bottomrule
\end{tabular}
}
\end{table}

\begin{table}[t]
  \centering
  \scriptsize
  \setlength{\tabcolsep}{2.5pt}
  \caption{Realized \TRLT{} (\%) by (collective, nodes) scenario on the held-out H100 contexts (\capsplit) for \AUTO{},
    UCB1 and the supervised GBM+cost selector. \capTRLT{} $n$: contexts; loss share: the scenario's share of \AUTO{}'s
    total test loss; mean and max over the scenario's contexts; $n_{>10}$: contexts above 10\%. 4-node rows are the
    Llama3 trace contexts; AG/AR/RS: all-gather/all-reduce/reduce-scatter. Bold: best per row among the three policies,
    separately for mean and max.}
  \label{tab:scenario-breakdown}
  \resizebox{\columnwidth}{!}{\begin{tabular}{@{}lrrrrrrrrr@{}}
\toprule
& & & \multicolumn{3}{c}{NCCL \AUTO{}} & \multicolumn{2}{c}{UCB1} & \multicolumn{2}{c}{GBM+cost} \\
\cmidrule(lr){4-6}\cmidrule(lr){7-8}\cmidrule(lr){9-10}
Scenario & $n$ & loss share & mean & max & $n_{>10}$ & mean & max & mean & max \\
\midrule
AG, 4 nodes & 3 & 3\% & 9.5 & 29 & 1 & 0.67 & 1.3 & \textbf{0.08} & \textbf{0.2} \\
AG, 16 nodes & 24 & 21\% & 8.1 & 131 & 3 & \textbf{1.08} & \textbf{3.3} & 1.83 & 39.2 \\
AG, 32 nodes & 23 & 12\% & 5.0 & 78 & 3 & \textbf{1.19} & \textbf{3.7} & 1.63 & 20.6 \\
AR, 4 nodes & 2 & 0\% & \textbf{0.0} & \textbf{0} & 0 & \textbf{0.00} & \textbf{0.0} & \textbf{0.00} & \textbf{0.0} \\
AR, 16 nodes & 26 & 18\% & 6.3 & 65 & 3 & 2.81 & \textbf{9.9} & \textbf{0.94} & 11.7 \\
AR, 32 nodes & 26 & 13\% & 4.5 & 50 & 3 & 2.84 & \textbf{13.5} & \textbf{2.35} & 25.8 \\
RS, 4 nodes & 3 & 0\% & \textbf{0.0} & \textbf{0} & 0 & 0.43 & 1.3 & 0.05 & 0.2 \\
RS, 16 nodes & 24 & 22\% & 8.5 & 134 & 3 & \textbf{0.43} & \textbf{3.0} & 2.91 & 40.1 \\
RS, 32 nodes & 23 & 12\% & 4.9 & 75 & 3 & \textbf{1.50} & \textbf{9.1} & 1.61 & 21.0 \\
\bottomrule
\end{tabular}
}
\end{table}

\subsection{Scenario-level tests and exceedance intervals}
\label{app:region-inference}
The primary test procedure is over the nine (collective, nodes) scenarios, which are fixed before looking at outcomes
(Table~\ref{tab:tail-region-tests}): an exact sign-flip test on the scenario means and on the per-scenario
difference in contexts above $10\%$, with a wild-cluster bootstrap-$t$ with all $2^9$ Rademacher sign patterns
enumerated as a check~\cite{cameron2008}.
They give $p{=}0.016$ for \AUTO{}$-$UCB1: the smallest two-sided $p$ nine clusters allow is $2/512{\approx}0.004$, and
one scenario favors \AUTO{} (Table~\ref{tab:scenario-per-scenario}).
The same tests cannot separate \AUTO{} from commit-once ETC ($p{\approx}0.4$), and on GB200 they are null for UCB1
against \AUTO{} ($p{=}0.91$, $0.81$).
We report about forty tests and intervals in this appendix without a multiplicity correction; only the four
confirmatory comparisons of Table~\ref{tab:confirmatory}, Holm-corrected, should be read as inferential.
Table~\ref{tab:exceedance-ci} gives exceedance probabilities twice: the exact Clopper--Pearson interval that treats
the 154 contexts as independent, which is too narrow because the failures cluster, and an exact interval on the
number of scenarios with an exceedance (\AUTO{} 7 of 9, UCB1 1 of 9), which is wide but counts the scenario as the unit.
Because the sample p99 of 154 contexts lies between the second- and third-largest values, a bootstrap interval for it is
bounded by the sample maximum.

\begin{table}[t]
  \centering
  \scriptsize
  \setlength{\tabcolsep}{3pt}
  \caption{Tests with the nine (collective, nodes) scenarios as the unit. ``Statistic'': mean difference (pp)
    or exceedance-count difference (contexts above $10\%$) of \AUTO{} minus the policy; sign-flip: exact
    two-sided over $2^9$ sign patterns.}
  \label{tab:tail-region-tests}
  \resizebox{\columnwidth}{!}{\begin{tabular}{@{}llrrr@{}}
\toprule
Test & Comparison & Units & Statistic & $p$ \\
\midrule
sign-flip, scenario means & \AUTO{}-UCB1 mean & 9 & 3.99 & 0.016 \\
wild-cluster bootstrap-$t$ & \AUTO{}-UCB1 mean & 9 & 5.09 & 0.016 \\
sign-flip, exceedance diff. & \AUTO{}-UCB1 $n_{>10}$ & 9 & 18.00 & 0.016 \\
sign-flip, scenario means & \AUTO{}-Softmax mean & 9 & 3.90 & 0.031 \\
wild-cluster bootstrap-$t$ & \AUTO{}-Softmax mean & 9 & 4.08 & 0.031 \\
sign-flip, exceedance diff. & \AUTO{}-Softmax $n_{>10}$ & 9 & 14.00 & 0.047 \\
sign-flip, scenario means & \AUTO{}-ETC mean & 9 & 1.34 & 0.402 \\
wild-cluster bootstrap-$t$ & \AUTO{}-ETC mean & 9 & 0.90 & 0.410 \\
sign-flip, exceedance diff. & \AUTO{}-ETC $n_{>10}$ & 9 & 8.00 & 0.359 \\
\bottomrule
\end{tabular}
}
\end{table}

\begin{table}[t]
  \centering
  \scriptsize
  \setlength{\tabcolsep}{3pt}
  \caption{Contexts above $10\%$ realized \TRLT{} per policy: count (share of 154), an exact interval that
    treats contexts as independent, and the number of the nine scenarios with at least one such context with an
    exact interval on that proportion. \capkey{} \capbold{}}
  \label{tab:exceedance-ci}
  \resizebox{\columnwidth}{!}{\begin{tabular}{@{}lrrr@{}}
\toprule
Policy & contexts $>10\%$ & independent-contexts interval & scenarios with $\ge1$ [interval] \\
\midrule
\AUTO{} & 19 (12.3\%) & [7.6, 18.6]\% & 7/9 [40, 97]\% \\
GBM+cost & 10 (6.5\%) & [3.2, 11.6]\% & 6/9 [30, 93]\% \\
UCB1 & \textbf{1} (0.6\%) & [0.0, 3.6]\% & 1/9 [0, 48]\% \\
Softmax & 5 (3.2\%) & [1.1, 7.4]\% & 2/9 [3, 60]\% \\
KL-UCB & 4 (2.6\%) & [0.7, 6.5]\% & 2/9 [3, 60]\% \\
ETC & 11 (7.1\%) & [3.6, 12.4]\% & 2/9 [3, 60]\% \\
\bottomrule
\end{tabular}
}
\end{table}

Table~\ref{tab:confirmatory} gives the confirmatory family.
The family is the four comparisons the claims rest on; it was chosen after the individual tests had been reported, so
Holm's step-down correction, applied within each scope, guards against multiplicity inside the family but not against
choosing it.
With nine scenarios the \AUTO{}$-$UCB1 and warm-versus-cold claims survive it; the six 16/32-node scenarios alone (the
4-node scenarios hold only the eight trace contexts) give each the smallest $p$ six clusters allow, $0.031$, which no
correction over four tests can keep below $0.05$.
UCB1's count of contexts above $10\%$ stays between one and three across three sampler seeds and the realized-objective
re-tune.
The split-pool rows choose each context's oracle arm on the odd steady iterations of every pool and price every policy
(selection-priced, from its pull counts) on the even ones, and the reverse; the ranking and the exceedance counts are
unchanged, and the \AUTO{}$-$UCB1 sign-flip stays at $p{=}0.016$ for mean and exceedance.

\begin{table}[t]
  \centering
  \scriptsize
  \setlength{\tabcolsep}{3pt}
  \caption{Confirmatory family, tuning-seed variation and split-pool check (held-out test). Top: exact two-sided
    sign-flip $p$ over scenario means (``mean'') or per-scenario exceedance-count differences (``contexts $>$10\%''),
    realized \TRLT{}, unadjusted and Holm-adjusted over the four rows. Middle: the UCB1 champion re-tuned with other TPE
    sampler seeds or on the realized objective, realized \TRLT{}. Bottom: selection-priced \TRLT{} with the oracle arm
    chosen on the odd or even steady iterations and every policy priced on the other half. \capkey{}}
  \label{tab:confirmatory}
  \resizebox{\columnwidth}{!}{\begin{tabular}{lrrrr}
\toprule
Comparison & \multicolumn{2}{c}{nine scenarios} & \multicolumn{2}{c}{six 16/32-node} \\
 & $p$ & Holm $p$ & $p$ & Holm $p$ \\
\midrule
\AUTO{}$-$UCB1, mean & 0.016 & 0.047 & 0.031 & 0.125 \\
\AUTO{}$-$UCB1, contexts $>$10\% & 0.016 & 0.047 & 0.031 & 0.125 \\
cold-tuned$-$warm UCB1, mean & 0.004 & 0.016 & 0.031 & 0.125 \\
tuned GBM+cost$-$UCB1, contexts $>$10\% & 0.062 & 0.062 & 0.062 & 0.125 \\
\midrule
UCB1 tuning & mean (\%) & p99 (\%) & $>$10\% & \\
\midrule
TPE seed 42 (headline) & 1.61 & 9.9 & 1 & \\
TPE seed 43 & 2.05 & 11.3 & 2 & \\
TPE seed 44 & 1.53 & 10.7 & 2 & \\
realized objective & 1.62 & 11.4 & 3 & \\
\midrule
Split pool (oracle on / policy) & mean (\%) & p99 (\%) & $>$10\% & \\
\midrule
odd / \AUTO{} & 5.95 & 103.4 & 18 & \\
odd / UCB1 & 0.99 & 8.0 & 0 & \\
odd / GBM+cost (fixed) & 1.74 & 35.9 & 10 & \\
even / \AUTO{} & 6.02 & 103.1 & 18 & \\
even / UCB1 & 0.99 & 8.0 & 0 & \\
even / GBM+cost (fixed) & 1.77 & 37.9 & 9 & \\
\bottomrule
\end{tabular}
}
\end{table}

\subsection{Tail statistics and scenario-clustered intervals}
\label{app:tail-stats}
Table~\ref{tab:tail-stats} gives the quantile profile, CVaR$_{5\%}$ (mean of the worst
$8$ contexts), and exceedance counts.
The paired exceedance comparison is exact: \AUTO{} versus UCB1 above $10\%$ (18 contexts only
\AUTO{}, none only UCB1, $p{<}10^{-4}$), $25\%$ (12 versus 0, $p{=}5\times10^{-4}$), and
$50\%$ (6 versus 0, $p{=}0.03$); GBM+cost versus UCB1 above $10\%$ (10 versus 1, $p{=}0.012$);
and ETC versus UCB1 above $10\%$ (11 versus 1, $p{=}0.006$).
The signed-rank test on means ($p{=}0.34$) and the sign test ($24$ versus $62$ contexts,
$p{=}10^{-4}$, in favor of \AUTO{} on the typical context) answer a different question from
the tail.
Table~\ref{tab:scenario-ci} resamples whole scenarios (nine clusters) and blocks of four
adjacent sizes (41 blocks) instead of contexts: the \AUTO{}$-$UCB1 mean, p99, and
CVaR$_{5\%}$ differences keep excluding zero, and the GBM+cost p99 and CVaR differences do too while its mean
difference includes zero.
Nine clusters is coarse, so these intervals are indicative only.

\begin{table}[t]
  \centering
  \scriptsize
  \setlength{\tabcolsep}{2.5pt}
  \caption{Realized \TRLT{} distribution (\%) over the 154 held-out H100 contexts, \capsplit{} \capTRLT{} p$q$: $q$-th
    percentile over contexts; CVaR$_{5\%}$: mean of the worst 5\% (8) contexts; $n_{>x}$: contexts with \TRLT{} above
    $x\%$. ETC commits after $m{=}1$ pull per arm. \capbold{}}
  \label{tab:tail-stats}
  \resizebox{\columnwidth}{!}{\begin{tabular}{@{}lrrrrrrrrr@{}}
\toprule
Policy & mean & p50 & p90 & p95 & p99 & CVaR$_{5\%}$ & $n_{>10}$ & $n_{>25}$ & $n_{>50}$ \\
\midrule
NCCL \AUTO{} & 6.08 & \textbf{0.00} & 17.7 & 39.6 & 103.3 & 77.2 & 19 & 12 & 6 \\
Supervised GBM+cost & 1.78 & \textbf{0.00} & \textbf{3.8} & 14.6 & 32.2 & 25.7 & 10 & 4 & \textbf{0} \\
ETC ($m{=}1$) & 4.74 & 2.53 & 8.3 & 12.1 & 63.0 & 35.1 & 11 & 3 & 3 \\
KL-UCB & 2.11 & 1.09 & 4.2 & \textbf{6.2} & 22.7 & 16.3 & 4 & 2 & \textbf{0} \\
Softmax & 1.75 & 0.87 & 4.0 & 6.4 & 11.9 & 11.1 & 5 & \textbf{0} & \textbf{0} \\
UCB1 & \textbf{1.61} & 0.98 & 4.1 & \textbf{6.2} & \textbf{9.9} & \textbf{9.1} & \textbf{1} & \textbf{0} & \textbf{0} \\
\bottomrule
\end{tabular}
}
\end{table}

\begin{table}[t]
  \centering
  \scriptsize
  \setlength{\tabcolsep}{2.5pt}
  \caption{Paired difference (percentage points, point estimate and 95\% bootstrap interval,
    10{,}000 draws) in realized \TRLT{} statistics under three resampling units. The first
    column resamples contexts as in the main text; the others resample whole (collective,
    nodes) scenarios or blocks of four adjacent message sizes. $^{\circ}$: the interval
    includes zero. \capkey{}}
  \label{tab:scenario-ci}
  \resizebox{\columnwidth}{!}{\begin{tabular}{@{}llrrr@{}}
\toprule
Contrast & statistic & iid contexts & scenarios (9) & size blocks (41) \\
\midrule
AUTO$-$UCB1 & mean & 4.5 [1.8, 7.8] & 4.5 [2.8, 6.3] & 4.5 [1.3, 8.3] \\
AUTO$-$UCB1 & p99 & 93.4 [36.8, 124.9] & 93.4 [53.4, 128.1] & 93.4 [36.9, 125.0] \\
AUTO$-$UCB1 & CVaR5\% & 68.1 [32.8, 102.0] & 68.1 [45.4, 92.2] & 68.1 [32.0, 102.3] \\
GBM+cost$-$UCB1 & mean & 0.2 [-0.8, 1.2]$^{\circ}$ & 0.2 [-0.8, 1.2]$^{\circ}$ & 0.2 [-0.7, 1.2]$^{\circ}$ \\
GBM+cost$-$UCB1 & p99 & 22.3 [7.3, 31.3] & 22.3 [10.9, 35.9] & 22.3 [8.5, 31.3] \\
GBM+cost$-$UCB1 & CVaR5\% & 16.6 [5.2, 25.6] & 16.6 [7.7, 27.3] & 16.6 [5.8, 25.7] \\
ETC$-$UCB1 & mean & 3.1 [1.9, 4.7] & 3.1 [1.4, 4.9] & 3.1 [1.5, 5.2] \\
ETC$-$UCB1 & p99 & 53.1 [4.8, 61.7] & 53.1 [0.6, 61.4] & 53.1 [0.7, 61.8] \\
ETC$-$UCB1 & CVaR5\% & 26.0 [5.6, 50.5] & 26.0 [2.3, 48.6] & 26.0 [2.2, 49.6] \\
\bottomrule
\end{tabular}
}
\end{table}

\subsection{Split and tuning robustness}
\label{app:split-robustness}
The main result uses one split and one sampler seed.
Table~\ref{tab:split-robustness} repeats the protocol for the reverse scale split (tune on
16/32 nodes, test on 4/8), three random 50/50 context splits (in-distribution), three
leave-one-collective-out splits, and two further Optuna sampler seeds on the main split.
Every row re-tunes UCB1, Softmax, KL-UCB, and ETC on that split's train contexts (151 trials
each, as in the main study), refits the supervised baselines on the same
train contexts, and replays each frozen champion once on test; AUTO needs no fitting.
Cells are realized test mean / p99 (\%).
The \AUTO{} and GBM+cost cells of the two Optuna-seed rows repeat the main
split, since those policies are deterministic, and ETC is not re-tuned in them.
The tuned UCB1 champions are warm and stable across all splits ($\alpha{=}0.068$--$0.107$,
pseudo-count $2.20$--$4.45$ against $0.093$ and $2.89$ on the main split), while Softmax
varies more (its tuned temperature ranges from $0.005$ to $0.149$) and moves its mean by up to
$0.8$\,pp across Optuna seeds.
With ten settings and few contexts per split, differences below about $1$\,pp in the mean
or $10$\,pp in the p99 between the leading policies should not be read as rankings.

\begin{table*}[t]
  \centering
  \scriptsize
  \caption{Realized test \TRLT{} (mean / p99, \%) on the H100 sweep per split and policy, with every learning family re-tuned
    and every supervised selector refit on that split's train contexts. \capTRLT{} Realized \TRLT{} charges the latencies
    actually served; p99 is the 99th percentile over test contexts; $n_{\mathrm{test}}$: number of test contexts. ETC is
    re-tuned only where listed. Bold: best mean and best p99 per row.}
  \label{tab:split-robustness}
  \begin{tabular}{@{}lrrrrrrr@{}}
\toprule
Split & $n_{\mathrm{test}}$ & \AUTO{} & GBM+cost & ETC & Softmax & KL-UCB & UCB1 \\
\midrule
scale-OOD (main) & 154 & 6.08 / 103.3 & 1.78 / 32.2 & 4.74 / 63.0 & 1.75 / 11.9 & 2.11 / 22.7 & \textbf{1.61} / \textbf{9.9} \\
reverse scale (train 16/32, test 4/8) & 154 & 3.96 / 60.7 & 1.99 / 23.8 & 4.04 / 16.9 & 1.51 / 16.5 & 1.38 / \textbf{11.7} & \textbf{1.20} / \textbf{11.7} \\
random 50/50, seed 42 & 150 & 4.43 / 74.2 & \textbf{0.75} / 13.3 & 5.07 / 51.3 & 2.46 / 27.2 & 1.56 / \textbf{9.5} & 1.29 / 11.8 \\
random 50/50, seed 43 & 150 & 5.80 / 104.0 & \textbf{0.93} / 13.0 & 4.66 / 63.2 & 1.80 / 16.0 & 2.14 / \textbf{12.7} & 1.69 / 15.2 \\
random 50/50, seed 44 & 150 & 4.62 / 74.2 & \textbf{1.10} / 10.0 & 3.73 / 18.2 & 1.32 / 15.6 & 1.37 / \textbf{9.0} & 1.24 / 9.5 \\
leave out all-gather & 98 & 4.67 / 79.9 & \textbf{0.45} / 8.2 & 2.73 / 9.0 & 0.88 / 14.6 & 1.26 / 5.5 & 0.89 / \textbf{4.5} \\
leave out all-reduce & 104 & 5.66 / 50.3 & \textbf{2.34} / 43.5 & 7.61 / 71.0 & 3.17 / 17.6 & 3.23 / 32.7 & 2.68 / \textbf{13.6} \\
leave out reduce-scatter & 98 & 4.81 / 77.1 & \textbf{0.37} / 8.4 & 2.85 / 8.0 & 0.87 / 16.0 & 1.33 / 5.6 & 0.76 / \textbf{4.6} \\
scale-OOD, Optuna seed 43 & 154 & 6.08 / 103.3 & 1.78 / 32.2 & -- & \textbf{1.60} / 19.6 & 2.02 / 16.3 & 2.05 / \textbf{11.3} \\
scale-OOD, Optuna seed 44 & 154 & 6.08 / 103.3 & 1.78 / 32.2 & -- & 2.38 / 11.9 & 2.10 / 14.5 & \textbf{1.53} / \textbf{10.7} \\
\bottomrule
\end{tabular}

\end{table*}

\subsection{A cost-model probe without a bandit, and other exceedance thresholds}
\label{app:probe}
UCB1 mostly plays \AUTO{}'s arm, so Table~\ref{tab:probe} asks whether the bandit is needed at all.
The probe plays \AUTO{}'s arm, then every other offered arm in cost-model order; each arm's first $k$ samples are
discarded and the next $m$ averaged, after which it commits to the fastest, or keeps \AUTO{}'s arm unless another beats
it by a margin.
A 24-point grid ($k\in\{0,1,2,5\}$, $m\in\{1,3,5\}$, margin $0$ or $5\%$) is scored on the train contexts and the
train-best point is reported on test.
Table~\ref{tab:exceedance-thresholds} repeats the primary scenario-level test at three exceedance thresholds, and
Table~\ref{tab:tuned-gbm-nvls} compares the tuned static fit with UCB1 when NVLS is excluded.
Under pull-count replay the train-best probe ($k{=}1$, $m{=}1$, no margin) serves exactly the samples of ETC with $m{=}1$
and one discarded sample, so its row equals ETC's.

\begin{table}[t]
  \centering
  \scriptsize
  \caption{Static fits against UCB1 with NVLS excluded, H100 test contexts. \capreal{} $\Delta_{\mathrm{scen}}$: the
    equal-weight scenario-mean difference to UCB1 that the scenario sign-flip tests; $p$ values are that test for the mean
    and for contexts above $10\%$.}
  \label{tab:tuned-gbm-nvls}
  \resizebox{\columnwidth}{!}{\begin{tabular}{lrrrrrr}
\toprule
Policy & mean (\%) & p99 (\%) & $n_{>10\%}$ & $\Delta_{\mathrm{scen}}$ (pp) & $p_{\mathrm{mean}}$ & $p_{>10\%}$ \\
\midrule
GBM+cost (tuned) & 1.46 & 38.6 & 9 & +0.38 & 0.422 & 0.031 \\
GBM+cost (fixed) & 1.79 & 32.2 & 12 & +0.31 & 0.406 & 0.031 \\
\AUTO{} & 6.27 & 111.4 & 20 & +4.36 & 0.016 & 0.016 \\
UCB1 (champion) & 1.23 & 11.2 & 2 & -- & -- & -- \\
\bottomrule
\end{tabular}
}
\end{table}

\begin{table}[t]
  \centering
  \scriptsize
  \caption{Cost-model probe against UCB1 on the H100 test contexts, both arm sets. \capreal{} ``train sel.'' is the
    selection-priced train \TRLT{} the probe point was chosen on; $p$ values are the scenario sign-flip against UCB1 for
    the mean and for contexts above $10\%$.}
  \label{tab:probe}
  \resizebox{\columnwidth}{!}{\begin{tabular}{llrrrrrr}
\toprule
Arm set & Policy & train sel. & mean (\%) & p99 (\%) & $n_{>10\%}$ & $p_{\mathrm{mean}}$ & $p_{>10\%}$ \\
\midrule
NVLS offered & \AUTO{} & 4.04 & 6.08$\pm$1.58 & 103.3 & 19 & 0.016 & 0.016 \\
 & Probe ($k{=}1$, $m{=}1$) & 2.42 & 4.74$\pm$0.75 & 63.0 & 11 & 0.004 & 0.500 \\
 & UCB1 (champion) & 0.86 & 1.61$\pm$0.19 & 9.9 & 1 & -- & -- \\
\midrule
NVLS excluded & \AUTO{} & 2.17 & 6.27$\pm$1.66 & 111.4 & 20 & 0.016 & 0.016 \\
 & Probe ($k{=}1$, $m{=}1$) & 2.36 & 3.21$\pm$0.22 & 11.6 & 2 & 0.004 & 1.000 \\
 & UCB1 (champion) & 0.56 & 1.23$\pm$0.18 & 11.2 & 2 & -- & -- \\
\bottomrule
\end{tabular}
}
\end{table}

\begin{table}[t]
  \centering
  \scriptsize
  \caption{Scenario-level comparisons at several exceedance thresholds on the H100 test contexts (realized \TRLT{},
    NVLS offered). Mean: difference of the equal-weight scenario means in pp; thresholds: contexts above it for $a$ vs
    $b$; scenario sign-flip $p$ in parentheses.}
  \label{tab:exceedance-thresholds}
  \resizebox{\columnwidth}{!}{\begin{tabular}{lrrrr}
\toprule
Comparison ($a-b$) & mean & $>$5\% & $>$10\% & $>$20\% \\
\midrule
\AUTO{}$-$UCB1 & +3.99 (0.016) & 25 vs 13 (0.062) & 19 vs 1 (0.016) & 15 vs 0 (0.016) \\
GBM+cost (tuned)$-$UCB1 & +0.07 (0.914) & 11 vs 13 (0.812) & 7 vs 1 (0.062) & 2 vs 0 (0.500) \\
GBM+cost (tuned)$-$Softmax & -0.03 (0.965) & 11 vs 13 (1.000) & 7 vs 5 (0.844) & 2 vs 1 (1.000) \\
GBM+cost (tuned)$-$KL-UCB & -0.25 (0.734) & 11 vs 12 (1.000) & 7 vs 4 (0.531) & 2 vs 2 (1.000) \\
\bottomrule
\end{tabular}
}
\end{table}

\begin{table}[t]
  \centering
  \scriptsize
  \caption{Paired tail differences of the tuned GBM+cost selector against the online champions on the H100 test
    contexts (realized \TRLT{}, NVLS offered; positive: $a$ loses more). 95\% percentile intervals from 10{,}000
    resamples of the nine (collective, nodes) scenarios, and of the 154 contexts. With nine clusters the scenario
    intervals are approximate. Fixed-setting GBM+cost and \AUTO{} against UCB1 are shown for reference; their intervals
    come from a separate Monte Carlo run from Table~\ref{tab:scenario-ci}, hence small differences.}
  \label{tab:tuned-gbm-tail}
  \resizebox{\columnwidth}{!}{\begin{tabular}{llrrr}
\toprule
Comparison ($a-b$) & Statistic & $\Delta$ (pp) & Scenario 95\% CI & Context 95\% CI \\
\midrule
GBM+cost (tuned)$-$UCB1 & mean & -0.24 & [-0.96, +0.65] & [-1.03, +0.73] \\
 & p99 & +17.46 & [+2.62, +35.88] & [+1.75, +31.12] \\
 & CVaR5\% & +11.31 & [+3.28, +21.58] & [+0.68, +22.57] \\
\midrule
GBM+cost (tuned)$-$Softmax & mean & -0.38 & [-1.65, +0.76] & [-1.24, +0.61] \\
 & p99 & +15.41 & [-6.12, +36.35] & [-5.79, +30.44] \\
 & CVaR5\% & +9.39 & [-0.51, +22.45] & [-1.94, +21.18] \\
\midrule
GBM+cost (tuned)$-$KL-UCB & mean & -0.74 & [-2.40, +0.56] & [-1.83, +0.37] \\
 & p99 & +4.64 & [-22.33, +35.85] & [-22.05, +29.60] \\
 & CVaR5\% & +4.11 & [-14.51, +22.13] & [-11.54, +18.15] \\
\midrule
GBM+cost (fixed)$-$UCB1 & mean & +0.17 & [-0.84, +1.22] & [-0.73, +1.20] \\
 & p99 & +22.32 & [+10.84, +35.86] & [+7.54, +31.22] \\
 & CVaR5\% & +16.56 & [+7.65, +27.14] & [+5.41, +25.41] \\
\midrule
\AUTO{}$-$UCB1 & mean & +4.46 & [+2.81, +6.30] & [+1.71, +7.79] \\
 & p99 & +93.39 & [+53.71, +128.48] & [+36.85, +125.09] \\
 & CVaR5\% & +68.09 & [+45.51, +91.13] & [+33.60, +101.94] \\
\bottomrule
\end{tabular}
}
\end{table}

\subsection{Supervised selectors tuned like the learners}
\label{app:supervised-tuning}
The online families get 151 TPE trials each, so Table~\ref{tab:supervised-tuning} gives the tree selectors the same
budget on every H100 split: 151 TPE trials (50 random start-up trials, sampler seed 42, trial 0 the fixed settings of
Section~\ref{sec:protocol}) over the sklearn hyperparameters (gradient boosting: trees, learning rate, depth, leaf size,
row and feature subsampling; random forest: trees, depth, leaf size, feature subsampling) and whether to use the
node-count feature.
The objective is the mean selection-priced \TRLT{} of the argmin choice on validation contexts inside the train
partition, with folds that copy the shift each split tests: leave one train node count out for the two scale splits,
leave one train collective out for the leave-one-collective-out splits, and 5-fold over contexts for the random splits.
The winner is refit on the whole train partition and replayed once on test; the online cells repeat
Table~\ref{tab:split-robustness}.
On the main split tuning cuts GBM+cost's validation loss from $2.05\%$ to $0.30\%$, mostly by dropping the node-count
feature.
Its test mean drops below UCB1's, but its tail stays heavier (7 against 1 contexts above $10\%$; $p{=}0.06$).
Tuning is not uniformly helpful: on one random split it raises GBM+cost's mean from $0.93\%$ to $1.62\%$, and
tuned gradient boosting without the cost feature reaches $6.79\%$ on the main split against $4.76\%$ fixed, because
validation one node doubling away does not predict a test two doublings away.
At seen scales, the best tuned fit has the lower mean on every random and leave-one-collective-out split, with a tail
comparable to the online policies' except when all-reduce is held out.

\begin{table*}[t]
  \centering
  \scriptsize
  \caption{Supervised selectors with fixed settings and tuned like the learners (151 TPE trials, shift-matched
    validation folds inside the train partition), against the online champions of Table~\ref{tab:split-robustness}, on
    every H100 split. \capTRLT{} Cells: realized test \TRLT{} mean (\%) / p99 (\%) / $n_{>10}$, the number of test
    contexts above $10\%$; $n_{\mathrm{test}}$: number of test contexts. Last column: exact sign-flip $p$ over the
    (collective, nodes) scenarios of tuned GBM+cost against UCB1, for the mean and for $n_{>10}$. Bold: best of each of
    the three numbers per row.}
  \label{tab:supervised-tuning}
  \resizebox{\textwidth}{!}{\begin{tabular}{@{}lrrrrrrrr@{}}
\toprule
& & \multicolumn{3}{c}{Supervised} & \multicolumn{3}{c}{Online} & \\
\cmidrule(lr){3-5}\cmidrule(lr){6-8}
Split & $n_{\mathrm{test}}$ & GBM$+$cost fixed & GBM$+$cost tuned & RF$+$cost tuned & Softmax & KL-UCB & UCB1 & $p$ mean / $n_{>10}$ \\
\midrule
scale-OOD (main) & 154 & 1.78 / 32.2 / 10 & \textbf{1.37} / 27.3 / 7 & 3.23 / 76.7 / 9 & 1.75 / 11.9 / 5 & 2.11 / 22.7 / 4 & 1.61 / \textbf{9.9} / \textbf{1} & 0.91 / 0.06 \\
reverse scale (train 16/32, test 4/8) & 154 & 1.99 / 23.8 / 12 & 1.97 / 30.7 / 7 & 1.30 / 16.0 / 6 & 1.51 / 16.5 / 10 & 1.38 / \textbf{11.7} / \textbf{3} & \textbf{1.20} / \textbf{11.7} / \textbf{3} & 0.38 / 0.12 \\
random 50/50, seed 42 & 150 & 0.75 / 13.3 / 3 & \textbf{0.64} / 13.3 / 3 & 0.69 / 16.7 / 3 & 2.46 / 27.2 / 8 & 1.56 / \textbf{9.5} / \textbf{1} & 1.29 / 11.8 / 2 & 0.01 / 1.00 \\
random 50/50, seed 43 & 150 & 0.93 / 13.0 / 5 & 1.62 / 17.7 / 11 & \textbf{0.76} / \textbf{12.7} / \textbf{4} & 1.80 / 16.0 / 10 & 2.14 / \textbf{12.7} / 5 & 1.69 / 15.2 / 6 & 0.80 / 0.25 \\
random 50/50, seed 44 & 150 & 1.10 / 10.0 / 2 & 0.77 / 9.6 / 1 & \textbf{0.75} / 9.6 / 1 & 1.32 / 15.6 / 5 & 1.37 / \textbf{9.0} / \textbf{0} & 1.24 / 9.5 / 2 & 0.10 / 1.00 \\
leave out all-gather & 98 & 0.45 / 8.2 / \textbf{0} & \textbf{0.06} / \textbf{0.4} / \textbf{0} & 0.49 / 9.1 / 1 & 0.88 / 14.6 / 3 & 1.26 / 5.5 / 1 & 0.89 / 4.5 / \textbf{0} & 0.12 / 1.00 \\
leave out all-reduce & 104 & 2.34 / 43.5 / 6 & \textbf{1.64} / 24.3 / 5 & 1.72 / 24.3 / \textbf{4} & 3.17 / 17.6 / 6 & 3.23 / 32.7 / 5 & 2.68 / \textbf{13.6} / 5 & 0.38 / 1.00 \\
leave out reduce-scatter & 98 & 0.37 / 8.4 / \textbf{0} & \textbf{0.21} / 5.9 / \textbf{0} & 0.66 / 15.2 / 2 & 0.87 / 16.0 / 3 & 1.33 / 5.6 / \textbf{0} & 0.76 / \textbf{4.6} / \textbf{0} & 0.12 / 1.00 \\
\bottomrule
\end{tabular}
}
\end{table*}

\subsection{Independently tuned cold champions}
\label{app:cold-tuned}
Table~\ref{tab:cold-tuned} compares, for each warm-capable family, the cold policy at the warm
champion's hyperparameters (the matched ablation of Table~\ref{tab:warm-cold}), the cold
policy with its own 151-trial search (warm start pinned off, same split and sampler seed), and
the warm champion.
Cold-tuned train objectives are $2.3$--$2.5\%$ for the five leading families against
$0.82$--$1.07\%$ warm (selection-priced).
Paired warm$-$cold-tuned differences are $-3.2$\,pp (UCB1), $-3.0$ (Softmax), $-2.8$
(Gaussian Thompson), $-2.9$ ($\varepsilon$-greedy), $-2.2$ (KL-UCB), and $-0.7$ (UCB-tuned),
all $p{<}0.001$; UCB-V is the exception ($+0.06$, $p{=}0.19$).

\begin{table}[t]
  \centering
  \scriptsize
  \setlength{\tabcolsep}{3pt}
  \caption{Realized test \TRLT{} (mean, p99; \%) of each warm-capable family cold at the warm
    champion's hyperparameters, cold with its own search, and the warm champion. \capkey{} \capbold{}}
  \label{tab:cold-tuned}
  \resizebox{\columnwidth}{!}{\begin{tabular}{lrrrrrr}
\toprule
& \multicolumn{2}{c}{Cold, warm knobs} & \multicolumn{2}{c}{Cold, own search} & \multicolumn{2}{c}{Warm (champion)} \\
\cmidrule(lr){2-3}\cmidrule(lr){4-5}\cmidrule(lr){6-7}
Family & mean & p99 & mean & p99 & mean & p99 \\
\midrule
KL-UCB & 4.94 & \textbf{40.8} & \textbf{4.27} & \textbf{33.9} & 2.11 & 22.7 \\
Softmax & 6.71 & 49.5 & 4.77 & 63.0 & 1.75 & 11.9 \\
UCB1 & 8.82 & 98.7 & 4.84 & 63.0 & \textbf{1.61} & \textbf{9.9} \\
Gaussian Thompson & \textbf{4.93} & 64.5 & 4.93 & 64.5 & 2.17 & 17.8 \\
$\varepsilon$-greedy & 6.63 & 75.0 & 4.97 & 63.5 & 2.12 & 23.3 \\
UCB-tuned & 7.07 & 45.6 & 7.07 & 45.6 & 6.35 & 30.1 \\
UCB-V & 8.65 & 51.7 & 8.80 & 52.2 & 8.86 & 49.8 \\
\bottomrule
\end{tabular}
}
\end{table}

\subsection{Is the prior effect a replay artifact? Commit-once and cold grids}
\label{app:transient}
Table~\ref{tab:transient} replays three grids on the main split with NVLS offered and with NVLS
excluded: ETC with $m\in\{1,2,3\}$ exploration rounds and $k\in\{0,1,2,3,5,10\}$ discarded first
pulls per arm (the tuned champion is $m{=}1$, $k{=}1$, so it already commits on each arm's
second sample); round-robin ETC of Madan et al.\ with $m\in\{1,3,5,10\}$ rounds and the 5\% (their
default) or 1\% (tuned) safety gate, no discard; and the five leading independently tuned cold
families with $k\in\{0,1,2,3,5,10\}$ and every other knob at its cold-tuned value ($k$ is selected
on train realized \TRLT{}).
Larger $k$ never helps in a 100-step horizon: a context has about $2.6$ arms, so $k{=}10$ spends roughly 30 of the 100 steps
exploring (ETC $m{=}1$: $14.3\%$ mean), and $k{=}2,3$ are worse than $k{=}1$ because the transient lasts
about ten pulls and cannot be skipped cheaply.
The only grid cell that looks past the tail is NVLS exclusion: with it, the tuned ETC
($3.21\%$, $11.6\%$ p99) and cold UCB1 and Softmax ($3.3\%$, $12.0\%$) have the tail of the warm
champions, and the warm advantage that remains is in the mean (Table~\ref{tab:transient-paired}).
The tuned-ETC row under NVLS exclusion ($3.21\%$, $11.6\%$ p99, 2 contexts above $10\%$) differs from
Table~\ref{tab:nvls-policy} ($3.26\%$, $13.8\%$, 3 contexts) only because the champion carries the reward-renormalization
wrapper (adaptive observed max) and this grid's entry does not; we confirmed this by replaying both: with NVLS offered
they agree ($4.74\%$, $63.0\%$, 11), and the difference under exclusion is one context above $10\%$, which is what a p99
over a handful of contexts looks like.
KL-UCB, whose cold search ended at $k{=}0$ and a large $c$, keeps a heavier tail ($27.0\%$) with NVLS excluded, as
does its warm champion ($27.5\%$).
We did not run a cold variant that visits arms in cost-model rank order: with $m$ full rounds the
order changes only when each arm is first seen, and a cost-model order would be a warm start by
another name.

\begin{table}[t]
  \centering
  \scriptsize
  \setlength{\tabcolsep}{2.5pt}
  \caption{Commit-once and cold policies under transient controls (test, 154 contexts, \%):
    realized mean and p99 and selection-priced mean with NVLS offered, realized mean and p99 with
    NVLS excluded (restricted oracle). ``Train-selected $k$'': the number of discarded first pulls
    chosen on train realized \TRLT{}, shown as (offered/excluded). Madan rows use $k{=}0$ as in the original method; the tuned Madan
    champion ($k{=}1$) gives the same test numbers as the tuned ETC champion. \capkey{} \capbold{}}
  \label{tab:transient}
  \resizebox{\columnwidth}{!}{\begin{tabular}{@{}lrrrrr@{}}
\toprule
& \multicolumn{3}{c}{NVLS offered} & \multicolumn{2}{c}{NVLS excluded} \\
\cmidrule(lr){2-4}\cmidrule(lr){5-6}
Policy & realized & p99 & selection & realized & p99 \\
\midrule
ETC, tuned champion ($m{=}1$, $k{=}1$) & 4.74 & 63.0 & \textbf{2.70} & \textbf{3.21} & \textbf{11.6} \\
ETC $m{=}1$, $k{=}0$ & 7.14 & 97.6 & 5.97 & 5.98 & 73.6 \\
ETC $m{=}1$, $k{=}10$ & 14.25 & 66.6 & 12.10 & 12.76 & 43.4 \\
ETC $m{=}3$, $k{=}10$ & 16.69 & 72.8 & 14.57 & 15.13 & 53.0 \\
ETC, $m{=}1$, train-selected $k$ (offered $k{=}1$; excluded $k{=}1$) & 4.74 & 63.0 & \textbf{2.70} & \textbf{3.21} & \textbf{11.6} \\
ETC, $m{=}3$, train-selected $k$ (offered $k{=}0$; excluded $k{=}0$) & 5.88 & 55.9 & 3.76 & 4.67 & 22.8 \\
Madan RR $m{=}1$, gate 1\% & 7.14 & 97.6 & 5.97 & 5.98 & 73.6 \\
Madan RR $m{=}5$, gate 5\% & 8.07 & 62.1 & 5.88 & 6.80 & 26.6 \\
Madan RR $m{=}10$, gate 5\% & 13.33 & 63.2 & 11.17 & 11.88 & 39.5 \\
\midrule
Cold KL-UCB, train-selected $k$ (0/0) & \textbf{4.27} & \textbf{33.9} & 2.78 & 3.53 & 27.0 \\
Cold Softmax, train-selected $k$ (1/1) & 4.77 & 63.0 & 2.71 & 3.25 & 12.0 \\
Cold UCB1, train-selected $k$ (1/1) & 4.84 & 63.0 & 2.78 & 3.31 & 12.0 \\
Cold Gauss.\ Thompson, train-selected $k$ (1/1) & 4.93 & 64.5 & 2.87 & 3.42 & 17.5 \\
Cold $\varepsilon$-greedy, train-selected $k$ (1/1) & 4.97 & 63.5 & 2.93 & 3.47 & 16.0 \\
\bottomrule
\end{tabular}
}
\end{table}

\begin{table}[t]
  \centering
  \scriptsize
  \setlength{\tabcolsep}{2.5pt}
  \caption{Warm champion versus the independently tuned cold champion (train-selected $k$), per
    family, with NVLS offered and excluded: selection-priced mean (\%), realized paired mean
    difference warm$-$cold (pp, 95\% bootstrap interval over contexts), realized p99 (\%), and
    contexts above $10\%$ (warm/cold). \capbold{}}
  \label{tab:transient-paired}
  \resizebox{\columnwidth}{!}{\begin{tabular}{@{}llrrrrrr@{}}
\toprule
& & \multicolumn{2}{c}{selection mean} & \multicolumn{3}{c}{realized} & \\
\cmidrule(lr){3-4}\cmidrule(lr){5-7}
NVLS & Family & warm & cold & warm$-$cold [95\% CI] & p99 w/c & $n_{>10}$ w/c \\
\midrule
offered & UCB1 & 1.00 & 2.78 & -3.22 [-4.78, -1.97] & 9.9/63.0 & 1/10 \\
offered & Softmax & 0.93 & 2.71 & -3.02 [-4.42, -1.93] & 11.9/63.0 & 5/10 \\
offered & KL-UCB & 1.18 & 2.78 & -2.16 [-2.94, -1.49] & 22.7/33.9 & 4/8 \\
\midrule
excluded & UCB1 & 0.77 & 2.24 & -2.08 [-2.58, -1.60] & 11.2/12.0 & 2/2 \\
excluded & Softmax & \textbf{0.70} & \textbf{2.19} & -2.04 [-2.44, -1.65] & 8.6/12.0 & 1/2 \\
excluded & KL-UCB & 1.37 & 2.63 & -1.62 [-2.41, -0.80] & 27.5/27.0 & 6/5 \\
\bottomrule
\end{tabular}
}
\end{table}

\subsection{Cold search range and the largest-scale holdout}
\label{app:largest-scale}
\paragraph{Cold search range.}
The main cold-tuned UCB1 sat at the lower edge of its $\alpha$ range.
Table~\ref{tab:cold-wide} repeats the cold searches for UCB1 and Softmax with lower bounds $50$--$100\times$ smaller and three
sampler seeds: five of six land on the same near-greedy champion (one discarded first sample, $\alpha$ or $\tau$ near
$0.0005$--$0.003$; $4.75$--$4.93\%$ mean and $63.0\%$ p99), and one Softmax search ends at $5.36\%$ / $33.1\%$ with 21
contexts above $10\%$.
The boundary was therefore not what limited the cold policies.

\paragraph{Largest-scale holdout.}
Fitting the static selectors on the 4/8/16-node contexts and testing on the 32-node contexts only
(Table~\ref{tab:largest-scale}; 72--78 contexts per setting; online policies keep their frozen hyperparameters) removes most of the
extrapolation: GBM+cost reaches $0.47\%$ / $7.1\%$ p99 on H100 against UCB1's $1.89\%$ / $11.0\%$, and
$1.40\%$ against $2.31\%$ on GB200 MNNVL-off.
With MNNVL on, \AUTO{} ($0.23\%$) is the best policy and GBM+cost ($6.94\%$, 13 contexts above $10\%$) the worst but for
linear+cost.
Our reading is that the online advantage over static selectors in the main protocol is largely the cost of extrapolating
two node doublings, though even with near-target data the static fit fails on one of three fabrics.
GBM+cost is the better of the two static fits in every setting, so it was not chosen after the fact; the online champions
were not re-tuned on 4/8/16 nodes.

\begin{table}[t]
  \centering
  \scriptsize
  \setlength{\tabcolsep}{3pt}
  \caption{Cold-tuned UCB1 and Softmax with an extended lower bound on $\alpha$ / $\tau$ and three sampler seeds
    (realized test \TRLT{}, \%). \capkey{} \capbold{}}
  \label{tab:cold-wide}
  \resizebox{\columnwidth}{!}{\begin{tabular}{@{}llrrrrr@{}}
\toprule
Family & sampler seed & $\alpha$ / $\tau$ & $k$ & train sel.\ obj.\ & test realized mean / p99 & $n_{>10}$ \\
\midrule
UCB1 & 42 & 0.0033 & 1 & 2.41 & \textbf{4.75} / 63.0 & \textbf{10} \\
Softmax & 42 & 0.0005 & 1 & 2.41 & 4.93 / 63.0 & 11 \\
UCB1 & 43 & 0.0030 & 1 & 2.41 & 4.76 / 63.0 & \textbf{10} \\
Softmax & 43 & 0.2072 & 0 & \textbf{2.26} & 5.36 / \textbf{33.1} & 21 \\
UCB1 & 44 & 0.0029 & 1 & 2.41 & 4.76 / 63.0 & \textbf{10} \\
Softmax & 44 & 0.0007 & 1 & 2.41 & 4.93 / 63.0 & 11 \\
\bottomrule
\end{tabular}
}
\end{table}

\begin{table*}[t]
  \centering
  \scriptsize
  \caption{Largest-scale holdout: supervised selectors fit on the 4/8/16-node contexts; every policy scored on the 32-node
    contexts only (72 H100, 78 GB200 MNNVL-off, 74 MNNVL-on); online policies keep their hyperparameters tuned on H100 4/8
    nodes. \capTRLT{} Realized \TRLT{} (\%, latencies actually served): mean, p99 (99th percentile over contexts), and
    $n_{>10}$, the number of contexts above 10\%. MNNVL: multi-node NVLink. \capbold{}}
  \label{tab:largest-scale}
  \begin{tabular}{@{}lrrrrrrrrr@{}}
\toprule
& \multicolumn{3}{c}{H100} & \multicolumn{3}{c}{GB200 MNNVL-off} & \multicolumn{3}{c}{GB200 MNNVL-on} \\
\cmidrule(lr){2-4}\cmidrule(lr){5-7}\cmidrule(lr){8-10}
Policy & mean & p99 & $n_{>10}$ & mean & p99 & $n_{>10}$ & mean & p99 & $n_{>10}$ \\
\midrule
AUTO & 4.80 & 76.2 & 9 & 3.38 & 51.4 & 8 & \textbf{0.23} & \textbf{5.4} & \textbf{0} \\
ETC & 4.73 & 33.7 & 7 & 4.63 & 28.1 & 5 & 2.72 & 12.4 & 1 \\
Softmax & 2.30 & 14.5 & 4 & 2.56 & \textbf{13.9} & \textbf{1} & 1.61 & 7.9 & \textbf{0} \\
UCB1 & 1.89 & 11.0 & \textbf{1} & 2.31 & 18.1 & 4 & 2.76 & 11.7 & 6 \\
KL-UCB & 3.06 & 34.7 & 3 & 2.19 & 16.6 & 2 & 2.47 & 10.7 & 2 \\
GBM$+$cost & \textbf{0.47} & \textbf{7.1} & \textbf{1} & \textbf{1.40} & 18.3 & 3 & 6.94 & 70.0 & 13 \\
linear$+$cost & 1.99 & 24.5 & 4 & 4.72 & 87.3 & 6 & 10.50 & 83.0 & 16 \\
\bottomrule
\end{tabular}

\end{table*}

\subsection{The recommended configuration, and the primary test on it}
\label{app:recommended}
Section~\ref{sec:lessons} recommends excluding the NVLS setup arm and, for a long-lived tuner, a bounded-memory mean.
The headline table uses neither, so we replayed the combination on the same split and harness.
Neither recommendation is pre-registered: we chose to exclude NVLS after seeing the live setup stall and the NVLS-offered
test results; only the gain below is selected, on train data alone.
\emph{Constant-gain mean:} $\hat r_a\leftarrow(1-\eta)\hat r_a+\eta r$, the constant step-size estimator for
non-stationary problems~\cite{sutton2018rl} that discounted and sliding-window UCB~\cite{garivier2011switching} build on (the first observation of an arm initializes
it; a warm-started prior counts as that observation), replacing the clipped-SGD mean of UCB1 and the sample means of
Softmax and KL-UCB; all other knobs stay at the champions' values, which were tuned with NVLS offered and the original
estimator.
$\eta\in\{0.05, 0.1, 0.2, 0.3, 0.5\}$ is chosen per family on the \emph{train} contexts (selection-priced train \TRLT{}, the
tuning objective of the main protocol), with and without the 1\,ms update flush of Appendix~\ref{sec:update-freq}.
Table~\ref{tab:recommended} gives the train-selected variants.
Train selects $\eta{=}0.1$ for UCB1 with and without the flush, so the recommended configuration is UCB1 with
$\eta{=}0.1$ and no flush: $1.21\%$ mean, $5.7\%$ p99 and no context above $10\%$, against $1.23\%$, $11.2\%$ and two
for the champion.
Train prefers the flush ($0.45\%$ against $0.58\%$ selection-priced \TRLT{}; $1.11\%$, $8.0\%$ and none on test), but that
objective charges no sync cost, and on H100-A the measured cost of a sync (Section~\ref{sec:lessons}) makes a 1\,ms flush,
one sync every 1--3 small collectives, far more expensive than its $0.13$\,pp train gain, so we report it without
recommending it.
The deployment default of our policy library ($\eta{=}0.2$) was not chosen this way: we picked it on the drift replay below,
which runs on the \emph{test} contexts, so it is not a held-out choice.
On this stationary test replay it is worse without the flush ($1.45\%$, $10.6\%$, two contexts) and better with it ($0.97\%$, $7.1\%$, one).
We therefore recommend the train-selected $\eta{=}0.1$, accepting a larger abrupt-drift loss than $\eta{=}0.2$ ($4.38\%$ against
$2.92\%$ below).
The ordering of the two gains also depends on the harness: in the drift replay's stationary column (bootstrap steady-pool
sampling, NVLS offered) $\eta{=}0.2$ beats $\eta{=}0.1$ ($0.96\%$ against $1.20\%$), and in this sequential replay with NVLS
excluded the reverse holds.
A plausible reason is that only sequential replay serves the lazy-initialization transient, so a gain chosen offline
depends on how that transient is modelled.
With the flush, train selects $\eta{=}0.3$ for Softmax ($0.92\%$, $5.9\%$, none); KL-UCB does not improve ($1.83$--$1.90\%$),
and the offline replay sees only how a batched update changes learning, not what its sync costs.
We did not isolate why the flush helps slightly at $T{=}100$; a batched update averages several samples, including
lazy-initialization samples, into one.

Table~\ref{tab:scenario-primary} applies the primary test of Section~\ref{sec:h100} to the comparisons the paper had tested
only per context.
The sign-flip statistic is the absolute unweighted mean of the nine per-scenario mean differences, compared with all $2^9$
sign patterns (two-sided); the wild-cluster bootstrap-$t$ instead tests the context-pooled mean, which weights scenarios by
their 2--26 contexts, with a cluster-robust standard error.
The two reach the same conclusion in every row.
GBM+cost against UCB1 is not separated on the mean in any variant ($p{=}0.28$--$0.88$; Table~\ref{tab:scenario-per-scenario}
gives the per-scenario differences), and its excess of contexts above $10\%$ is ($p{=}0.031$ in every variant, the smallest
value its five or six differing scenarios allow).
With NVLS excluded, \AUTO{} versus UCB1 still rejects ($p{=}0.016$ for the mean and for the exceedance count, with the
champion or the constant-gain UCB1).
ETC and the cold-tuned UCB1 are worse than the warm champion in all nine scenarios on the mean ($p{=}0.004$, the smallest
value nine clusters allow) but not on the exceedance count ($p{=}1$).
With NVLS offered the warm champions beat their independently cold-tuned counterparts in all nine scenarios for UCB1,
Softmax and KL-UCB ($p{=}0.004$), so the prior claim of Section~\ref{sec:prior} holds at the scenario level for the mean;
its effect on the exceedance count is not significant at that level ($p{=}0.5$).

Under drift (Table~\ref{tab:ewma-drift}; the best$\leftrightarrow$worst swap of Appendix~\ref{sec:drift}, NVLS offered,
bootstrap steady-pool replay, so the stationary column is not the sequential one of Table~\ref{tab:drift-ablation}), the
constant gain is what the drift appendix's horizon caveat asks for: UCB1 with $\eta{=}0.2$ pays $2.92\%$ dynamic \TRLT{}
under the abrupt swap against $5.88\%$ for the discounted champion and $6.60\%$ for the champion, with p99 $9.6\%$ against
$30.6\%$ and $34.4\%$.
With $\eta{=}0.1$ it pays $4.38\%$.
Under the gradual crossfade the UCB1 variants are within 2\,pp of each other ($9.1$--$10.9\%$).
This is the same single-swap, $T{=}100$ test as the drift appendix, with gains read off on test contexts.
The recommended configuration (NVLS excluded, constant gain) has been evaluated only in this offline replay, which
updates after every collective and charges no sync cost, never live.

\begin{table}[t]
  \centering
  \scriptsize
  \setlength{\tabcolsep}{3pt}
  \caption{The recommended configuration on the H100 sweep, \capsplit{} NVLS is excluded from every policy and from the
    oracle. \capTRLT{} Test columns are realized \TRLT{} (\%, latencies actually served): mean, p99 (99th percentile over
    contexts) and $n_{>10}$, the number of contexts above 10\%. ``gain'' is the constant-gain $\eta$ selected on train;
    ``train sel.'' is the selection-priced train \TRLT{} (each step charged the chosen arm's steady mean) used to select it,
    which for GBM+cost is its fitting error. Bold: best per test column (lower is better).}
  \label{tab:recommended}
  \resizebox{\columnwidth}{!}{\begin{tabular}{@{}lrrrrr@{}}
\toprule
Policy (NVLS excluded) & gain & train sel. & test mean & p99 & $n_{>10}$ \\
\midrule
\AUTO{} & -- & 2.17 & 6.27 & 111.4 & 20 \\
Supervised GBM+cost & -- & 0.04 & 1.79 & 32.2 & 12 \\
UCB1 champion & -- & 0.56 & 1.23 & 11.2 & 2 \\
Softmax champion & -- & 0.68 & 1.22 & 8.6 & 1 \\
KL-UCB champion & -- & 0.60 & 1.90 & 27.5 & 6 \\
\midrule
KL-UCB constant gain & 0.1 & 0.73 & 1.90 & 19.1 & 3 \\
Softmax constant gain & 0.1 & 0.52 & 1.16 & 11.0 & 3 \\
UCB1 constant gain & 0.1 & 0.58 & 1.21 & \textbf{5.7} & \textbf{0} \\
KL-UCB constant gain, 1\,ms flush & 0.1 & 0.54 & 1.83 & 19.1 & 3 \\
Softmax constant gain, 1\,ms flush & 0.3 & 0.46 & \textbf{0.92} & 5.9 & \textbf{0} \\
UCB1 constant gain, 1\,ms flush & 0.1 & 0.45 & 1.11 & 8.0 & \textbf{0} \\
\bottomrule
\end{tabular}
}
\end{table}

\begin{table*}[t]
  \centering
  \scriptsize
  \caption{Scenario-level tests with the primary procedure (realized \TRLT{}, held-out test): exact sign-flip over the nine (collective,
    nodes) scenario means of $a{-}b$ and over the per-scenario differences in contexts above $10\%$, and the wild-cluster
    bootstrap-$t$ on the context-pooled mean, all two-sided; ``pooled'' is the context-weighted mean difference (pp),
    ``scen.-mean'' the unweighted mean of the nine scenario means that the sign-flip tests; ``scen.\ $a$/$b$ better'' counts
    scenarios by the sign of their mean difference (ties omitted). ``Recommended'' is UCB1 with the train-selected constant
    gain $\eta{=}0.1$ (Table~\ref{tab:recommended}). \capkey{}}
  \label{tab:scenario-primary}
  \begin{tabular}{@{}llrrrrrrr@{}}
\toprule
Arm set & $a$ vs.\ $b$ & pooled $a{-}b$ & scen.-mean $a{-}b$ & scen.\ $a$/$b$ better & $p$ (sign-flip) & $p$ (wild-$t$) & $n_{>10}$ $a$/$b$ & $p$ ($n_{>10}$) \\
\midrule
NVLS excluded & \AUTO{} vs.\ UCB1 (champion) & +5.04 & +4.36 & 1/7 & 0.016 & 0.016 & 20/2 & 0.016 \\
NVLS excluded & GBM+cost vs.\ UCB1 (champion) & +0.56 & +0.31 & 3/5 & 0.406 & 0.141 & 12/2 & 0.031 \\
NVLS excluded & \AUTO{} vs.\ UCB1 (recommended) & +5.06 & +4.34 & 1/7 & 0.016 & 0.016 & 20/0 & 0.016 \\
NVLS excluded & GBM+cost vs.\ UCB1 (recommended) & +0.58 & +0.28 & 4/4 & 0.461 & 0.195 & 12/0 & 0.031 \\
NVLS excluded & UCB1 (champion) vs.\ UCB1 (recommended) & +0.02 & -0.03 & 5/2 & 0.891 & 0.922 & 2/0 & 1.000 \\
NVLS excluded & GBM+cost vs.\ UCB1 (recommended, 1\,ms flush) & +0.68 & +0.38 & 3/5 & 0.281 & 0.195 & 12/0 & 0.031 \\
NVLS excluded & UCB1 (recommended) vs.\ UCB1 (recommended, 1\,ms flush) & +0.10 & +0.10 & 3/5 & 0.203 & 0.312 & 0/0 & 1.000 \\
NVLS excluded & ETC ($m{=}1$) vs.\ UCB1 (champion) & +2.03 & +1.92 & 0/9 & 0.004 & 0.004 & 3/2 & 1.000 \\
NVLS excluded & UCB1 cold-tuned ($k{=}1$) vs.\ UCB1 (champion) & +2.08 & +1.95 & 0/9 & 0.004 & 0.004 & 2/2 & 1.000 \\
NVLS offered & \AUTO{} vs.\ UCB1 (champion) & +4.46 & +3.99 & 1/7 & 0.016 & 0.016 & 19/1 & 0.016 \\
NVLS offered & GBM+cost vs.\ UCB1 (champion) & +0.17 & +0.05 & 4/4 & 0.883 & 0.719 & 10/1 & 0.031 \\
NVLS offered & UCB1 cold-tuned vs.\ UCB1 (champion) & +3.22 & +2.71 & 0/9 & 0.004 & 0.004 & 10/1 & 0.500 \\
NVLS offered & Softmax cold-tuned vs.\ Softmax (champion) & +3.02 & +2.57 & 0/9 & 0.004 & 0.004 & 10/5 & 0.500 \\
NVLS offered & KL-UCB cold-tuned vs.\ KL-UCB (champion) & +2.16 & +1.92 & 0/9 & 0.004 & 0.004 & 8/4 & 0.500 \\
NVLS offered & UCB1 cold (warm knobs) vs.\ UCB1 warm (warm knobs) & +7.20 & +5.64 & 1/8 & 0.008 & 0.008 & 24/1 & 0.031 \\
\bottomrule
\end{tabular}

\end{table*}

\begin{table*}[t]
  \centering
  \scriptsize
  \caption{Per-scenario mean differences $a{-}b$ (percentage points of realized \TRLT{}, H100 held-out test) behind the
    primary test, for \AUTO{} and the supervised GBM+cost selector against UCB1 on both arm sets (``offered'' is the
    headline arm set, ``excl.''\ excludes NVLS); ``rec.'' is the recommended constant-gain UCB1, ``+flush'' with the 1\,ms
    update flush. Negative favours $a$. AG/AR/RS: all-gather/all-reduce/reduce-scatter.
    $n$ is the number of contexts; the 4-node rows are the Llama3 trace contexts. The last row is the sign-flip statistic. \capkey{}}
  \label{tab:scenario-per-scenario}
  \begin{tabular}{@{}lrrrrrrr@{}}
\toprule
Scenario & $n$ & \AUTO{}$-$UCB1, offered & GBM+cost$-$UCB1, offered & \AUTO{}$-$UCB1, excl. & GBM+cost$-$UCB1, excl. & GBM+cost$-$rec., excl. & GBM+cost$-$rec.+flush, excl. \\
\midrule
AG, 16 nodes & 24 & +6.98 & +0.75 & +6.98 & +0.75 & +0.27 & +0.14 \\
AG, 32 nodes & 23 & +3.84 & +0.43 & +3.84 & +0.43 & -0.19 & -0.11 \\
AG, 4 nodes & 3 & +8.86 & -0.60 & +8.86 & -0.60 & -0.83 & -0.88 \\
AR, 16 nodes & 26 & +3.50 & -1.87 & +4.78 & +0.10 & +0.62 & +0.92 \\
AR, 32 nodes & 26 & +1.69 & -0.49 & +3.80 & -0.14 & +0.98 & +1.29 \\
AR, 4 nodes & 2 & +0.00 & +0.00 & +0.00 & +0.00 & +0.00 & +0.00 \\
RS, 16 nodes & 24 & +8.07 & +2.48 & +8.07 & +2.48 & +2.35 & +2.21 \\
RS, 32 nodes & 23 & +3.38 & +0.10 & +3.38 & +0.10 & -0.31 & -0.21 \\
RS, 4 nodes & 3 & -0.43 & -0.38 & -0.43 & -0.38 & -0.38 & +0.05 \\
\midrule
unweighted mean & & +3.99 & +0.05 & +4.36 & +0.31 & +0.28 & +0.38 \\
\bottomrule
\end{tabular}

\end{table*}

\subsection{A size gate: leaving small messages to the cost model}
\label{app:size-gate}
Below 1\,MiB the live learners play \AUTO{}'s arm on $93$--$100\%$ of timed collectives and still pay the plugin's
profiling and sync cost (Section~\ref{sec:live}).
A size gate removes that cost where learning has nothing to find: contexts below a message-size threshold keep \AUTO{}'s
choice with the plugin passive, and the bandit runs only at or above it.
Offline, a gated context takes \AUTO{}'s realized and selection-priced \TRLT{} and an ungated one the UCB1 champion's
(headline arm set, NVLS offered).
The threshold is chosen from $\{$none, 64\,KiB, 256\,KiB, 1, 4, 16, 32\,MiB$\}$ by mean selection-priced \TRLT{} on the
train contexts, the tuning objective of the main protocol; train selects 1\,MiB (Table~\ref{tab:size-gate}).
On the held-out H100 test contexts \AUTO{} is exact on all 58 contexts below 1\,MiB, where UCB1 pays only exploration,
and all 19 of \AUTO{}'s contexts above $10\%$ are at 4\,MiB or more.
The gated policy has a $1.15\%$ mean, $9.5\%$ p99 and one context above $10\%$, against $1.61\%$, $9.9\%$ and one for UCB1
everywhere.
At the scenario level it is better in five scenarios, worse in none and tied in four, so the mean improvement has
$p{=}0.06$ under the sign-flip test.
Against the tuned GBM+cost selector ($1.37\%$, $27.3\%$, seven contexts; Appendix~\ref{app:supervised-tuning}) the
gated policy is not separated on the mean ($p{=}0.41$) and has six fewer contexts above $10\%$ ($p{=}0.06$); against
\AUTO{} it is better on both ($p{=}0.016$).
Test would have preferred 4\,MiB ($1.03\%$); we report the train choice.

The gate does not transfer.
On GB200 \AUTO{} is not exact at small sizes (a mean of $3.2$ and $2.1\%$, worst contexts $56$ and $72\%$, below 1\,MiB on
MNNVL off and on), so the frozen H100 threshold raises UCB1's p99 from $23.1$ to $46.2\%$ and from $15.2$ to $31.3\%$.
Chosen on each fabric's own train contexts, as the frozen-hyperparameter protocol otherwise forbids, the threshold falls to
64 and 256\,KiB and the gate changes little ($2.50\%$ and $26.3\%$ against $2.59\%$ and $23.1\%$ on MNNVL off; $3.01\%$
and $15.2\%$ against $3.54\%$ and $15.2\%$ on MNNVL on).
A deployment would therefore need the threshold from a sweep of the target platform, at the scale that is available.
That the gate removes the plugin cost is a design property, not a measurement: NCCL still calls the profiler for gated
collectives, the plugin has to return without timing or counting them, and no live run with a gate exists.

\begin{table}[t]
  \centering
  \scriptsize
  \caption{Size gate: \AUTO{} below the threshold, the UCB1 champion at or above it. \capsplit{} \capTRLT{} ``train sel.''\ is
    the mean selection-priced train \TRLT{} used to choose the threshold ($\star$: chosen); test columns are realized
    \TRLT{} (\%): mean, p99 (99th percentile over contexts) and $n_{>10}$, contexts above 10\%. GB200 rows replay the
    frozen H100 hyperparameters; ``own-train gate'' chooses the threshold on that fabric's train contexts. Bold: best per
    column within each block of policies (the threshold grid is not bolded).}
  \label{tab:size-gate}
  \begin{tabular}{@{}lrrrr@{}}
\toprule
Gate threshold & train sel.\ mean & test mean & test p99 & $n_{>10}$ \\
\midrule
none & 0.86 & 1.61 & 9.9 & 1 \\
64\,KiB & 0.82 & 1.46 & 9.9 & 1 \\
256\,KiB & 0.80 & 1.35 & 9.9 & 1 \\
1\,MiB$^\star$ & 0.63 & 1.15 & 9.5 & 1 \\
4\,MiB & 1.02 & 1.03 & 9.5 & 1 \\
16\,MiB & 1.29 & 1.36 & 20.8 & 5 \\
32\,MiB & 1.84 & 1.64 & 23.3 & 6 \\
\midrule
\multicolumn{5}{@{}l}{\emph{Held-out test, H100 (gate at the starred threshold)}} \\
NCCL \AUTO{} & & 6.08 & 103.3 & 19 \\
GBM$+$cost (tuned) & & 1.37 & 27.3 & 7 \\
UCB1 & & 1.61 & 9.9 & \textbf{1} \\
UCB1, size-gated & & \textbf{1.15} & \textbf{9.5} & \textbf{1} \\
\midrule
\multicolumn{5}{@{}l}{\emph{Held-out test, GB200 MNNVL off}} \\
NCCL \AUTO{} & & 3.80 & 60.1 & 16 \\
UCB1 & & 2.59 & \textbf{23.1} & \textbf{8} \\
UCB1, H100 gate (1\,MiB) & & 2.53 & 46.2 & 10 \\
UCB1, own-train gate (64\,KiB) & & \textbf{2.50} & 26.3 & \textbf{8} \\
\midrule
\multicolumn{5}{@{}l}{\emph{Held-out test, GB200 MNNVL on}} \\
NCCL \AUTO{} & & \textbf{2.65} & 56.3 & \textbf{10} \\
UCB1 & & 3.54 & \textbf{15.2} & 13 \\
UCB1, H100 gate (1\,MiB) & & 3.24 & 31.3 & 12 \\
UCB1, own-train gate (256\,KiB) & & 3.01 & \textbf{15.2} & 12 \\
\bottomrule
\end{tabular}

\end{table}

\subsection{A labelled target-scale sweep for the static selectors}
\label{app:label-budget}
The remaining argument for online selection is that it needs no labelled sweep at the deployment scale, while the static
selectors are fit on 4/8 nodes and extrapolate.
Table~\ref{tab:label-budget} gives the static selectors a labelled mini-sweep at the target scale and scores them on the
target-scale contexts that were not labelled, against the frozen UCB1 champion on the same contexts.
The mini-sweep labels every $j$-th message size of each 16/32-node scenario ($j{=}8,4,2$), every measured arm, with $k$
post-warm-up samples ($k{=}1,10,89$; iterations $11..10{+}k$, what a sweep with warm-up would record), averaged over the $j$
offsets of the labelled sizes; the models, features and fit are those of the harness, refit on train plus labels.
Without labels the script reproduces the harness's GBM+cost choices on the 16/32-node contexts.
GBM$+$cost uses the labels well.
With one sample per arm at every fourth size it beats UCB1's mean on H100 and GB200 MNNVL-on; on MNNVL-off it matches
UCB1 only with every second size labelled with all 89 samples ($2.31\%$ against $2.63\%$).
Its p99 stays heavier than UCB1's at every budget except on H100 with every second size labelled ($8\%$ against
$10\%$).
Three caveats: only these strides and budgets were tried, the static models use fixed settings, and the mini-sweep's own cost (a
warm-up per arm, including the NVLS setup stall, at every labelled size) is not charged.
The bandit's \TRLT{} charges nothing for labels either, since its exploration is part of its served steps.

\begin{table*}[t]
  \centering
  \scriptsize
  \caption{The supervised GBM+cost selector refit on the 4/8-node train contexts plus a labelled target-scale mini-sweep,
    against the frozen UCB1 champion (no labels), both scored on the unlabelled 16/32-node contexts. Cells: realized \TRLT{}
    mean / p99 (\%; \TRLT{} is the latency served over $T{=}100$ steps in excess of the measured-best arm's, relative to
    it; p99 over contexts). ``$1/j$, $k$'' labels every $j$-th size of each target-scale scenario, all measured arms, $k$
    post-warm-up samples each; values average the $j$ offsets. The cell counts give labelled against all target-scale
    (context, arm) cells. Bold: the better of the two policies per platform and column, for mean and p99 separately.}
  \label{tab:label-budget}
  \begin{tabular}{@{}llrrrrr@{}}
\toprule
& target-scale labels (sizes, samples per arm) & none & 1/8, $k{=}1$ & 1/4, $k{=}1$ & 1/2, $k{=}1$ & 1/2, $k{=}89$ \\
\midrule
H100 & labelled (context, arm) cells & 0/590 & 74/590 & 148/590 & 295/590 & 295/590 \\
 & GBM$+$cost & 1.87 / 33 & \textbf{1.25} / 28 & \textbf{1.01} / 25 & \textbf{0.67} / \textbf{8} & \textbf{0.68} / \textbf{8} \\
 & UCB1 (no labels) & \textbf{1.68} / \textbf{10} & 1.68 / \textbf{10} & 1.68 / \textbf{10} & 1.68 / 10 & 1.68 / 10 \\
\midrule
GB200 MNNVL-off & labelled (context, arm) cells & 0/672 & 84/672 & 168/672 & 336/672 & 336/672 \\
 & GBM$+$cost & 4.80 / 83 & 3.78 / 49 & 3.40 / 39 & 2.96 / 35 & \textbf{2.31} / 28 \\
 & UCB1 (no labels) & \textbf{2.63} / \textbf{25} & \textbf{2.64} / \textbf{27} & \textbf{2.64} / \textbf{28} & \textbf{2.63} / \textbf{26} & 2.63 / \textbf{26} \\
\midrule
GB200 MNNVL-on & labelled (context, arm) cells & 0/602 & 75/602 & 150/602 & 301/602 & 301/602 \\
 & GBM$+$cost & \textbf{3.02} / 57 & \textbf{2.98} / 40 & \textbf{2.45} / 30 & \textbf{2.30} / 33 & \textbf{2.32} / 33 \\
 & UCB1 (no labels) & 3.59 / \textbf{15} & 3.59 / \textbf{15} & 3.59 / \textbf{15} & 3.58 / \textbf{19} & 3.58 / \textbf{19} \\
\bottomrule
\end{tabular}

\end{table*}

\begin{table}[t]
  \centering
  \scriptsize
  \caption{Constant-gain means under the drift replay (dynamic \TRLT{} mean, p99 in parentheses, \%; 154 test contexts,
    10 seeds, $T{=}100$, NVLS offered). Gains are a fixed grid, not train-selected. \capkey{} \capbold{}}
  \label{tab:ewma-drift}
  \begin{tabular}{@{}lrrr@{}}
\toprule
Policy & stationary & abrupt & gradual \\
\midrule
\AUTO{} & 5.99 (131.7) & 45.01 (200.4) & 44.60 (199.8) \\
UCB1 champion & 0.87 (6.5) & 6.60 (34.4) & 10.93 (47.7) \\
UCB1, $\gamma{=}0.995$ & 1.05 (6.3) & 5.88 (30.6) & 10.68 (46.8) \\
UCB1, gain 0.1 & 1.20 (7.9) & 4.38 (15.5) & 9.35 (\textbf{37.4}) \\
UCB1, gain 0.2 & 0.96 (6.5) & \textbf{2.92} (\textbf{9.6}) & \textbf{9.13} (41.3) \\
Softmax champion & \textbf{0.75} (9.4) & 31.57 (134.6) & 31.99 (139.4) \\
Softmax, $\gamma{=}0.85$ & 0.84 (\textbf{6.2}) & 5.96 (19.1) & 13.11 (77.5) \\
Softmax, gain 0.1 & 1.13 (16.0) & 8.10 (24.8) & 11.15 (39.7) \\
Softmax, gain 0.2 & 0.86 (9.0) & 4.97 (22.1) & 9.79 (39.2) \\
\bottomrule
\end{tabular}

\end{table}

\subsection{Tuning on the realized metric}
\label{app:realized-tuned}
The main study tunes on selection-priced train \TRLT{} but reports realized \TRLT{}.
We repeated the search for five families with the realized train \TRLT{} as the objective (same
split, harness, sampler seed, search spaces, and 151 trials), replayed the frozen winners once on
test, and report them beside the main champions in Table~\ref{tab:realized-tuned}.
The three leaders are not separated by the change: UCB1 stays at $1.62\%$ mean (p99 $11.4\%$,
three contexts above $10\%$ against one), Softmax moves to $1.63\%$ ($19.6\%$), Gaussian Thompson
improves ($2.17\%\to1.80\%$) and KL-UCB is flat ($2.11\%\to2.20\%$).
The realized search did not beat the main champions on train realized \TRLT{} for Softmax and
KL-UCB ($1.57\%$ against $1.43\%$, $1.68\%$ against $1.35\%$).
This shows that the 151-trial TPE search is noisy, not that either objective is better.
Every realized-tuned winner is warm.
ETC is unchanged ($m{=}1$ with the first sample discarded).
Because the leaders stay within noise, the featured-policy choice does not depend on the tuning
metric, but UCB1's one-context tail (Table~\ref{tab:realized}) is partly a property of the
selection-priced champion we chose.

\begin{table}[t]
  \centering
  \scriptsize
  \setlength{\tabcolsep}{3pt}
  \caption{Champions tuned on selection-priced train \TRLT{} (main study) and on realized train
    \TRLT{}: train realized mean, held-out test realized mean and p99, and contexts above $10\%$ (\%). \capkey{} \capbold{}}
  \label{tab:realized-tuned}
  \resizebox{\columnwidth}{!}{\begin{tabular}{@{}lrrrrrrrr@{}}
\toprule
& \multicolumn{4}{c}{Tuned on selection-priced train \TRLT{} (main)} & \multicolumn{4}{c}{Tuned on realized train \TRLT{}} \\
\cmidrule(lr){2-5}\cmidrule(lr){6-9}
Family & train & test & p99 & $n_{>10}$ & train & test & p99 & $n_{>10}$ \\
\midrule
UCB1 & 1.39 & \textbf{1.61} & \textbf{9.9} & \textbf{1} & \textbf{1.17} & \textbf{1.62} & \textbf{11.4} & \textbf{3} \\
Softmax & 1.43 & 1.75 & 11.9 & 5 & 1.57 & 1.63 & 19.6 & 6 \\
Gaussian Thompson & 1.66 & 2.17 & 17.8 & 11 & 1.57 & 1.80 & 16.3 & 5 \\
KL-UCB & \textbf{1.35} & 2.11 & 22.7 & 4 & 1.68 & 2.20 & 13.0 & \textbf{3} \\
ETC & 4.16 & 4.74 & 63.0 & 11 & 4.15 & 4.74 & 63.0 & 11 \\
\bottomrule
\end{tabular}
}
\end{table}

\subsection{Margin-gated hybrid}
\label{app:margin-gate}

\begin{figure}[!htbp]
  \centering
  \includegraphics[width=\linewidth]{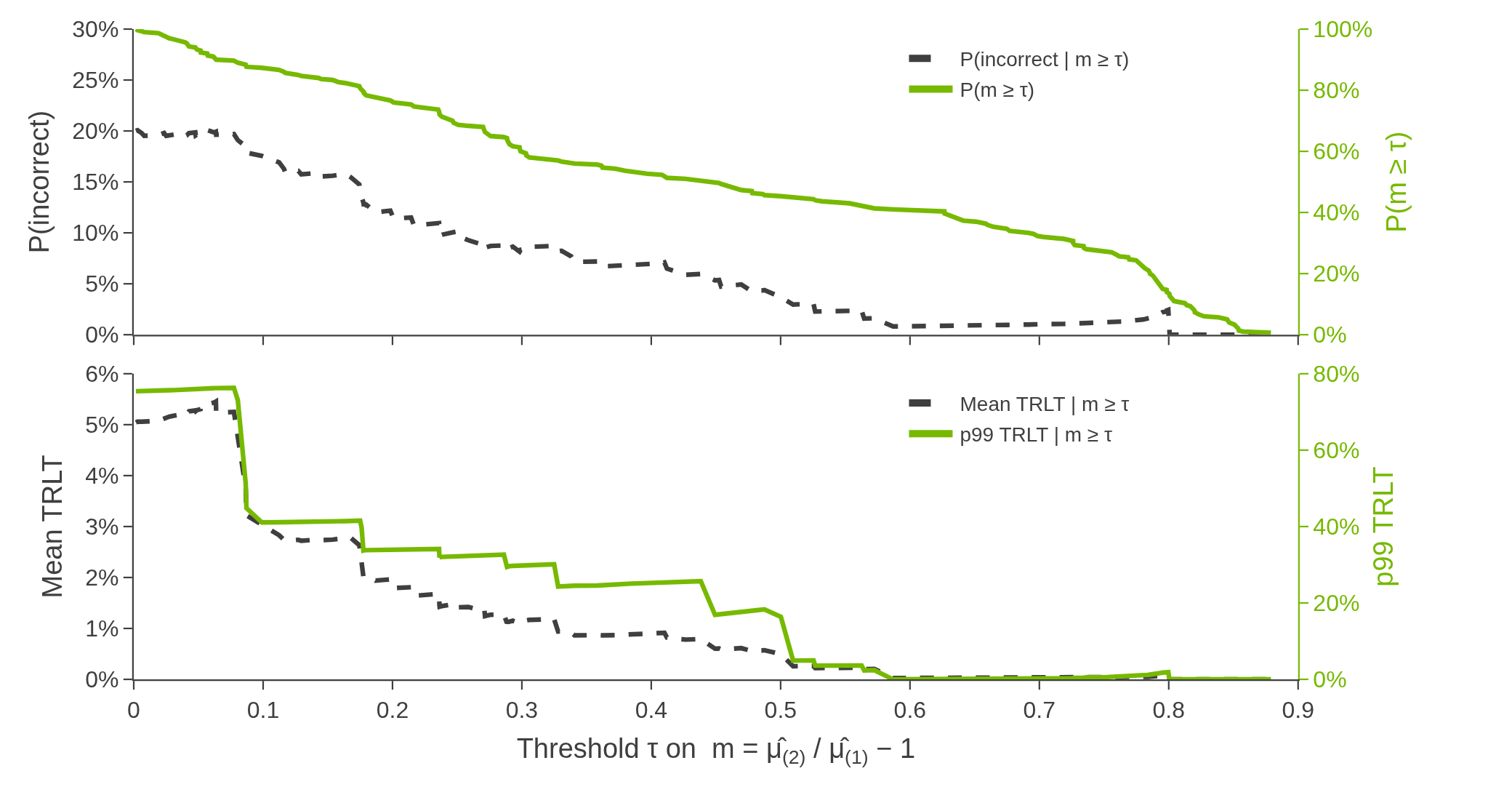}
  \caption{\AUTO{} confidence audit with the observable runner-up margin $m$. Threshold
    $\tau$ retains contexts with $m\ge\tau$. \textbf{Top:} conditional mis-ranking
    probability (dashed, left) and retained fraction $P(m\ge\tau)$ (green, right).
    \textbf{Bottom:} mean (dashed) and p99 (green) \AUTO{} \TRLT{} over the retained
    contexts. Over all contexts the error rate is $20.0\%$, the mean \TRLT{}
    $5.04\%$, and the p99 $75.4\%$; at $\tau\simeq0.50$ the retained $45.3\%$ have
    $3.7\%$, $0.49\%$, and $16.4\%$. H100 sweep, $n{=}300$ contexts, measured-best on
    iterations 11--99; selection-priced.}
  \label{fig:cost-model-margin-error-rate}
\end{figure}

Figure~\ref{fig:cost-model-margin-error-rate} suggests a cheap policy: trust \AUTO{} when
its predicted runner-up margin is at least $\tau$ and run the bandit only on the rest.
Because bandits are independent per context, the hybrid's per-context result is exactly
\AUTO{}'s or UCB1's, so Table~\ref{tab:margin-gate} evaluates it post hoc.
Choosing $\tau$ on the train contexts gives $\tau{=}0.5$, which hands $52\%$ of test contexts
to the bandit and reaches $1.43\%$ mean but a $17.9\%$ p99 and three contexts above $10\%$,
a tail between UCB1's and \AUTO{}'s.
Margin alone does not separate the contexts \AUTO{} gets wrong, since its catastrophic contexts have margins up to $0.57$.
A larger $\tau{=}0.75$, which the train contexts did not select, matches UCB1 ($1.40\%$,
$9.9\%$, one context above $10\%$) while handing $72\%$ of test contexts to the bandit; this
is a post-hoc observation, not a recommended threshold.
A gated ETC does not help (best train $\tau{=}0.3$: $3.09\%$ mean, $63.0\%$ p99).

\begin{table}[t]
  \centering
  \scriptsize
  \caption{Hybrid that plays \AUTO{} when its predicted runner-up margin $m{=}\hat\mu_{(2)}/\hat\mu_{(1)}-1$ is at least
    $\tau$ and UCB1 otherwise, H100 sweep, \capsplit{} ``to bandit'': share of test contexts handed to UCB1. \capTRLT{}
    Realized \TRLT{} (\%, latencies actually served): mean, p99 (99th percentile over contexts), max, and $n_{>10}$,
    contexts above 10\%. $^{\ast}$: $\tau$ selected on train+eval. \capbold{}}
  \label{tab:margin-gate}
  \begin{tabular}{@{}rrrrrrr@{}}
\toprule
& & \multicolumn{1}{c}{train} & \multicolumn{4}{c}{test (154 contexts)} \\
\cmidrule(lr){3-3}\cmidrule(lr){4-7}
$\tau$ & to bandit & mean & mean & p99 & max & $n_{>10}$ \\
\midrule
$0$ (\AUTO{}) & 0\% & 3.98 & 6.08 & 103.3 & 134 & 19 \\
0.1 & 13\% & 2.44 & 2.83 & 41.4 & 50 & 13 \\
0.2 & 24\% & 1.57 & 2.01 & 37.6 & 50 & 7 \\
0.3 & 40\% & 1.08 & 1.59 & 27.6 & 43 & 4 \\
0.5$^{\ast}$ & 52\% & \textbf{1.01} & 1.43 & 17.9 & 33 & 3 \\
0.75 & 72\% & 1.22 & \textbf{1.40} & \textbf{9.9} & \textbf{13} & \textbf{1} \\
$1.0$ (UCB1) & 100\% & 1.39 & 1.61 & \textbf{9.9} & \textbf{13} & \textbf{1} \\
\bottomrule
\end{tabular}

\end{table}

\subsection{Full realized ranking}
\label{app:realized-full}
Table~\ref{tab:realized-full} lists every policy on realized \TRLT{}, with the train mean, the
p99 interval, and exceedance counts.
The train-realized means rank KL-UCB ($1.35\%$), UCB1 ($1.39\%$), and Softmax ($1.43\%$) as the
selection-priced train objective did, within noise; for the supervised rows the
train column is an in-sample fit.

\begin{table}[t]
  \centering
  \scriptsize
  \setlength{\tabcolsep}{2.5pt}
  \caption{All policies on realized \TRLT{} (\%), H100 sweep, \capsplit{} \capTRLT{} Realized \TRLT{} charges the
    latencies actually served. Columns: train (train+eval) mean, an in-sample fit for supervised rows; test mean $\pm$
    \SEM{} over contexts; test p99 (99th percentile over contexts) with its 95\% bootstrap interval; $n_{>10}$ and
    $n_{>50}$, the contexts above 10\% and 50\%. Supervised rows without ``+cost'' lack the cost-model prediction as a
    feature. Rows sorted by test mean; bold: best per test column (Random excluded).}
  \label{tab:realized-full}
  \resizebox{\columnwidth}{!}{\begin{tabular}{@{}lrrrrrr@{}}
\toprule
& train & \multicolumn{3}{c}{test, realized \TRLT{} (\%)} & & \\
\cmidrule(lr){3-5}
Policy & mean & mean $\pm$ \SEM{} & p99 & p99 95\% CI & $n_{>10}$ & $n_{>50}$ \\
\midrule
Supervised GBM+cost, tuned & -- & \textbf{1.37}$\pm$0.43 & 27.3 & [12.7, 40.1] & 7 & \textbf{0} \\
UCB1 & 1.39 & 1.61$\pm$0.19 & \textbf{9.9} & [7.6, 13.5] & \textbf{1} & \textbf{0} \\
Softmax & 1.43 & 1.75$\pm$0.22 & 11.9 & [7.5, 22.2] & 5 & \textbf{0} \\
Supervised GBM+cost & 0.04 & 1.78$\pm$0.50 & 32.2 & [18.3, 40.1] & 10 & \textbf{0} \\
KL-UCB & 1.35 & 2.11$\pm$0.36 & 22.7 & [8.1, 38.5] & 4 & \textbf{0} \\
$\varepsilon$-greedy & 1.42 & 2.12$\pm$0.46 & 23.3 & [13.1, 53.4] & 7 & 1 \\
Supervised linear+cost & 2.96 & 2.14$\pm$0.49 & 29.3 & [18.0, 42.8] & 8 & \textbf{0} \\
Gaussian Thompson & 1.66 & 2.17$\pm$0.34 & 17.8 & [15.1, 22.9] & 11 & \textbf{0} \\
Supervised random forest+cost & 0.29 & 3.97$\pm$1.44 & 103.3 & [27.3, 134.2] & 9 & 4 \\
Hierarchical EXP3/Softmax & 4.97 & 4.60$\pm$0.72 & 42.4 & [16.8, 72.5] & 21 & 2 \\
Madan RR-ETC & 4.15 & 4.74$\pm$0.75 & 63.0 & [15.3, 71.2] & 11 & 3 \\
ETC & 4.16 & 4.74$\pm$0.75 & 63.0 & [15.3, 71.2] & 11 & 3 \\
Supervised GBM & 0.19 & 4.76$\pm$1.36 & 76.7 & [39.7, 129.8] & 18 & 5 \\
NCCL \AUTO{} & 3.98 & 6.08$\pm$1.58 & 103.3 & [46.3, 134.2] & 19 & 6 \\
UCB-tuned & 6.32 & 6.35$\pm$0.54 & 30.1 & [14.4, 66.8] & 16 & 1 \\
Supervised random forest & 0.17 & 6.85$\pm$1.87 & 102.5 & [67.9, 184.5] & 21 & 8 \\
UCB-V & 9.54 & 8.86$\pm$0.77 & 49.8 & [22.2, 77.9] & 41 & 2 \\
Tsallis-Inf & 9.72 & 9.77$\pm$0.86 & 57.4 & [27.4, 82.6] & 37 & 2 \\
EXP3 & 9.99 & 10.57$\pm$0.89 & 57.6 & [27.4, 81.3] & 55 & 2 \\
FPL & 11.27 & 12.05$\pm$1.04 & 66.3 & [37.3, 83.8] & 72 & 3 \\
EXP3++ & 28.92 & 29.58$\pm$2.01 & 101.9 & [80.6, 104.1] & 122 & 34 \\
Random & 42.30 & 46.85$\pm$3.79 & 190.3 & [145.7, 197.3] & 125 & 53 \\
Supervised linear & 77.54 & 65.35$\pm$12.30 & 596.9 & [549.9, 705.0] & 43 & 33 \\
\bottomrule
\end{tabular}
}
\end{table}

\clearpage
\makeatletter
\immediate\write\@auxout{\string\newlabel{LastPage}{{}{\the\numexpr\value{page}-1\relax}{}{}{}}}
\makeatother
\end{document}